\documentclass{aa}  
\usepackage{natbib}
\bibpunct{(}{)}{;}{a}{}{,}
\usepackage{graphicx}
\usepackage{caption}
\usepackage[skip=0.5ex]{subcaption}

\usepackage{txfonts}
\usepackage{lscape}

\usepackage{siunitx}

\newcommand{\OIII}{O\,{\footnotesize \textit{\rm III}}}
\newcommand{\NaI}{Na\,{\footnotesize \textit{\rm I}}}
\newcommand{\degree}{\ensuremath{^\circ}}
\newcommand{\ocl}{\textit{OCL}}
\newcommand{\pv}{\textit{pv}}
\newcommand{\lwm}{\textit{lwm}}

\newcommand{\kms}{km~s$^{-1}$}

\usepackage{amsmath}
\usepackage{enumitem}
\usepackage{times}
\usepackage{xcolor}

\usepackage{soul}

\usepackage{natbib,twoopt}
\usepackage[breaklinks=true]{hyperref} 
\bibpunct{(}{)}{;}{a}{}{,}             

 \usepackage{hyperref}

\begin{document} 

   \title{Frequency and nature of central molecular outflows in nearby star-forming disk galaxies}

   \author{Sophia K. Stuber
          \inst{\ref{i1}}
          \and
          Toshiki Saito\inst{\ref{i1}}
          \and 
          Eva Schinnerer\inst{\ref{i1}}
          \and 
          Eric Emsellem\inst{\ref{i2},\ref{i3}}
          \and 
          Miguel Querejeta\inst{\ref{i4}}
          \and 
          Thomas G. Williams\inst{\ref{i1}}
          \and 
          Ashley~T.~Barnes\inst{\ref{i5}}
          \and 
          Frank Bigiel\inst{\ref{i5}}
          \and 
          Guillermo Blanc\inst{\ref{i6}, \ref{i7}}
          \and 
          Daniel~A.~Dale\inst{\ref{i8}}
          \and 
          Kathryn Grasha\inst{\ref{i9}}
          \and 
          Ralf Klessen\inst{\ref{i10}}
          \and 
          J.~M.~Diederik Kruijssen\inst{\ref{i11}}
          \and 
          Adam~K.~Leroy\inst{\ref{i12},\ref{i13}}
          \and 
          Sharon Meidt\inst{\ref{i14}}
          \and 
          Hsi-An Pan\inst{\ref{i1}}
          \and 
          Erik Rosolowsky\inst{\ref{i15}}
          \and 
          Andreas Schruba\inst{\ref{i16}}
          \and 
          Jiayi Sun\inst{\ref{i12}}
          \and 
          Antonio Usero\inst{\ref{i4}}
          }

   \institute{
            Max-Planck-Institut für Astronomie, Königstuhl 17, 69117 Heidelberg Germany\label{i1}\email{stuber@mpia.de}
        \and 
            European Southern Observatory, Karl-Schwarzschild st 2\label{i2}
        \and 
            Univ Lyon, Univ Lyon1, ENS de Lyon, CNRS, Centre de Recherche Astrophysique de Lyon UMR5574, F-69230 Saint-Genis-Laval France\label{i3}
        \and 
            Observatorio Astron{\'o}mico Nacional (IGN), C/Alfonso XII 3, Madrid E-28014, Spain\label{i4}
        \and 
            Argelander-Institut für Astronomie, Universität Bonn, Auf dem Hügel 71, 53121 Bonn, Germany\label{i5}
        \and 
            The Observatories of the Carnegie Institution for Science, 813 Santa Barbara Street, Pasadena, CA 91101, USA\label{i6}
        \and 
            Departamento de Astronomía, Universidad de Chile, Casilla 36-D, Santiago, Chile\label{i7}
        \and 
            Department of Physics \& Astronomy, University of Wyoming, Laramie, WY 82071\label{i8}
        \and
            Research School of Astronomy and Astrophysics, Australian National University, Canberra, ACT 2611, Australia\label{i9}
        \and 
            Universit\"{a}t Heidelberg, Zentrum f\"{u}r Astronomie, Albert-Ueberle-Str. 2, 69120 Heidelberg, Germany\label{i10}
        \and 
            Astronomisches Rechen-Institut, Zentrum f\" ur Astronomie der Universit\"at Heidelberg, M\"onchhofstra\ss e 12-14, D-69120 Heidelberg, Germany\label{i11}
        \and 
            Department of Astronomy, The Ohio State University, 140 West 18th Avenue, Columbus, Ohio 43210, USA\label{i12}
        \and 
            Center for Cosmology and Astroparticle Physics, 191 West Woodruff Avenue, Columbus, OH 43210, USA\label{i13}
        \and 
            Sterrenkundig Observatorium, Universiteit Gent, Krijgslaan 281 S9, B-9000 Gent, Belgium\label{i14}
        \and 
            4-183 CCIS, University of Alberta, Edmonton, Alberta, Canada\label{i15}
        \and 
            Max-Planck-Institut f{\"u}r Extraterrestrische Physik, Giessenbachstra{\ss}e 1, D-85748 Garching bei M{\"u}nchen, Germany\label{i16}
             }

   \date{Received 15 April 2021; accepted 9 July 2021}

  \abstract{
    Central molecular outflows in spiral galaxies are assumed to modulate their host galaxy's star formation rate by removing gas from the inner region of the galaxy.  Outflows consisting of different gas phases appear to be a common feature in local galaxies, yet, little is known about the frequency of molecular outflows in main sequence galaxies in the nearby universe.
    We develop a rigorous set of selection criteria, which allow the reliable identification of outflows in large samples of galaxies. Our criteria make use of central spectra, position-velocity diagrams and velocity-integrated intensity maps (line-wing maps).
    We use this method on high-angular resolution \mbox{CO\,(2--1)} observations from the PHANGS-ALMA survey, which provides observations of the molecular gas for a homogeneous sample of 90 nearby main sequence galaxies at a resolution of ${\sim}100\,$pc.
    We find correlations between the assigned outflow confidence and stellar mass or global star formation rate (SFR). We determine the frequency of central molecular outflows to be $25\pm2$\% considering all outflow candidates, or $20\pm2$\% for secure outflows only. Our resulting outflow candidate sample of $16{-}20$ galaxies shows an overall enhanced fraction of active galactic nuclei (AGN) (50\%) and bars (89\%) compared to the full sample (galaxies with AGN: 24\%, with bar: 61\%). We extend the trend between mass outflow rates and SFR known for high outflow rates down to lower values ($\log_{10}{\dot{\rm M}_{\rm out}}\,[\mathrm{M}_\odot~\mathrm{yr}^{-1}]<0$). 
    Mass loading factors are of order unity, indicating that these outflows are not efficient in quenching the SFR in main sequence galaxies.}


    \titlerunning{Frequency of central molecular outflows}
    \authorrunning{S.K. Stuber}
   \maketitle
%

\section{Introduction} \label{sec:intro}


Understanding the mechanisms driving galactic outflows and their role in galaxy evolution is one of the challenges in modern astronomy and astrophysics. 
Recent theoretical works have suggested that galactic outflows are essential for regulating galaxy evolution \citep[e.g., the shape of the galaxy stellar mass function;][]{Scannapieco2004,Silk2012} by quenching and morphologically transforming ``star-forming'' late-type galaxies to ``red and dead'' early-type galaxies \citep[e.g.,][]{DiMatteo2005,Sijacki2007}.
Thus, outflows are thought to be capable of shutting down star formation by heating and dragging away molecular clouds that are ready to form stars (negative feedback; e.g., \citealt{Sturm2011,Fabian2012}).
On the other hand, outflows are also assumed to be able to enhance star formation by shock-driven compression and fragmentation of the surrounding interstellar medium (positive feedback; e.g., \citealt{vanBreugel993,Silk2013,Maiolino2017}). 


Driven by active star formation, active galactic nuclei (AGN) or both, galactic outflows are considered a common feature of these galaxies. 
Galactic outflows driven by AGN activity are considered to play a crucial role in the coevolution of supermassive black holes and their host galaxies \citep[e.g.][]{Magorrian1998,KormendyHo2013,Costa2014}.
Also, they are the primary mechanism that pollute the circum- and intergalactic medium with metals. Therefore, they are strongly related to the chemical enrichment across cosmic history \citep[e.g.,][]{Veilleux2005}.


Galactic outflows are known to consist of multiple gas phases: hot X-ray gas \citep[e.g.,][]{Komossa2003}, warm ionized gas ([\OIII], H$\alpha$; \citealt{Venturi18,LopezCoba2020, Hogarth21}), neutral atomic gas (e.g., the \NaI\ doublet; \citealt{Sato09}), warm and cold molecular gas (OH, CO; \citealt{Feruglio2010,Veilleux2013, Zschaechner2018, Lutz2020}), and dense molecular gas (HCN, HCO$^+$; \citealt{Michiyama2018,Krieger2019}). 
The molecular gas phase may dominate the outflow mass in certain active galaxies as revealed by multiphase observations of nearby galaxies, such as NGC~0253 \citep{Krieger2019}, quasar hosts at redshift $z\sim$2.4 \citep{Carniani2015} or AGN hosts when comparing the molecular to their ionized gas phase \citep{Fluetsch2019}.
This motivates a detailed characterization of the molecular phase of outflows.


Progress is being made with both numerical simulations and observational surveys to further quantify galactic outflows.
As an example, recent cosmological simulations have predicted the statistical properties of galactic outflows in star-forming ``main sequence'' (MS) galaxies at different redshifts and their relation to observable properties of the host galaxies such as star formation rate (SFR) and stellar mass in the context of galaxy evolution \citep[e.g.,][]{Nelson2019}.

Observationally, thanks to advances in recent instrumentation, galactic outflow studies are now reaching statistically significant sample sizes of several tens or even several hundreds of galaxies \citep[e.g.,][]{Veilleux2013,Concas2019,ForsterSchreiber2019,Roberts-Borsani2020}. 
These statistical works now provide typical outflow properties at given galaxy properties and redshift, although many of these lack spatial resolution sufficient to 
separate the non-circular motions that signpost outflowing gas from the rotation-dominated kinematics of the gas settled in the disk, except for very massive outflows with large velocities of several $100$\,\kms.
Therefore, resolved and multi-phase studies are still rare and have been limited to individual galaxies: a few local MS galaxies \citep[e.g.,][]{Combes2013,Lopez2019, Hogarth21}, local starburst galaxies \citep[e.g.,][]{Bolatto2013Nature,Leroy2015}, nearby luminous and ultraluminous infrared galaxies (LIRGs and ULIRGs; e.g., \citealt{Feruglio2010,Cicone2014}), as well as the Fermi Bubbles of the Milky Way \citep[e.g.,][]{Su2010, Bordoloi2017}.
A more detailed overview on galactic outflows and recent studies is provided by \citet{Veilleux2020}.

This implies that our understanding of the kinematic and morphological structures of galactic outflows is largely biased by nearby luminous galaxies.
A major next advance will come in understanding the frequency and effect of nuclear outflows in ``normal'' star-forming galaxies. 
Even if these molecular outflows are not reaching the circumgalactic medium (CGM), they may play a crucial role determining the fate of the starbursts or activity in galaxy centers.
Thus, in order to fully understand the relation between central outflows and their host galaxy's star formation rate, high angular resolution observations are needed to see the interaction of the outflowing gas and its surrounding, as well as a large sample to understand statistical properties.
Therefore, determining these properties for a representative sample of nearby ``normal'' galaxies, like MS galaxies, is a key objective.
In this paper, we make a first step by estimating the frequency of molecular outflows in nearby MS galaxies.


For the purpose of a systematic search for central molecular outflows in nearby main sequence galaxies, $^{12}$CO~($J=2{-}1$) (hereafter \mbox{CO\,(2--1)} line) is suitable, which is a fair tracer of bulk molecular gas.
We use sensitive, high (${\sim}100$\,pc) resolution \mbox{CO\,(2--1)} ALMA data for a large fraction of nearby, face-on, relatively massive MS galaxies obtained by the Physics at High Angular resolution in Nearby GalaxieS (PHANGS) survey\footnote{\url{http://www.phangs.org}} \citep{Leroy20b}. 

The PHANGS-ALMA \mbox{CO\,(2--1)} data products include position-position-velocity 3D data cubes as well as line-integrated 2D moment maps \citep{Leroy20a}. Fully utilizing this 2D and 3D information allows us to identify molecular outflows. 
With these data we can search for outflows in galaxies spanning two orders of magnitude in stellar mass and derive the frequency of outflows based on a representative galaxy sample.
Most important for this study, the sample selection does not contain any prior information about the presence of molecular outflows and nuclear activity, making the PHANGS sample unbiased for studying molecular outflows (Section~\ref{sec:data}).
The PHANGS survey also includes an IFU survey with MUSE for 19 galaxies, which allows possible future studies to compare molecular and ionized gas outflow properties. 

In this paper, we employ three popular methods to identify outflows that have been frequently used in the literature: 1D spectra \citep[e.g.,][]{Feruglio2010}, 2D position-velocity diagrams \citep[e.g.,][]{Oosterloo17}, and channel-selected 2D integrated intensity maps (hereafter line-wing maps; e.g., \citealt{Sakamoto14}). We describe both the strength and weakness of each method based on our evaluation experiences.

This paper is structured as follows: 
In Section~\ref{sec:data}, we briefly describe the PHANGS sample properties and ALMA data.
Section~\ref{sec:methods} summarizes the methods used to identify outflows and our evaluation process. 
We present the results of the evaluation process in Section~\ref{sec:Results}, including the final outflow candidates. 
In Section~\ref{sec:Discussion}, we compare these candidates to the full sample, discuss possible biases and derive outflow rates which are compared to other samples.
Finally, Section~\ref{sec:Summary} provides a brief summary and conclusion of the main results. 

\section{Data and sample} 
\label{sec:data}


The PHANGS-ALMA survey allows for the first time a systematic search for central molecular outflows in a representative sample of nearby galaxies on the star-forming main sequence.
Here we briefly describe the PHANGS-ALMA survey and the characteristics of the data we use, and highlight information related to our molecular outflow study. Detailed descriptions are provided in other papers (\citealt{Leroy20a, Leroy20b}; see also \citealt{Herrera20} and Section~2 of \citealt{Sun20} for a comprehensive summary).

\subsection{PHANGS-ALMA sample}

PHANGS-ALMA is an ALMA Large Program (PI E.~Schinnerer) to map the \mbox{CO\,(2--1)} line in the disks of 90 nearby galaxies when including its pilot and extension projects \citep{Leroy20b}.
This sample comprises close to all nearby ($\mathrm{d} < 24$\,Mpc), massive ($9.3 \leq \log_{10}(\mathrm{M}_{\star}/\mathrm{M}_{\odot}) \leq 11.1$), relatively face-on ($i < 75^{\circ}$), 
star-forming ($\log_{10}(\mathrm{sSFR}/\mathrm{yr}^{-1}) > -11$) galaxies that are observable by ALMA ($-75^{\circ} < \delta < +25^{\circ}$). 
Although PHANGS includes galaxies with high specific SFR, most starburst galaxies are excluded as they are rare in the nearby ($z =0$) universe. For more details on completeness and data selection we refer to \citet{Leroy20b}.

We use 80 (out of a final sample of 90) PHANGS galaxies that were processed and available by September 2020 (i.e., internal data release version~3.4).
An updated data release with improved continuum subtraction is publicly available on the homepage\footnote{\url{http://www.phangs.org}}.
Global properties of the sample galaxies\footnote{The compilation of this information is taken from the internal sample table version~1.6.} are taken from \citet{Leroy20b} with distances from \citet{Anand20} and galaxy geometries from \citet{Lang20} and are listed in Table~\ref{tab:PHANGSPropertyTable}. Galaxies are classified as AGN based on \citet{VeronCatalogue13thedition} and/or barred using the compilation by \citet{Querejeta2021}, largely based on the work of \citet{Herrera-Endoqui2015} and \citet{Menendez-Delmestre2007}.

\subsection{PHANGS-ALMA data}

The PHANGS-ALMA observations used the extended 12m array, compact 7m array, and total power antennas in order to recover emission arising from all spatial scales within nearby galaxies. The data calibration and imaging is generally following standard procedures, we refer the reader for details on data reduction and product generation to \citet{Leroy20a}. 
The delivered data cubes and images have a median native angular resolution of $1.3\arcsec$ translating into a spatial resolution of ${\sim}100$\,pc, a spectral resolution of ${\sim}2.5$\,\kms, and a typical rms sensitivity of ${\sim}85$\,mK at their native resolution \citep{Leroy20b}. This translates into typical molecular gas mass surface densities of $3\sigma \approx 6\,\mathrm{M}_{\odot}\,\mathrm{pc}^{-2}$ per channel (assuming a MW-like \mbox{CO\,(1--0)}-to-H$_2$ conversion factor of $\alpha_{CO} = 4.35 \mathrm{M}_{\odot} \mathrm{pc}^{-2} \left( \mathrm{K\, km \, s}^{-1}\right)^{-1}$ and a \mbox{CO\,(2--1)}-to-\mbox{CO\,(2--1)} line ratio of $\mathrm{R}_{21} = 0.65$ \citep{Bolatto2013Nature,Leroy2013,denBrok2021}.

We degraded the spectral resolution from $2.5$\,\kms\ to $5$\,\kms\ by smoothing the data to increase the signal to noise. 
We do so by applying Hanning smoothing using \texttt{CASA} (version~5.5) task \texttt{specsmooth}\footnote{\url{https://casa.nrao.edu/docs/TaskRef/specsmooth-task.html}} 
using \texttt{dmethod=copy} to avoid channel-to-channel correlation. 

\section{Methods} 
\label{sec:methods}

In this Section, we present our procedure to 
systematically identify central molecular outflows in a large sample of galaxies with similar properties. 
Rather than case studies tailored to individual galaxies \citep[e.g., NGC~0253;][]{Krieger2019}, we present a generic approach to identify galaxies which are likely to possess molecular outflows. 
This allows us to statistically compare properties of outflows (e.g., frequency, outflow masses, and outflow rates) as well as the non-detection rate of the sample in a robust, quantitative way. 
Such comparisons will provide a better understanding of which properties are necessary for galaxies to launch outflows. 

We begin with a short review of previous identification methods in the literature (Section~\ref{sec:methods:literature}), before providing a detailed description of our applied methodology (Section~\ref{sec:methods:thispaper}).

\subsection{Outflow identification in the literature}
\label{sec:methods:literature}

Identifications of outflows in the literature have used both spatial and kinematic information. As the use of spatial information requires 
favorable geometries of the outflowing gas, such as the presence of radio jets with a large and therefore more easily detectable extent of the outflowing gas, or a favorable disk inclination, \citep[for instance, as in the case of M82 studied by e.g.,][]{Walter2002, Leroy2015}, we focus on the kinematic approach.
This approach leverages the fact that
despite being less massive and thus less luminous than the brighter spiral arms, disk or central region, outflowing gas possesses a relatively high outwards velocity component, such that at a given projected position in a galaxy it is kinematically decoupled from the local rotational velocity. 
In order to leave the central disk of the galaxy, outflows must have velocities that are higher than the circular velocities of gas within the galaxy disk, but may not necessarily reach escape velocity of the galaxy halo. 
The outflow candidates we identify in this work are therefore consistent with (central) fountain flows.
Tracing such low-mass gas components with anomalous velocities can be done by using one dimensional information (spectra for different apertures) or also considering spatial information in the form of position-velocity (\pv) diagrams or line-wing maps (\lwm) \citep[e.g.,][]{Cicone2014, Fluetsch2019, Lutz2020}.

Outflows in several galaxies have been found via evidence for broad components in spectra from central regions observed via molecular or ionized gas tracers \citep[e.g.,][]{Veilleux2013, Davies2020, Wylezalek2020}. 
Often, Gaussian fitting is used to distinguish between the single- or double-horned spectral line expected for gas in rotating disks and an underlying broader extended component that indicates an outflow.
However, bar-induced streaming motions, 
as well as fast-rotating central disks can result in a similar broadening of the spectrum \citep[e.g., high velocity dispersions due to bar streaming motions are described in][]{Kormendy1982}.
Also, other non-circular motions (e.g., from spiral arms)
may have a major contribution to the observed spectra, which emphasizes the importance of combining this method with the below listed methods.
Furthermore, the movement of gas perpendicular to the line of sight or absolute projected outflow velocities that are consistently smaller than the velocities of more luminous components in the disk will 
not be visible in the spectrum taken over a central region, nor distinguishable from a spectrum taken over the full galaxy disk. 

Finally, when using only the integrated spectrum, the extent and exact location of the outflow cannot be determined. These quantities are crucial to estimating timescales and mass outflow rates.

Another method to identify the high-velocity components arising from an outflow is via \pv--diagrams along the kinematic major and minor axes. 
Although observations are limited to projected velocity components only (resulting in possible underestimation of the true three dimensional velocities), this method provides a rough estimate of the position and velocity of the outflowing gas.

The third commonly used method to detect outflows is via line-wing maps (\lwm), which is widely used in identifying outflows from protostars \citep[e.g.][]{Dunham2014, Maud2015}.
The data cube is integrated over a previously determined velocity range -- the velocities where the outflows appear distinguishable from the emission of the galaxy -- and then presented as a map of integrated intensity (moment~0). 
This technique provides two advantages: First, the integration of intensity over several selected channels allows for locating the usually fainter gas, and secondly, this integrated intensity is then displayed in a two dimensional map, making it easier to see the outflow as a distinct component.

In this paper we applied the three methods described above to identify outflow signatures. The combination then provides
the best possible inspection of data cubes for identifying outflow candidates.

\subsection{Methodology applied in this paper}
\label{sec:methods:thispaper}

As we expect that most of our identifications would require follow-up observations to truly confirm the presence of an outflow, we will use the term \textit{outflow candidates} instead
of referring to them as outflows.

The finding charts of the three methods for each PHANGS galaxy (examples are shown in Figures~\ref{fig:method:spectraexample}, \ref{fig:methods:pvexample}, and~\ref{fig:methods:linewingmapexample}) were independently evaluated by three authors (S.K.S., T.S., and E.S., hereafter inspector A, B, C) and a ranking was generated, as defined in Table~\ref{tab:method:ranking}, for the possible presence of molecular outflows.
This ranking number represents the \textit{outflow confidence label} (\ocl) and ranges from~0 (no detected velocity signature) to~3 (velocity signature found, interpreted as very likely outflow signature).
An \ocl\ of~3 is assigned when most of the criteria for each method (listed below) are fulfilled. On the other hand, an \ocl\ of~0 is assigned when none of the criteria are fulfilled. \ocl{}s of~1 or~2 correspond to intermediate cases, where only one to two criteria are fulfilled or the velocity signature is weak (for example, if the wings in the spectra are slightly broader than the usual double-horn profile, but are not very pronounced).

We assign an uncertainty of $\pm1$ to each inspector's \ocl\ assignment, as in some cases the inspectors found it difficult to decide between two neighboring \ocl{}, as a result of the evaluation criteria listed below, which allow for some deviations. 
A full list of inspector-averaged \ocl{}s per method and per galaxy are provided in Table~\ref{tab:PHANGSOCLtable} of Appendix~\ref{sec:Appendix:PHANGS_OCL_ALL}.

In the following sections, we describe the evaluation criteria for each of the methods used (spectra in Section~\ref{sec:methods:Spectrum}, \pv--diagrams in Section~\ref{sec:methods:Pv}, and \lwm\ in Section~\ref{sec:methods:lwm}).

\begin{table}
\caption{Outflow confidence labels (\ocl) \label{tab:method:ranking}}
\centering
\begin{small}
\begin{tabular}{cl}
\hline\hline
\noalign{\smallskip}
Ranking& Description\\
\noalign{\smallskip}
\hline
\noalign{\smallskip}
0 & no detected velocity signature, or insufficient S/N\tablefootmark{a}\\
1 & velocity signature, unlikely to be an outflow\\
2 & velocity signature, possibly an outflow\\
3 & velocity signature, (very) likely an outflow\\
\noalign{\smallskip}
\hline
\end{tabular}
\end{small}
\tablefoot{
\tablefoottext{a}{Insufficient signal to noise ratio (S/N) in the sense that, e.g., the wings are not clearly visible in the spectra and therefore a broad component could not be distinguished from noise.}
}
\end{table}

\subsubsection{Spectra}
\label{sec:methods:Spectrum}

\paragraph{Generation of the spectra}

For the 80 galaxies that we consider, we extracted integrated spectra from circular regions (within the plane of the sky) centered on the galaxy's center following previous works \citep[e.g.,][]{Fluetsch2019, Roberts-Borsani2020,Roberts-Borsani2020_ObsConstr}.
The adopted center positions and systemic velocities are taken from \citet{Lang20} as described in \citet{Leroy20b} and listed in Table~\ref{tab:PHANGSPropertyTable} of Appendix~\ref{sec:Appendix:PHANGSproperties}.

Given that outflows are reported to extend between a few $100$\,pc to kpc-scales, we choose two apertures for the integrated spectra. 
One is a circular aperture on the sky with a diameter of $300$\,pc, the other one is an annulus with an inner diameter of $300$\,pc and an outer diameter of $2000$\,pc (in short hereafter ``$300$\,pc spectrum'' and ``$2$\,kpc spectrum'').
This should allow us to detect outflowing gas at the center of a galaxy, as well as to trace fainter outflows at kpc-scales without central emission.  
For comparison, we also extracted spectra integrated over the full disk. These spectra tend to show relatively higher noise levels than the spectra integrated over smaller areas, as we did not apply any masking, that could include additional biases.

We use circular apertures on the sky in this analysis because (1) galaxy inclinations are mostly low by survey selection criteria, thus the actual projected aperture in the galaxy plane will have a small ellipticity, and (2) gas emission is often asymmetrically structured, which will have a larger impact on the analysis than selecting the perfect ellipticity.

\paragraph{Evaluation of the spectra}
\label{sec:methods:Spectrum:Eval}

The criteria for visual evaluation of the integrated spectra are as follows: 

\begin{itemize}[noitemsep,topsep=0pt,leftmargin=2\parindent]
\item \textbf{Broad wings:} We visually identify broad wings in the spectra as clear deviations from the expected shape of either a double-horned spectrum, a single-Gaussian spectrum, or shapes in-between. 
We do not fit the profiles, but even without fitting, all three investigators find \ocl{}s with good agreement (see Section~\ref{sec:Results:OCLs}). 

\item \textbf{Comparison to the spectrum of the whole galaxy:} 
We compare the spectrum of the central region to the spectrum obtained for the whole galaxy as defined above.
If the line wings of the central spectrum exceed the velocity extent of the whole galaxy spectrum at 5\% normalized peak intensity, we take this as evidence for the existence of gas that is moving at enhanced velocities.
We chose a threshold of 5\% of the normalized peak intensity as it traces the spectral wings well, does not directly depend on the noise level, and exceeds the 3$\sigma$ noise level for most galaxies.

\item \textbf{Absolute velocities:} The absolute velocities reached in the line wings are used to identified gas with large velocities as could be generated by outflows. 
We visually estimate the velocity offset between the midpoint at 20\% normalized peak intensity and the 5\% normalized peak intensity of the central spectra. If this offset reaches $\gtrsim 100\,$\kms, this is considered unusual and taken as a hint for the existence of an outflow.
For most lower mass galaxies in our sample we do not expect to find gas moving at 100\,\kms\ or more, as the peak velocities of their (observed) rotation curves typically do not exceed this value \citep{Lang20}. 

\end{itemize}

We assigned an individual \ocl\ for both apertures (see Section~\ref{sec:methods:Spectrum}), as the spatial extent of the central outflow can vary and, thus, outflow signatures might only be visible in the central $300$\,pc spectrum or the $2$\,kpc spectrum, but not necessarily in both spectra.

We experimented with using Gaussian fits for an automated classification (e.g., \citealt{Saito18}), but we found that the analysis of the results varied significantly due to large variations in spectral shapes seen in our sample of galaxies.

\begin{figure*}[t!]
    \centering
    \includegraphics[width = 0.8\textwidth]{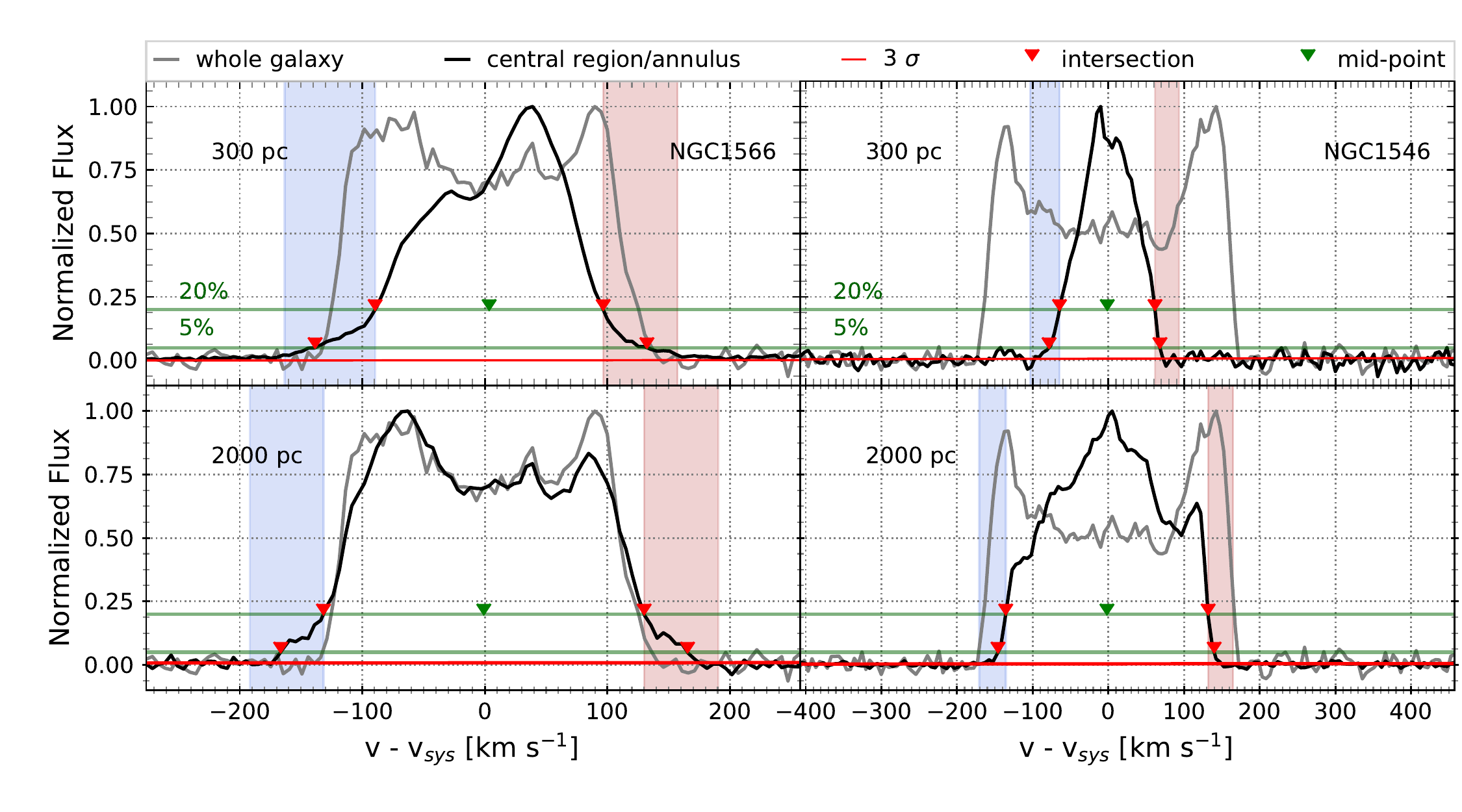}
    \caption{Spectra of NGC~1566 and NGC~1546 used for visual classification of outflow candidates. Integrated and normalized spectra (black solid line) of the central regions, i.e., the inner $300$\,pc in diameter (top panels), and using annuli covering diameter with $300\,{\rm pc} < d_{\rm gal} < 2\,{\rm kpc}$ (bottom panels). For comparison, the integrated spectrum from the whole galaxy normalized to a peak of unity is shown by a gray line.  
    The red solid lines represent the interpolated $3\sigma$ noise level for the central region/annulus.
    The mid-points of the 20\% intersections of the peak (green triangle) are shown for comparison.
    Velocity has the systemic velocity subtracted.
    The transparent blue and red areas represent the velocity ranges over which the blue- and red-shifted line-wing maps (Section~\ref{sec:methods:lwm}) are integrated over. These areas are selected to correspond to the velocity at 20\% of the peak from the central spectrum out to the velocity at 5\% of the peak plus an empirically determined additional $25$\,\kms\ to take into account faint emission at low significance. The green solid lines mark the 20\% and 5\% peak levels, respectively. Intersections with the central spectrum are given as red triangles. 
    The spectra of NGC~1566 received \ocl{}s of 3,\,3,\,2 for the $300$\,pc spectrum and 2,\,2,\,3 for the $2$\,kpc-annulus spectrum from all inspectors due to the pronounced line wings. 
    Both images of NGC~1546 received an \ocl\ of~0 from all inspectors for both apertures as no broad wings are evident.
    }
    \label{fig:method:spectraexample}
\end{figure*}

Two example spectra are shown in Figure~\ref{fig:method:spectraexample}, displaying the effect of different apertures. NGC~1566 (left panels) shows broad wings that exceed the spectrum of the whole galaxy at 5\% normalized peak intensity up to high offset-velocities in the $300$\,pc and $2$\,kpc spectra resulting in high \ocl{}s.
However, this galaxy is not part of our outflow candidates as it lacks outflow signatures in the other methods (see Section~\ref{sec:Results:OCLs}) demonstrating the importance of combining all three methods.
The right panels of the figure display spectra of NGC~1546, where no broad component or other features typical for outflows could be visually identified resulting in \ocl{}s of~0 from all inspectors.

\subsubsection{\textit{Pv}--diagrams}
\label{sec:methods:Pv}

\paragraph{Generation of \pv--diagrams}
\label{sec:methods:Pv:generation}

\begin{figure*}
    \centering
    \textbf{NGC~3627}
    \includegraphics[width = 1\textwidth]{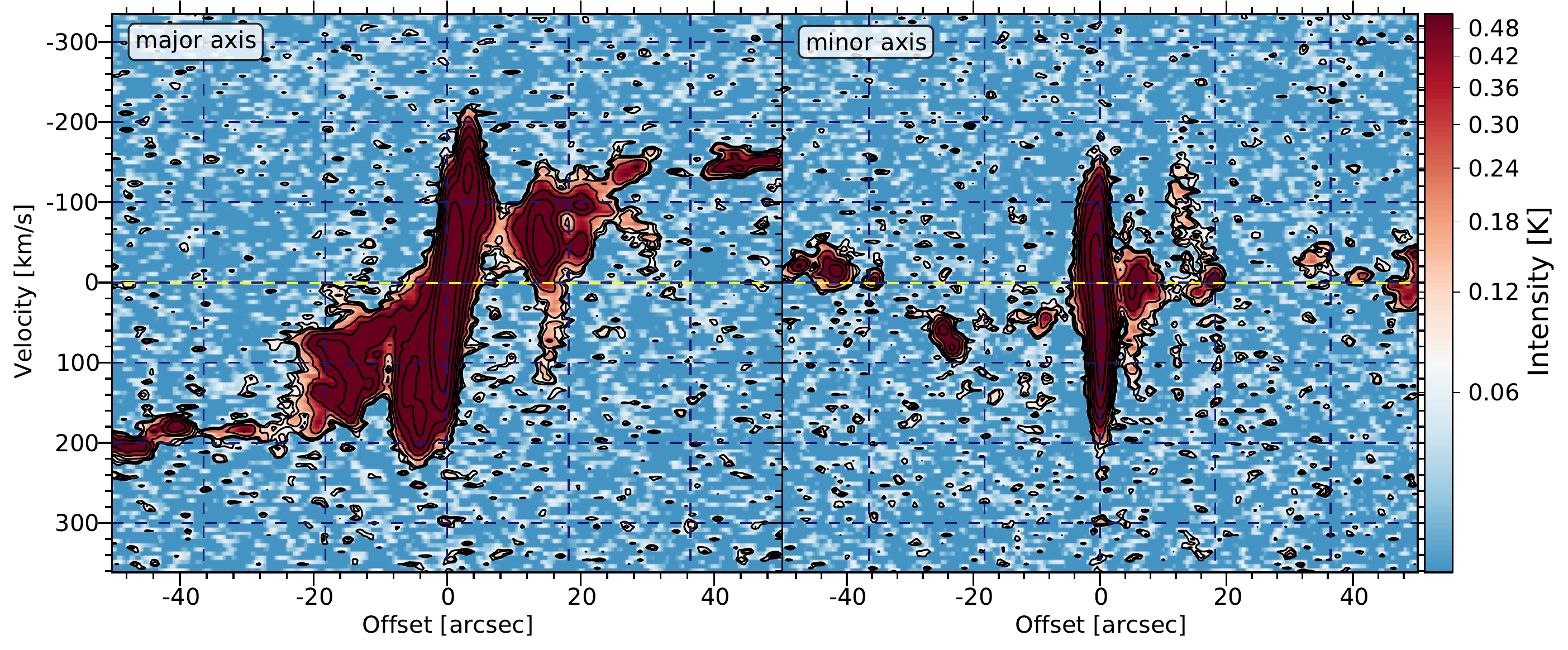}
    \textbf{NGC~4654}
    \includegraphics[width = 1\textwidth]{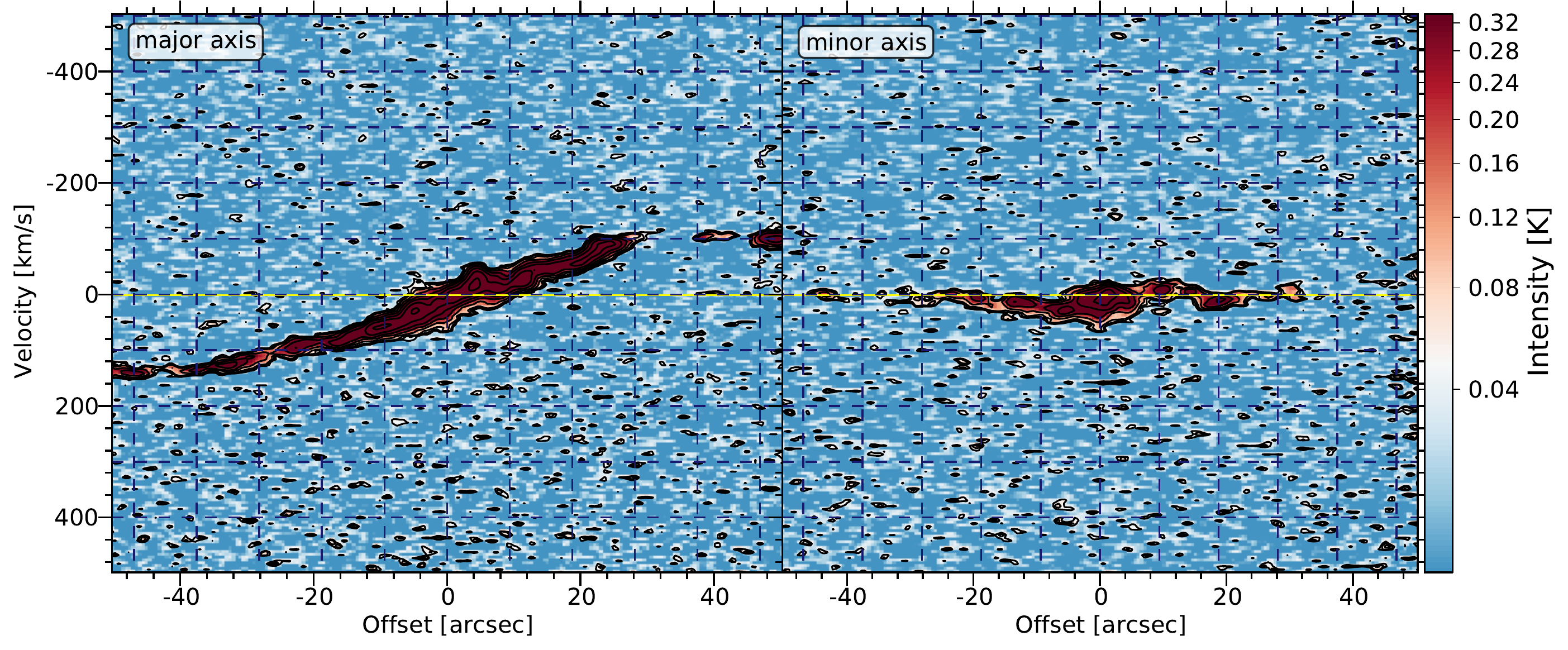}
    \caption{Example \pv--diagrams along the major and minor kinematic axes of NGC~3627 (top) and NGC~4654 (bottom) used to search for outflows. The color scale represents the intensity at a given position along the offset averaged over a width of 3~pixels. The contours are multiples of the noise at $2.5\sigma$, $3.5\sigma$, followed by $n_{i}\sigma$ with $n_{i} = 2n_{i-1}$, for $n_{0} = 2.5\sigma$ and $i = 1,2,3,\dots$. 
    The color range is saturated to emphasize the faint emission we are interested in. 
    The vertical (horizontal) dashed lines are separated in steps of 1\,kpc ($100$\,\kms). 
    The \pv--diagrams of NGC~3627 (NGC~4654) correspond to \ocl{}s of 3,\,2,\,2 (0,\,0,\,0) from the three different inspectors.
    The systemic velocity (yellow horizontal line) is added for comparison and is subtracted from the velocity axis.}
    \label{fig:methods:pvexample}
\end{figure*}

\textit{\pv}--diagrams are created using the \texttt{CASA} tool \textit{impv} along the major and minor kinematic axes of all PHANGS galaxies using the position angles from \citet[][see also Table~\ref{tab:PHANGSPropertyTable} in Appendix~\ref{sec:Appendix:PHANGSproperties}]{Lang20}.
Two examples are given in Figure~\ref{fig:methods:pvexample}. 
The width of the slit should be approximately a beam width, and so we used a width of
3$\times$ the pixel size. 
We set the length of the slit up to $100$\,arcsec (${\sim} 1.5{-}12.5$\,kpc). 
We note that the noise in the data cubes varies mildly as function of frequency and towards the edges of the surveyed area (which can be seen in some \pv--diagrams).
As comparison of the central part to signal further out is most critical for \pv--diagrams, the exact physical length of the slit is less important than for aperture-integrated spectra.

\paragraph{Evaluation of the \pv--diagrams}
\label{sec:methods:Pv:Eval}

Outflowing gas can be identified in a \pv--diagram as a high velocity component at certain spatial locations. 
For a simple rotating disk, we expect a constant velocity (centered at systemic velocity) along the minor kinematic axis and an \textit{\mbox{S-shaped}} rotation curve along the major kinematic axis. 
Deviations could be due to outflowing gas. 
However, galactic structures such as bars, spiral arms, or central disks can also modify the motions of the gas from simple circular rotation, making it difficult to distinguish between actual outflows and deviations due to other kinematic features \citep[e.g., position angle dependence on bar observations;][]{Koda2002kinematicaxis}.
We assigned a single \ocl\ for major and minor axes \pv--diagrams, as for most viewing orientations outflow signatures should be seen along both axes, due to its expected opening angle which would allow for projected velocity components.

We used the following criteria to assign the \ocl{}s listed in Table~\ref{tab:method:ranking} to all PHANGS galaxies based on the evaluation of their \pv--diagrams:

\begin{itemize}[noitemsep,topsep=0pt,leftmargin=2\parindent]
    \item \textbf{Faint emission with high velocity dispersion}: We expect the outflow signature to be a faint component with a large extent in velocity.
    We consider emission at $\gtrsim~100\,$\kms\ from the mean value of the rotation curve at a given location as a strong sign for outflowing gas.
    We consider signals between $40$ and $100$\,\kms\ to be still significant, while components with even lower absolute velocities are not considered.
    \footnote{In some cases, (residual) continuum emission leftover from imaging is evident as a nearly perfect vertical line (along the velocity axis), often crossing through the nucleus. The inspectors were aware of this effect during evaluation. 
    To distinguish between this effect and actual outflow signatures, we required the emission to exhibit at least some spatial variation over the velocity range examined for a positive outflowing identification.}
    
    \item \textbf{Comparing to galactic rotation and other rotation features}: Presence of (faint) emission at anomalous velocities compared to the \mbox{S-shaped} rotation curve (along major axis) or a constant velocity (along minor axis) are telling signs for an outflow.
    Rigid rotation (e.g., from a stellar bar) can be visible as straight lines of strong emission up to high velocities in the center of the galaxy and must be considered.

    \item \textbf{Relevance}: As for this work we are only searching for central outflows, the potential outflow signatures should lie within the central 1\,kpc (in radius). 
\end{itemize}

Two examples are presented in Figure~\ref{fig:methods:pvexample}: The \mbox{S-shaped} rotation curve along the major kinematic axis (left panel) as well as the constant velocity along the minor kinematic axis (right panel) of NGC~3627 (upper) and NGC~4654 (lower) can be seen.
In NGC~3627, multiple features suggest an outflow, such as, faint red-shifted emission at roughly $+16$\,arcsec along the major axis extending almost to ${\sim}150$\,\kms\ and faint red- and blue-shifted emission at $+5$ and $+14$\,arcsec along the minor axis with a velocity extent of ${\sim}100$ and ${\sim}150$\,\kms, respectively, possibly moving against the galaxy's primary sense of rotation.
These velocities are impossible to reach for gas in a purely rotating disk.
The features found in NGC~3627 fulfill most of the criteria listed above making it one of our outflow candidates (see Section~\ref{sec:Results:OCLs}). 
NGC~4654's \pv--diagrams fulfill none of the criteria resulting in \ocl{}s of~0.
We show \pv--diagrams of all outflow candidates in Appendix~\ref{sec:Appendix:AdditionalPlots}.

\subsubsection{Line-wing maps}
\label{sec:methods:lwm}

\paragraph{Generation of \lwm{}s}
\label{sec:methods:lwm:generation}

The \lwm{}s allow us to localize emission at a certain velocity (range) in the galaxy plane. 
For this, we overplotted the emission integrated over a selected velocity range on the moment~0 map (which is integrated over the full velocity axis and has a threshold of $5\sigma$) in order to compare to the overall gas distribution. 
The selection of the velocity range is based on the wings of the central spectra.
Assuming that outflowing gas will be present at the tails of the line wings, we set a lower limit to the offset velocities at the intersection at 20\% of the peak of the central spectrum. 
Lower velocities are considered to originate from galactic rotation.   
We set the upper limit to the offset velocities at the 5\% intersections of the central spectrum, which for most of our data still lies above the noise level. 
To this velocity range we add an additional $25$\,\kms\ 
to include faint emission.
An example of the velocity ranges used for \lwm{}s can be seen in Figure~\ref{fig:method:spectraexample} as blue and red shaded areas, respectively. 

This method is insensitive to outflows with velocities close to the systemic one \citep[e.g. the outflow in the nearby galaxy NGC~253, see ][]{Bolatto2013Nature,Walter2017,Krieger2019}, as the spectra are insensitive to such velocities as well, and thus the selected velocity ranges miss these velocities, but works well for outflows with a different velocity.
This also provides additional information about the spatial location of the outflowing gas contributing to the high-velocity wings.

\begin{figure*}[ht!]
    \centering
    \textbf{NGC~1566 - velocity range from 300\,pc spectrum:}
    \includegraphics[width= 0.9\textwidth]{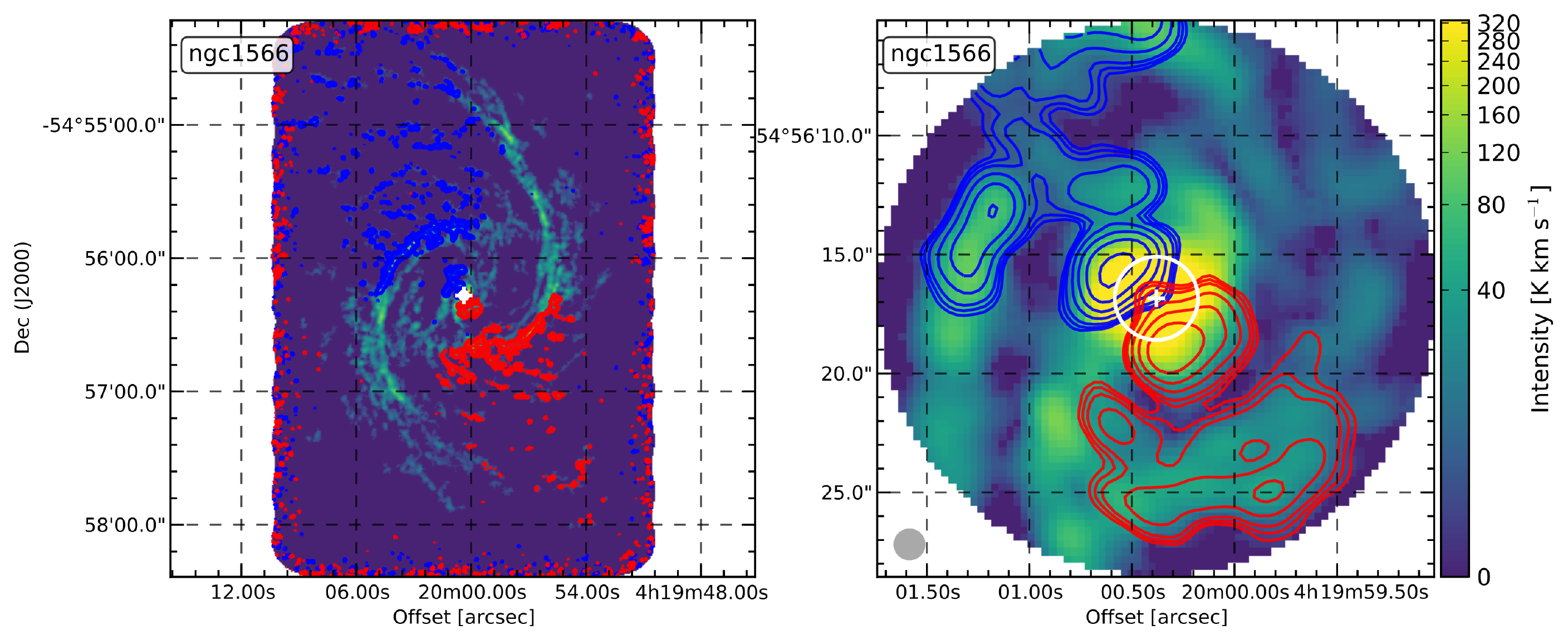}
    
    \textbf{NGC~1566 - velocity range from 2000\,pc spectrum:}
    \includegraphics[width= 0.9\textwidth]{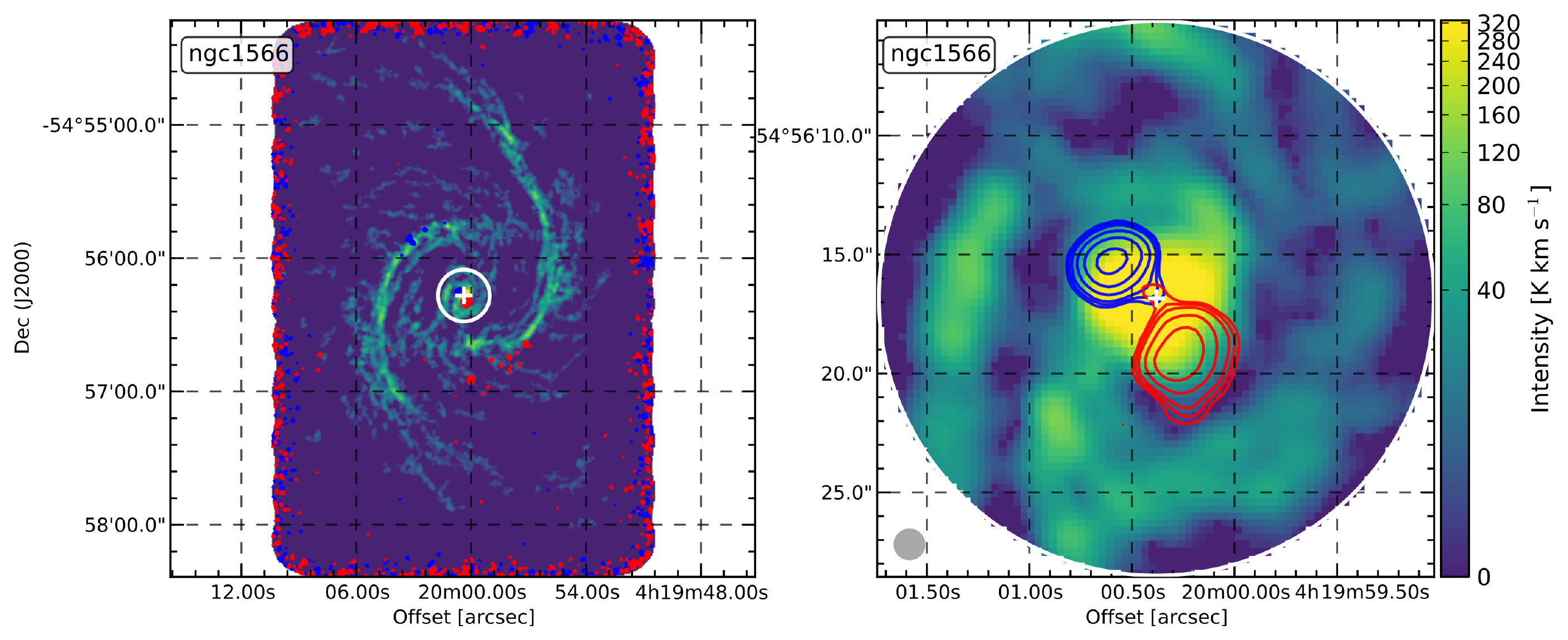}
    \caption{Line-wing maps for NGC~1566. The left panels show the full moment-0 map available with the integrated \lwm{}s (red- and blue-shifted contours, respectively) overplotted. The right panels show a zoom-in of the central region with 2\,kpc diameter. The beam size is shown as a gray circle in the lower left corner of the zoom images. Center positions are indicated with a white plus. The \lwm{}s shown in the upper panels are derived by integrating over the velocity range selected from the 300\,pc spectrum as described in Section~\ref{sec:methods:lwm}, the lower panels use the 2\,kpc annulus spectrum. A white circle indicates the size over which the corresponding spectrum is taken. These 4~panels together received an $\ocl_{\lwm}$ of 1,\,1,\,0 from the three inspectors.   
    The contour levels are selected to be the same as for the \pv--diagrams, with the noise being calculated for the zoom images, not for the full maps. The underestimation of the noise at the edges of the full map (left panels) is evident, however, it does not affect our search for central outflows.}
    \label{fig:methods:linewingmapexample}
\end{figure*}

This procedure is applied to both the 300\,pc and 2\,kpc spectra. 
An example of the resulting two images and their close-ups for one galaxy is presented in Figure~\ref{fig:methods:linewingmapexample}.
We add the integrated emission for the whole map, used to identify whether the high velocity emission is, for instance, something unique in the galaxy or rather corresponds to some other larger feature in the galaxy.
For example, if the emission does not arise from outflowing gas, the high velocity emission simply traces the high-velocity part of the rotation curve and we might see emission in the form of a spider diagram in our map. 
Thus emission that is located on both sides of the center on two positions on a central ring and elongated in oval shapes along the central ring might not trace an outflow. 
Also, molecular gas co-located with a stellar bar may be experiencing bar streaming motions, rather than outflowing. Alternatively, there could be a narrow outflowing structure with a different inclination, but the same position angle as the bar. This second possibility is statistically rare, and so we do not consider this a possibility in our analysis.

\paragraph{Evaluation}
\label{sec:methods:lwm:eval}

We assigned a single \ocl\ based on the \lwm{}s for each galaxy.
As the spatial extent and location of the outflowing gas can vary between a couple and several hundreds of pc,
we do not necessarily expect that outflowing signatures are present in both spectra. Therefore, we also do not necessarily expect that outflowing signatures are present in both \lwm{}s, as we use the velocity ranges from the two spectra. This means that not all of the automatically generated \lwm{}s will cover all of the outflowing gas and due to overlap in velocity range the resulting maps are also not independent.

The criteria for an outflow signature are as follows:

\begin{itemize}[noitemsep,topsep=0pt,leftmargin=2\parindent]

\item \textbf{No overlap with galaxy structures (location of emission)}: Comparing the location of the blue- and red-shifted emission to the overall galaxy structure such as inner ring, gas lanes in bars, and spiral arms allows us to exclude high velocity emission arising from non-circular motions within these galactic structures. For example, bars can induce strong shocks in their gas lanes which might be mistaken as an outflow.

\item \textbf{Comparing with overall rotation (velocity of emission)}: 
If the high velocity emission lies at ``unexpected'' velocities, such as finding red-shifted emission where blue-shifted emission is expected from the rotation curve, then this signature provides good evidence for the presence of an outflow.

\item \textbf{Spatial extent of the emission}: Given that our sample galaxies only host low-luminosity AGN and have only moderate global SFRs, we expect low-mass outflows (visible as faint emission relative to our bright galaxy disks) with small to moderate extents.

\item \textbf{Relevance}: If the emission of the potential outflow is located outside the central 1\,kpc (in radius), it is not considered in this study.

\end{itemize}

\section{Results}
\label{sec:Results}

\subsection{Resulting outflow confidence labels}
\label{sec:Results:OCLs}

We compare the \ocl{}s assigned by different inspectors (A,~B,~C) for each method individually, and find an overall good agreement with Spearman rank correlation coefficients of $\rho_\mathrm{S} > 0.67$.
All coefficients are about equally correlated and coincide within $3\sigma$.
This allows us to derive reasonable average \ocl{}s for each galaxy and each method by averaging across the inspectors.

We find the uncertainties of the Spearman rank correlation coefficients to be ${\leq} 0.07$. We estimate these uncertainties as follows:
First we perturb the \ocl{}s per inspector per method randomly within their adopted uncertainty of $+1$, $0$, or $-1$. Next, we calculate Spearman rank correlation coefficients for the perturbed set of \ocl{}s for all combinations of two inspectors for each method (jackknife). Then we repeat this a thousand times and calculate the standard deviation of the resulting Spearman values.
The Spearman rank correlation coefficients for the different combinations of inspectors per method and their uncertainties are listed in Table~\ref{tab:method:spearmanR}.

\begin{table}
\caption{Spearman rank correlation coefficient $\rho_\mathrm{S}$ of the \ocl{}s per method and per pair of inspectors. \label{tab:method:spearmanR}}
\centering
\begin{tabular}{llll}
\hline\hline
\noalign{\smallskip}
Method  & $\rho_\mathrm{S}$(A--B) &  $\rho_\mathrm{S}$(B--C)& $\rho_\mathrm{S}$(A--C)\\
\noalign{\smallskip}
\hline
\noalign{\smallskip}
Spectrum 300\,pc & 0.85 $\pm$ 0.04 & 0.81 $\pm$ 0.06 & 0.75 $\pm$ 0.07\\ 
Spectrum 2\,kpc & 0.77 $\pm$ 0.06 & 0.73 $\pm$ 0.06 & 0.67 $\pm$ 0.06\\ 
\pv--diagram & 0.90 $\pm$ 0.06 & 0.73 $\pm$ 0.06 & 0.75 $\pm$ 0.06\\ 
Line-wing map& 0.93 $\pm$ 0.06 & 0.87 $\pm$ 0.06 & 0.77 $\pm$ 0.06\\ 
\noalign{\smallskip}
\hline
\end{tabular}
\tablefoot{Spearman rank correlation coefficient $\rho_\mathrm{S}$ for the different combinations of \ocl{}s assigned by inspectors A,~B,~C to each galaxy and each method.
Overall there is good agreement between the different inspectors. 
Uncertainties are obtained by perturbing the values by their errors, see text for more details.}
\end{table}

The agreement of \ocl{}s between different methods is slightly less good, but within $3\sigma$. This is expected, as the different methods are sensitive to different outflow characteristics (see Section \ref{sec:methods}) which can result in deviating \ocl{}s. 
Next, we define $\overline{\ocl}$ as the method-average of the inspector-averaged \ocl{}s of the \pv--diagram, the \lwm, and the better ranked one of the spectra (see motivation in Section~\ref{sec:methods:Spectrum}). 
To ensure consistent treatment of the spectra-\ocl{}s with the other two methods, we require agreement of the inspectors on the same aperture and use the higher ranked inspector-averaged spectrum-\ocl. 

The resulting inspector-averaged \ocl{}s for each method as well as the $\overline{\ocl}$ for each galaxy are reported in Table~\ref{tab:PHANGSOCLtable} of Appendix~\ref{sec:Appendix:PHANGS_OCL_ALL}.
Figure~\ref{fig:Results:average_rankings} displays the $\overline{\ocl}$ (black cross) as well as the inspector-only-averaged \ocl{}s per method per galaxy (colored filled symbols).
Galaxies with $\overline{\ocl} \geq 2$ are considered \textit{outflow candidates} according to the definition of the \ocl{}s. 
Uncertainties for the \ocl\ and $\overline{\ocl}$ are derived by perturbing the \ocl\ per method per inspector randomly within their adopted error of $+1$, $0$, or $-1$, and performing jackknifing for each method. When averaging, this results in different inspector-averaged \ocl{}s per method as well as different inspector- and method-averaged $\overline{\ocl}$ for each perturbation.  We repeat this 1000 times to get the standard deviation of both the \ocl{}s per each galaxy per each method as well as for $\overline{\ocl}$.

\begin{figure}[h!]
    \centering 
    \includegraphics[height=0.65\textheight]{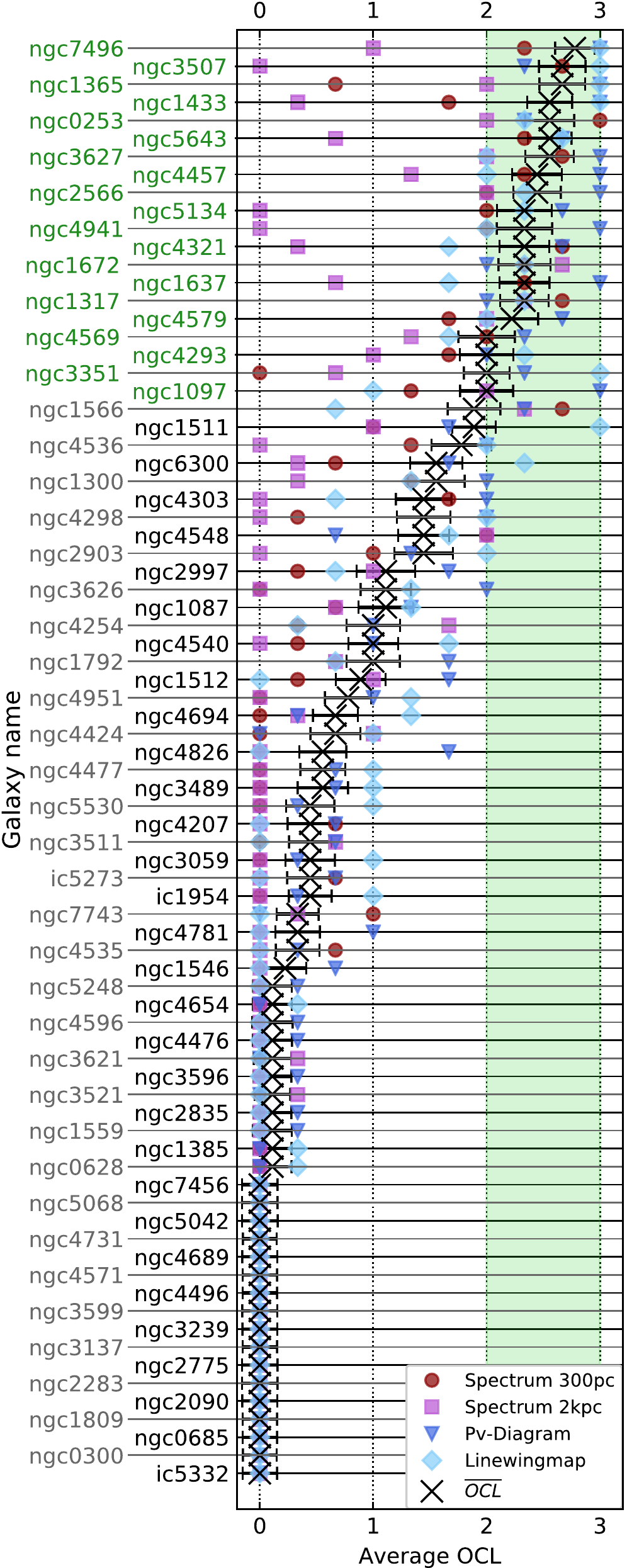}
    \caption{Average outflow candidate \ocl{}s for PHANGS galaxies. Black crosses represent the  $\overline{\ocl}$ per galaxy, which is the average of the inspector-averaged \ocl\ of the \pv--diagram, \lwm, and the higher ranked spectrum. 
    Galaxies with $\ocl \gtrsim 2$ are highly likely to exhibit strong central outflow signatures, whereas $\ocl \lesssim 1$ correspond to an absence of any outflowing gas signatures. 
    For each galaxy the inspector-averaged \ocl{}s of the individual methods are shown as red circle (central spectrum with 300\,pc aperture), purple square (spectrum of annulus with 2\,kpc outer diameter), blue triangle (\pv--diagram), and light-blue diamond (\lwm). 
    The green-shaded area marks all $2 \leq \ocl{} \leq 3$. A galaxy with $\overline{\ocl} \geq 2$  (and \lwm-\ocl{}$\geq2$) is considered an (secure) outflow candidate and emphasized with green galaxy names (see Section~\ref{sec:Results:Abundance}).
    Uncertainties are derived by perturbing and jackknifing the \ocl\ per method per inspector, as described in the text.
    }
    \label{fig:Results:average_rankings}
\end{figure}

\subsection{Frequency of central molecular outflows}
\label{sec:Results:Abundance}

In order to determine the frequency of central outflows, we need to establish a set of outflow candidates based on their \ocl\ values. Given our initial definition for assigning an \ocl\ (Table~\ref{tab:method:ranking}), we require that the average $\ocl \geq 2$, to ensure that possible outflow signatures are visible in all methods (see below). 
A total of 20 galaxies, or outflow candidates hereafter, meet this requirement resulting in an frequency~$F$ of $F\left( \overline{\ocl}\geq 2 \right) = 25\pm2\%$. 
The statistical uncertainty has been obtained similarly to the uncertainty of the inspector-averaged \ocl\ and the $\overline{\ocl}$. During each of the 1000 perturbations of the \ocl{}s, we additionally calculate the frequency according to our criteria. From the resulting 1000 frequency estimates, we can take the standard deviation as uncertainty.

For all outflow candidates, their $\overline{\ocl}$ as well as host galaxy properties (stellar mass, SFR, central SFR (cSFR), presence of bar and/or AGN) are listed in Table~\ref{tab:outflowcandidatelist}.

\begin{table*}
\caption{Outflow candidate host galaxies and their properties \label{tab:outflowcandidatelist}}
\centering
\begin{small}
\begin{tabular}{llrrrcccccc}
\hline\hline
\noalign{\smallskip}
Galaxy  & $\overline{\ocl}$ & $\log_{10}\,M_{*}$ & $\log_{10}$\,SFR& $\log_{10}$\,SFR$_\mathrm{center}$ &  Bar &  AGN &  $R^{\rm max}_{\rm outflow}$  & $R^{\rm weighted}_{\rm outflow}$ & v$^{\rm weighted}_{\rm outflow}$ &$\log_{10} \, \dot{M}_{\rm outflow}$\\
\noalign{\smallskip}
& & (M$_{\odot}$) &(M$_{\odot}$\,yr$^{-1}$) & (M$_{\odot}$\,yr$^{-1}$) & y/n & y/n & (") &(pc)  & (\kms) & M$_{\odot}$\,yr$^{-1}$)\\
\noalign{\smallskip}
(1) & (2) & (3) & (4) & (5) & (6) & (7) & (8) & (9) & (10) & (11)\\ 
\noalign{\smallskip}
\hline
\noalign{\smallskip}
NGC~7496 & 2.78 $\pm$ 0.18 & 9.997 & 0.354 & -0.437 $\pm <$ 0.001 & 1 & 1 &   8 & 316 & 83 & -0.033\\ 
NGC~3507 & 2.67 $\pm$ 0.20 & 10.396 & -0.004 & -1.506 $\pm$ 0.007 & 1 & 0 & 4 & 243 & 109 & -0.326\\ 
NGC~1365 & 2.67 $\pm$ 0.20 & 10.990 & 1.228 & 0.296 $\pm <$ 0.001 & 1 & 1 & 10 & 719 & 149 & 1.338\\ 
NGC~1433 & 2.56 $\pm$ 0.20 & 10.866 & 0.055 & -1.823 $\pm$ 0.002 & 1 & 0 & 6 & 333 & 111 & -0.715\\ 
NGC~0253 & 2.56 $\pm$ 0.21 & 10.637 & 0.699 & -0.145 $\pm <$ 0.001 & ... & 1 & 20 & 241 & 165 & 0.680\\ 
NGC~5643 & 2.56 $\pm$ 0.19 & 10.336 & 0.414 & -0.569 $\pm <$ 0.001 & 1 & 1 & 6 & 168 & 96 & 0.072\\ 
NGC~3627 & 2.56 $\pm$ 0.21 & 10.833 & 0.585 & -1.086 $\pm$ 0.001 & 1 & 1 & 6 & 208 & 170 & 0.491\\ 
NGC~4457 & 2.44 $\pm$ 0.22 & 10.415 & -0.515 & -1.282 $\pm$ 0.002 & 0 & 0 & 4 & 174 & 130 & 0.097\\ 
NGC~2566 & 2.44 $\pm$ 0.20 & 10.709 & 0.941 & 0.099 $\pm <$ 0.001 & 1 & 0 & 8 & 512 & 121 & 0.859\\ 
NGC~5134 & 2.33 $\pm$ 0.24 & 10.411 & -0.344 & -1.960 $\pm$ 0.010 & 1 & 0 & 4 & 241 & 106 & -0.797\\ 
NGC~4941 & 2.33 $\pm$ 0.23 & 10.175 & -0.355 & -1.229 $\pm$ 0.002 & 1 & 1 & 4 & 199 & 135 & -0.315\\ 
NGC~4321\tablefootmark{$\star$} & 2.33 $\pm$ 0.21 & 10.745 & 0.551 & -0.827 $\pm$ 0.001 & 1 & 0 & 4 & 142 & 107 & 0.077\\ 
NGC~1672 & 2.33 $\pm$ 0.22 & 10.729 & 0.881 & -0.086 $\pm <$ 0.001 & 1 & 1 & 4 & 199 & 201 & 0.089\\ 
NGC~1637\tablefootmark{$\star$} & 2.33 $\pm$ 0.21 & 9.946 & -0.195 & -1.061 $\pm$ 0.001 & 1 & 0 & 6 & 202 & 75 & -0.352\\ 
NGC~1317 & 2.33 $\pm$ 0.22 & 10.620 & -0.321 & -1.457 $\pm$ 0.001 & 1 & 0 & 4 & 224 & 65 & -0.159\\ 
NGC~4579 & 2.22 $\pm$ 0.24 & 11.146 & 0.336 & -1.126 $\pm$ 0.002 & 1 & 1 & 4 & 232 & 225 & 0.220\\ 
NGC~4569\tablefootmark{$\star$} & 2.00 $\pm$ 0.23 & 10.806 & 0.122 & -0.821 $\pm$ 0.001 & 1 & 1 & 6 & 338 & 151 & 0.432\\ 
NGC~4293 & 2.00 $\pm$ 0.24 & 10.506 & -0.289 & -0.942 $\pm$ 0.001 & 0 & 0 & 4 & 146 & 135 & 0.120\\ 
NGC~3351 & 2.00 $\pm$ 0.21 & 10.361 & 0.122 & -0.869 $\pm$ 0.001 & 1 & 0 & 4 & 152 & 98 & -0.352\\ 
NGC~1097\tablefootmark{$\star$} & 2.00 $\pm$ 0.23 & 10.760 & 0.676 & -0.230 $\pm<$ 0.001 & 1 & 1 & 6 & 264 & 238 & 0.609\\ 
\noalign{\smallskip}
\hline
\end{tabular}
\end{small}
\tablefoot{Outflow candidates and their host galaxy properties: (1) galaxy name, (2) average \ocl, (3) stellar mass with $0.112$\,dex statistical uncertainty, (4) SFR with $0.112$\,dex statistical uncertainty, (5) central SFR of the inner 2\,kpc in diameter, (6) bar flag (1=yes, 0=no), (7) AGN flag  (1=yes, 0=no), (8) maximum projected visually determined outflow extent, (9) average CO weighted outflow radius of the blue- and red-shifted outflow component, (10) average CO weighted outflow velocity of the blue- and red-shifted outflow component, (11) estimated mass outflow rates.
\tablefoottext{$\star$}{Galaxy does not fulfill the additional requirement of $\mbox{\lwm-\ocl} \geq 2$ (see Section~\ref{sec:Results:Abundance})}.
}
\end{table*}

\paragraph{Secure outflow candidates}
\label{sec:Results:AbundanceII}

In some cases, like NGC~4321 or NGC~1097, the putative outflow emission seen in the spectra and \pv--diagrams could also be explained via shocked regions or bar flows evident in the \lwm. 
Thus by adding a second criterion, we can increase the weight of the \lwm-\ocl{}s, which might be more sensitive to the contribution of streaming motions. Thus for \textit{secure outflow candidates} we require in addition to average $\overline{\ocl} \ge 2$ that the \lwm-\ocl\ is ${\geq}2$.

These two criteria are fulfilled by 16 out of 80 galaxies and result in a secure frequency estimate of $F_\mathrm{secure} \left(\overline{\ocl} \, \& \, \ocl_{\lwm} \geq 2 \right) = 20 \pm 2\,\%$. 
The four outflow candidates with $\mbox{\lwm-\ocl} < 2$ are indicated with a star ($^\star$) in Table~\ref{tab:outflowcandidatelist}.
We determine the statistical uncertainty using the same approach as before, but adding the second criteria to the calculations.

Further, we additionally inspect the data cubes of the outflow candidates for the presence of outflow emission without integrating or further reducing the available information per axis confirming the importance of the second criterion for the selection:
In all cases where $\mbox{\lwm-\ocl} <2$, emission in the data cube likely represents bar emission or shock emission rather than outflow-like emission. 

To unambiguously identify actual outflows among our outflow candidates, an individual analysis of the putative outflow geometry and the galaxy disk will be carried out in a future work. 
This involves analysing whether the putative outflowing gas can be found in counter-rotating regions, that could only be due to a bulk motion out of the plane.
Also, we plan on comparing the outflow position angles to those of galactic stellar bars, which could reveal more insights on the driving mechanisms, as AGN-driven outflows are generally randomly orientated compared to star-formation driven outflows which are usually orientated along the minor axes.
This is, however, beyond the scope of this work that focuses on statistical results rather than individual galaxy properties.

\section{Discussion}
\label{sec:Discussion}

Using standard outflow detection methods based on visual classification, we find the frequency of outflows to be $F = 20{-}25\% \pm 2\%_\mathrm{stat}$ depending on the number of criteria applied.
We divide the sample into outflow candidates and non-outflow galaxies using an average $\overline{\ocl}$-threshold (and additional an $\mbox{\lwm-\ocl}$) of 2 and higher. The outflow frequency as function of $\overline{\ocl}$-threshold in steps of $\Delta \overline{\ocl} = 0.1$ is given in Appendix~\ref{sec:Appendix:Frequency_threshold} for interested readers.

In Section~\ref{sec:Discussion:CandidatesVSPHANGS}, we discuss possible differences between outflow candidates and the remaining PHANGS sample galaxies. 
We test possible relations between the \ocl{}s and host galaxy properties and the reliability of the determined frequency 
in Section~\ref{sec:Discussion:OCLReliability}.
In Section~\ref{sec:Discussion:Outflowrates}, we estimate outflow masses and mass outflow rates for our candidates and compare them to literature results (see Section~\ref{sec:Discussion:Literature}).

\subsection{Outflow candidates versus non-outflow galaxies}
\label{sec:Discussion:CandidatesVSPHANGS}

Among our 80 targets, we find 20 outflow candidates with 16 classified as secure candidates.
We compare the outflow candidates to the full sample with respect to their host galaxy properties.

\begin{figure}[t]
    \centering
    \includegraphics[width = 0.4\textwidth]{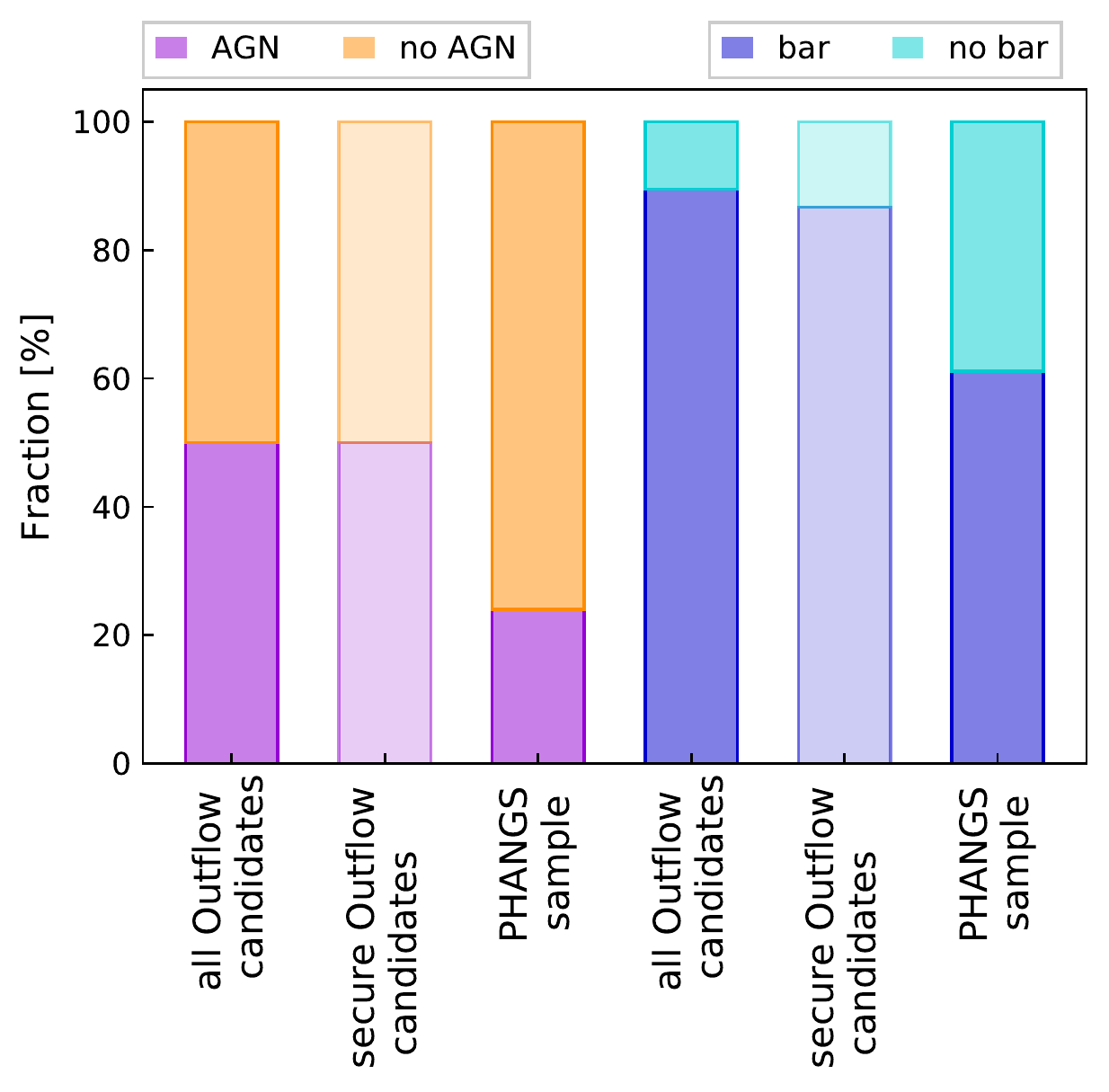}
    \caption{Fraction of AGN (violet, left) and bars (blue, right) among PHANGS outflow candidates and the secure outflow candidates, as well as the entire PHANGS sample.}
    \label{fig:Hist_CandidatesvsPHANGS}
\end{figure}

Figure~\ref{fig:Hist_CandidatesvsPHANGS} displays the fraction of AGN and bars among all and the secure outflow candidates, and the full sample.
Out of 20 (16~secure) outflow candidates, $50\%$ ($50\%$) contain an AGN, and $89\%$ ($87\%$) host a stellar bar, which are both significantly higher than the global fraction ($24\%$ AGN out of 79 galaxies with AGN/\linebreak[0]{}\mbox{no-AGN} classification, $61\%$ barred out of 72 galaxies with good morphological bar/\linebreak[0]{}\mbox{no-bar} classification).
Thus, we find an enhanced fraction of (secure) outflow candidates among our 19 AGN host galaxies of 53\% (42\%) compared to 17\% (13\%) for the 60 galaxies without an AGN. 
The same applies for barred galaxies, where a higher fraction of (secure) outflow candidates among our 44 barred galaxies of 39\% (30\%) can be to a fraction of 7\% (7\%) for the 28 unbarred galaxies.

This large fraction of AGN, with about half of all AGN being an outflow candidate, agrees well with the results of \citet{Rakshit2018}, who found that 45\% of Type~2 AGN as well as 89\% of Type~1 AGN possess a broad wing component in their ionized gas spectrum. Our AGN sub-sample consists of mainly Type~2 AGN.

The preference for barred galaxies among our molecular outflow candidates has not been reported before. 
This could be due to covariance with galaxy mass, given that both the bar fraction as well as the AGN fraction increase with stellar mass \citep[e.g.,][]{Rosas-Guevara2020TNG100, Consolandi2016, Gavazzi2015}.
This higher fraction is likely to be related to the findings that barred galaxies show a higher concentration of molecular gas in their central ${\sim}1$\,kpc region compared to unbarred galaxies \citep[e.g.,][]{Sakamoto1999, Sun20}, resulting in a reservoir for the outflow that otherwise might not be present. 

To analyse this further, we search for a possible influence of host galaxy properties such as stellar mass, SFR, and central SFR.

\subsubsection{Role of host galaxy on outflow candidate selection}
\label{sec:Discussion:HostgalCandidates}

\begin{figure*}[t!]
    \centering
    \includegraphics[width = 0.9\textwidth]{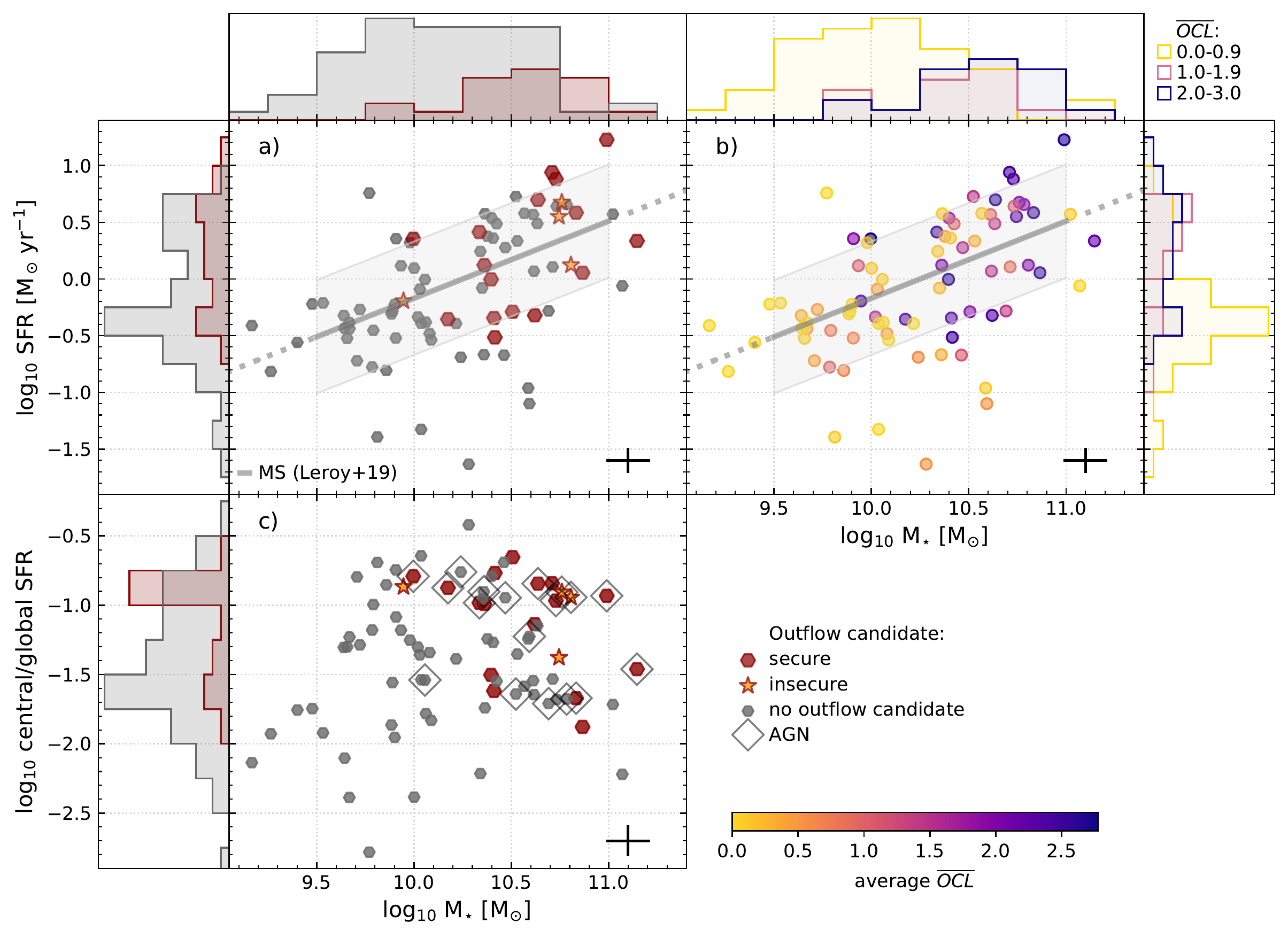}
    \caption{a) Distribution of PHANGS outflow candidates (secure candidate: red hexagon, otherwise: orange star) and non-outflow galaxies (gray hexagon) along the star-forming main sequence. The histograms show the distribution over stellar mass (top) and SFR (left) for outflow candidates (red, all candidates combined) and galaxies without outflows (gray).
    The main sequence according to \citet{Leroy2019} is added for comparison (gray solid line) with a width of 0.5\,dex (gray shaded area) and extrapolated to larger and smaller stellar masses (dotted gray line). 
    b) Distribution of PHANGS galaxies along the main sequence of star-forming galaxies. The $\overline{\ocl}$ is shown by color. A dark color represents a large $\overline{\ocl}$ (e.g., class~3), thus showing host galaxies with high probabilities to possess molecular outflows.
    Light colors represent low $\overline{\ocl}$ (e.g., class~0), thus no or only tentative outflow signatures are present in these galaxies.  
    The histograms show the distribution over stellar mass (top) and SFR (right) for three $\overline{\ocl}$ bins.
    c) Ratio of central-to-global SFR as a function of stellar mass for PHANGS galaxies likely to host a central molecular outflow (secure candidate: red hexagon, otherwise: orange star) or not (gray hexagon). 
    Histograms show the distribution of all outflow candidates (red) compared to galaxies without outflow signatures (gray) in stellar mass (top) and SFR ratio (left). 
    AGN are marked by an open black diamond. 
    For all plots the median uncertainty for individual data points is shown in the lower right corner.}
    \label{fig:Monsterplot}
\end{figure*}

As the PHANGS-ALMA sample is selected to represent normal main sequence galaxies at $z=0$ (see Section~\ref{sec:data}), we check for trends with SFR, central SFR (within 2\,kpc in diameter), and stellar mass (global properties from \citet{Leroy20b}, see same Section~\ref{sec:data}) among the outflow candidates and in the $\overline{\ocl}$ ranking. 
Figure~\ref{fig:Monsterplot} displays the distribution of $\overline{\ocl}$s and (secure) outflow candidates on the star-forming main sequence. 
It also shows the ratio of central-to-global SFR for (secure) outflow candidates as a function of stellar mass and highlights galaxies hosting an AGN.

\begin{figure}
    \centering
    \includegraphics[width = 0.45\textwidth]{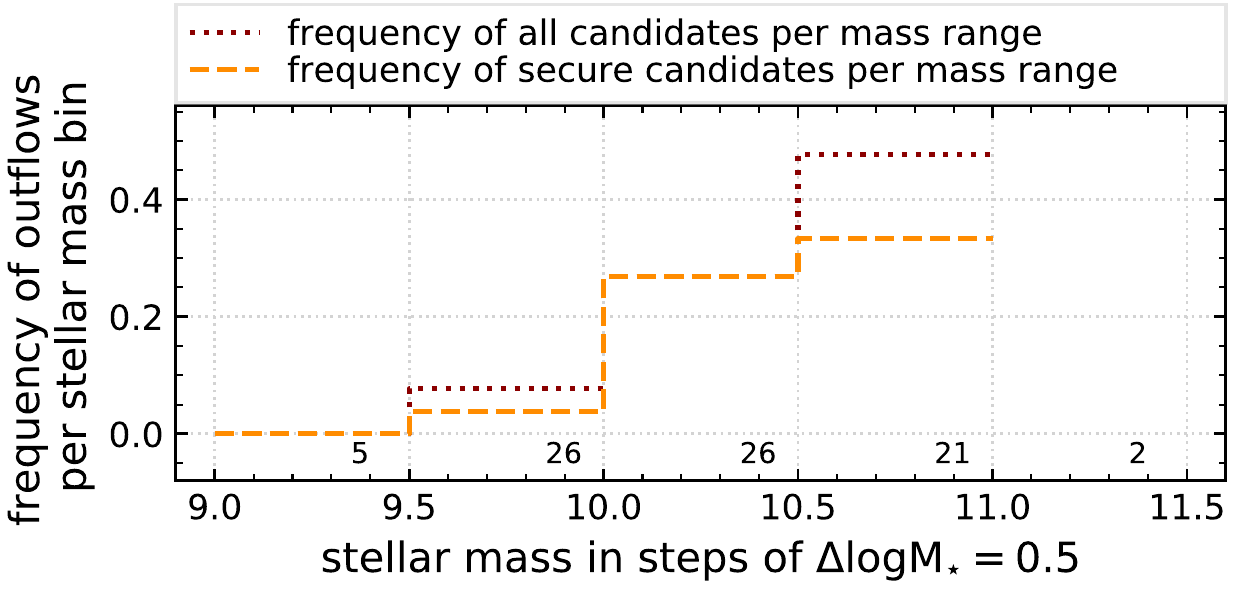}
    \caption{Fraction of (secure) outflow candidates as a function of stellar mass. The fraction is determined per stellar mass bin with widths of $\Delta \log_{10} (M_{\star}/\mathrm{M}_\odot) = 0.5$ that contain a minimum of 3 galaxies. Each level represents the fraction of outflow candidates relative to all galaxies in the mass bin. The total amount of galaxies per mass bin is given at the bottom of each bin.}
    \label{fig:Fraction_Massdependence}
\end{figure}

Our findings are as follows: 
\begin{itemize}[noitemsep,topsep=0pt,leftmargin=2\parindent]

    \item[a)] 
    All molecular outflow candidate galaxies have stellar masses of $\log_{10} (M_{\star}/\mathrm{M}_{\odot}) \gtrsim 10$ and SFRs of $\log_{10} \, \mathrm{SFR}/(\mathrm{M}_{\odot}\,\mathrm{yr}^{-1}) \gtrsim -0.5$, covering roughly the upper halves of the respective ranges spanned by the full sample. 
    This is also clearly evident from the histograms (outflow candidates (red) versus remaining galaxies (gray)).
    We discuss the possible existence of a minimum threshold in stellar mass and/or SFR, before a molecular outflow can be launched in a main sequence galaxy, in Section~\ref{sec:Discussion:OCLReliability}.
    Figure~\ref{fig:Fraction_Massdependence} shows the frequency of outflows in bins of stellar mass which we find to increase monotonically as a function of stellar mass over the stellar mass range probed by our sample.
    We use stellar mass bins of width $\Delta \log_{10}(M_{\star}) = 0.5$ here, but the same trend can be seen for smaller mass bins (e.g., $0.33$ or $0.25$). 
    Unfortunately, we want to emphasize that the sample size at each bin is small so that possible influences from AGN and bars cannot be investigated at this point.
    As galactic molecular outflows are thought to play a critical role in regulating the galaxy stellar mass function, the possible correlation between molecular outflow frequency and stellar mass found in local main sequence galaxies represents a future direction for studying galactic molecular outflows.
    
    \item[b)] We observe a weak trend in the SFR versus stellar mass plane where galaxies with high stellar mass and high SFR appear more likely to possess a higher $\overline{\ocl}$ value than galaxies with low stellar mass and low SFR indicated by Spearman rank correlation coefficients for SFR (stellar mass) and $\overline{\ocl}$s of $\rho_\mathrm{SFR} = 0.40 \pm 0.03$ ($\rho_{M_\star} = 0.56 \pm 0.03 $).
    Uncertainties of the Spearman rank correlation coefficient are determined via jackknifing.
    We perturb the \ocl{}s as well as the corresponding stellar mass and SFR values by their uncertainties, and calculate the Spearman rank correlation coefficient. We repeat this 1000 times and adopt the scatter among the Spearman rank correlation coefficients as uncertainty.
    In Section~\ref{sec:Discussion:OCLReliability}, we will come back to this topic and search for secondary dependencies, namely the presence or absence of a stellar bar or AGN.

    \item[c)]  
    There is a strong preference for outflow candidates (${\sim}60$\%) which have a significant ($>$10\%) fraction of their total SFR confined to the central 2\,kpc in diameter (Figure~\ref{fig:Monsterplot}). 
    Such a trend that the hosts of our outflow candidates exhibit higher fractional central SFR compared to the galaxies without outflows is not surprising as central SFR tends to correlate with global SFR (see point above).
    Another possible explanation might be that the central SFR is the more fundamental quantity that enables outflows to be launched compared to the global SFR. 
    Similarly to the global SFR, we find a threshold in the central-to-global SFR ratio fulfilled by our outflow candidates (see discussion in Section~\ref{sec:Discussion:OCLReliability}).
\end{itemize}

\subsubsection{Correlation of \ocl\ and host galaxy properties}
\label{sec:Discussion:OCLReliability}

We explore here how much the trends seen pertain to outflow host galaxies or present biases in our classification methodology.
Overall, we see a trend along the main sequence with significant ($\rho>3 \sigma$) Spearman rank correlation coefficients between $\overline{\ocl}$ and global SFR as well as stellar mass.

Since we utilize CO to identify outflows in our sample galaxies, host galaxy properties affecting the amount of molecular gas might lead to a potential bias.
A higher stellar mass (implying a higher central stellar mass surface density),
likely leads to higher molecular gas mass surface density ($\rho=0.67$ for stellar mass and molecular gas mass surface density) and therefore results in higher signal-to-noise observations which increase the chance to identify outflowing gas.
In fact, we find a strong correlation between $\overline{\ocl}$ and molecular gas mass surface density ($\rho=0.81$).

Similarly, if the conversion factor of CO luminosity to total molecular mass, $\alpha_\mathrm{CO}$, is different in the outflowing gas relative to the galaxy, it may possibly lead to a correlation between $\overline{\ocl}$ and stellar mass.
However, $\alpha_\mathrm{CO}$ in outflowing gas is currently unconstrained.
On the other hand, a higher molecular gas mass surface density can increase the incidence of molecular outflows as they require a molecular gas reservoir to be present.
As an example, \citet[][]{Hogarth21} find that edge-on galaxies that host molecular outflows seem to have their molecular gas more centrally distributed than in their control sample.
The correlation between $\overline{\ocl{}}$ and stellar mass is furthermore supported by recent studies that reported higher stellar mass galaxies to have more and faster outflows (if an AGN is present, e.g., at $z = 0.6{-}2.7$; \citealt{ForsterSchreiber2019}), thus a higher stellar mass can actually increase the chance of (detectable) outflows to be launched.

Similarly, \citet{Cicone2016} suggest that for galaxies with stellar masses comparable to those in our sample, a higher SFR increases the incidence of ionized gas outflows. 
This may suggest an increase of ionized gas outflow incidence with offset from the main sequence, for example, as it was found for galaxies at $z = 0.6{-}2.7$ \citep{ForsterSchreiber2019}.
Although ionized outflows might be more common when molecular gas is rare or not present, the underlying launching mechanisms out of the galaxy will have some similarities, making such ionized gas studies complementary.
We do not find an increase in outflow velocity with larger offset from the main sequence ($\rho_{\Delta\mathrm{MS}}=0.24$, $p$-value $> 0.2$) as they do for their candidates. This might be a consequence of the PHANGS sample not extending far above or below the main sequence.

The lack of correlation between galaxy distance and $\overline{\ocl}$ ($\rho_\mathrm{dist} = 0.164$, $p$-value $\sim 0.15$)
argues against any resolution dependency for the \ocl{}s or the relations found.

The observed trend between $\overline{\ocl}$ and both stellar mass and SFR highlights the importance to extend outflow studies to a larger range of main sequence galaxies. 
Up to now, outflows have been preferentially identified in starburst galaxies with stellar masses and SFRs in excess to those of the PHANGS galaxies \citep[Figure~\ref{fig:Results:LitcompCandidates}; e.g.][]{Cicone2014, Fluetsch2019, Lopez2019, Roberts-Borsani2020_ObsConstr, Roberts-Borsani2020}.
Although we see strong evidence for a minimum threshold in stellar mass and/or SFR for a main sequence galaxy to possess a (detectable) molecular outflow (e.g., Figure~\ref{fig:Monsterplot}), our data do not exclude the possibility of a continuous relation between frequency of outflows and stellar mass and/or SFR.
We infer a stellar mass threshold of $M_{\star} \gtrsim 10^{10.0}\, \mathrm{M}_{\odot}$ and SFR threshold of $\mathrm{SFR} \gtrsim 10^{-0.5}\, \mathrm{M}_{\odot}~\mathrm{yr}^{-1}$, or fractional central SFR of $\mathrm{cSFR}/\mathrm{SFR} \gtrsim 10^{-1.9}$.
Such thresholds have been found for ionized gas outflows as well, with a limiting stellar mass of $\mathrm{M}_{\star} \sim 10^{10.9}\, \mathrm{M}_{\odot}$ \citep[e.g.,][]{Genzel2014} below which the occurrence of outflow signatures drops sharply.
If a stellar mass threshold exists, it would be more sensible to
infer the actual outflow frequency for the stellar mass range where outflows can occur.  For our sample with $\log_{10} \left(M_{\star} / \mathrm{M}_{\odot} \right) \geq 10.0$, this corresponds to a frequency of outflow candidates of 36\%.
Further, this would imply that mechanisms capable of launching molecular outflows are less effective in galaxies of intermediate stellar mass. 

Regarding the correlations found with the ratio of central to global SFR, there is a longstanding idea in the literature that outflow velocity correlates strongly with a higher SFR per area, $\Sigma_\mathrm{SFR}$. This idea, that instead of global SFR the SFR per area has a larger influence on the presence of outflows was introduced by \citet[][]{Heckman1990, Heckman2015}. 
In addition, \citet[][]{Heckman2003} suggested a threshold of SFR per unit area above which outflows appear ubiquitous.
This might be similar to the threshold for central to global SFR found above.

There is still the possibility that our threshold is only a coincidence and outflows appear in galaxies spanning a wider range of stellar mass and SFR, 
(e.g., over smaller area and/or with smaller outflow velocities)
and our outflow detection methods are only sensitive to the high end of SFR and stellar mass. 
Studies such as \citet[e.g.][]{Lopez2019} analysing ionized gas emission from 273 highly inclined galaxies find no stellar mass threshold, but outflows over their full stellar mass range $9.5 < \log_{10} \left( M_{\star} /\mathrm{M}_{\odot} \right) < 11 $. 

To investigate whether other host galaxy properties affect the correlations of $\overline{\ocl}$ with stellar mass and SFR, we analyse the presence/\linebreak[0]{}absence of a bar or AGN. The full analysis is provided in Appendix~\ref{sec:Appendix:OCLreliabilityBarAGN}, here we briefly summarize our findings: 
Overall, barred galaxies show a stronger correlation between $\overline{\ocl}$ and SFR compared to non-barred galaxies, and a similar trend is seen for AGN compared to inactive galaxies. 
It could be plausible that these trends arise due an enhanced bar and AGN fraction at higher SFR in the sample, and the relevant correlation is between our $\overline{\ocl}$ and SFR.
In contrast, we find no significant difference between stellar mass and $\overline{\ocl}$ for galaxies with or without AGN and/or bar. 
Therefore, the role of bars and AGN in enhancing the outflow incidence above the dependence on stellar mass remains ambiguous for our galaxy sample.

To test if all methods are equally impacted by stellar mass, SFR, and the presence of either bar or AGN, we calculate Spearman rank correlation coefficients for \ocl{}s of the individual methods (300\,pc spectra, 2\,kpc  spectra, \pv--diagrams, and \lwm{}s, each averaged over the different inspectors) and stellar mass (SFR) for all galaxies (for details and $\rho$ values see Appendix~\ref{sec:Appendix:OCLMethodsplit}). 
Overall, we do not find a significant difference to the previously found correlations. 
As all methods trace gas kinematics in a different way, this is reassuring and lowers the chance of a possible inherent bias. 
This increases our confidence that the trends between stellar mass and outflow incidence or SFR and outflow incidence are real.

\subsection{Outflow masses, rates and other properties}
\label{sec:Discussion:Outflowrates}

In this paper we have identified outflow candidates in PHANGS using three different diagnostic tools. We treat them as candidates only, as a number of dynamical processes may mimic outflow signatures; follow-up with multi-wavelength data is required to unambiguously confirm the presence of outflowing gas. 
Nevertheless, here we derive outflow rates ($\dot{M}_\mathrm{outflow}$ in Table~\ref{tab:outflowcandidatelist}) based on some simple assumptions for an initial characterization of these potential outflows. 

First, we define an outer boundary to the outflow radial extent ($R^\mathrm{max}_\mathrm{outflow}$) in the plane of the sky from visual inspection of the different diagnostic plots and list them in Table~\ref{tab:outflowcandidatelist}. Next, we construct moment maps considering emission in the blue and red line wings of the ALMA cube separately, following the automated criteria described in Section~\ref{sec:methods:lwm} for a $300$\,pc aperture. For the range of channels in each wing (blue or red), we integrate the total molecular mass inside $R_\mathrm{outflow}^\mathrm{max}$ (i.e., we sum the moment~0 map constructed without any thresholding over these channels). We also derive the CO-weighted average radius ($R_\mathrm{outflow}^\mathrm{weighted}$) in the plane of the galaxy, and the CO-weighted moment~1 velocity ($v_\mathrm{outflow}^\mathrm{weighted}$), again inside $R^\mathrm{max}_\mathrm{outflow}$ and over each line wing. 
We discuss the connection between 3D outflow geometry and the here defined projected radius and line-of-sight velocity below.
Both for the weighting and the computation of the moment~1 map, we use a thresholded version of the ALMA cube ($4\sigma$ clipping).

With these quantities, we estimate a characteristic mass outflow rate as: 
\begin{equation}\label{eq:Outflowrate}
\dot{M}_\mathrm{outflow} \sim v_\mathrm{outflow}^\mathrm{weighted} \times M_\mathrm{mol,\,tot} / R_\mathrm{outflow}^\mathrm{weighted}~,
\end{equation}
for each of the wings, and add up both contributions ($\dot{M}_\mathrm{outflow} = \dot{M}^\mathrm{red}_\mathrm{outflow} + \dot{M}^\mathrm{blue}_\mathrm{outflow}$). 
We note that this is representative of an outflow made up of thin expelled shells or clumps averaged over time \citep{Veilleux2005, Fluetsch2019}. 
For a spherical or multi-conical geometry, the outflow rates would be three times higher \citep[e.g.,][]{Maiolino2012, Cicone2015}. 
In the conversion from CO luminosity to total molecular mass, we adopt a value of $\alpha_\mathrm{CO} = 0.8$ M$_\odot$\,(K\,km\,s$^{-1}$\,pc$^2)^{-1}$, typical of starburst galaxies and often assumed for molecular outflows \citep{Bolatto2013ConversionFactor,Lutz2020, Veilleux2020}.

\emph{The mass outflow rates in Table~\ref{tab:outflowcandidatelist} are solely based on the velocity projected along the line of sight, whereas their extent is measured in the plane of the sky.} 
Generally, constraining the angle of the outflow is not straightforward, and we do not attempt to make that correction here (thereby matching most literature), as it requires a dedicated case-by-case study. 
Since we measure the outflow radius vector and velocity vector as projected onto the plane of the sky and the line of sight, respectively, correction factors of $\sin \alpha$ and $\cos \alpha$ need to be applied to recover deprojected values, where $\alpha$ denotes the angle between the line of sight and radius vector or velocity vector.
The correction factor on the outflow rate is then $\sin \alpha \, / \cos \alpha = \tan \alpha$ \citep[see e.g.,][]{Krieger2019}.
For example, deviations of up to $30\degree$ from $\alpha = 45\degree$ result in correction factors of up to ${\sim} 4$.
Therefore, in the most optimistic case, the individual outflow rates could be considered accurate within a factor of a few.
In pathological cases where the outflow is extremely close to the line of sight or nearly perpendicular to it, the corrections can be much larger. 
Fortunately, when averaging mass outflow rates over an ensemble of galaxies, the projection corrections should cancel out (because the orientation of the outflows are random relative to the line of sight; \citealt{Cicone2015}). When averaging over the 20 outflow candidates in Table~\ref{tab:outflowcandidatelist}, we find a modest mean mass outflow rate of $\langle\dot{M}_\mathrm{outflow}\rangle \sim 3\,\mathrm{M}_\odot~\mathrm{yr}^{-1}$.

Another concern when studying such modest outflows is the difficulty to separate all of the outflowing gas from the gas rotating in the galaxy disk. While our definition of line wings attempts to do this, there could be a range of velocities in the outflow (e.g., due to projection), where some CO emission associated with the outflow might be indistinguishable from the rotating disk. 
In such a case, our estimates from the line wings should be regarded as lower limits. 
Finally, we also checked the sensitivity of our results to the outer radial outflow boundary, which we estimate by eye. 
Perturbing $R^\mathrm{max}_\mathrm{outflow}$ by ${\pm}2\arcsec$, roughly twice the typical resolution of the ALMA data, results in outflow rates that differ on average by 30\%, which is a modest variation. For outflow rates, we therefore assign a statistical uncertainty of $0.2$\,dex.
However, the actual uncertainty may be larger given the uncertainty contributions discussed above and might be closer to $\sim1\,$dex.

\begin{figure}
    \centering
    \includegraphics[width=0.47\textwidth]{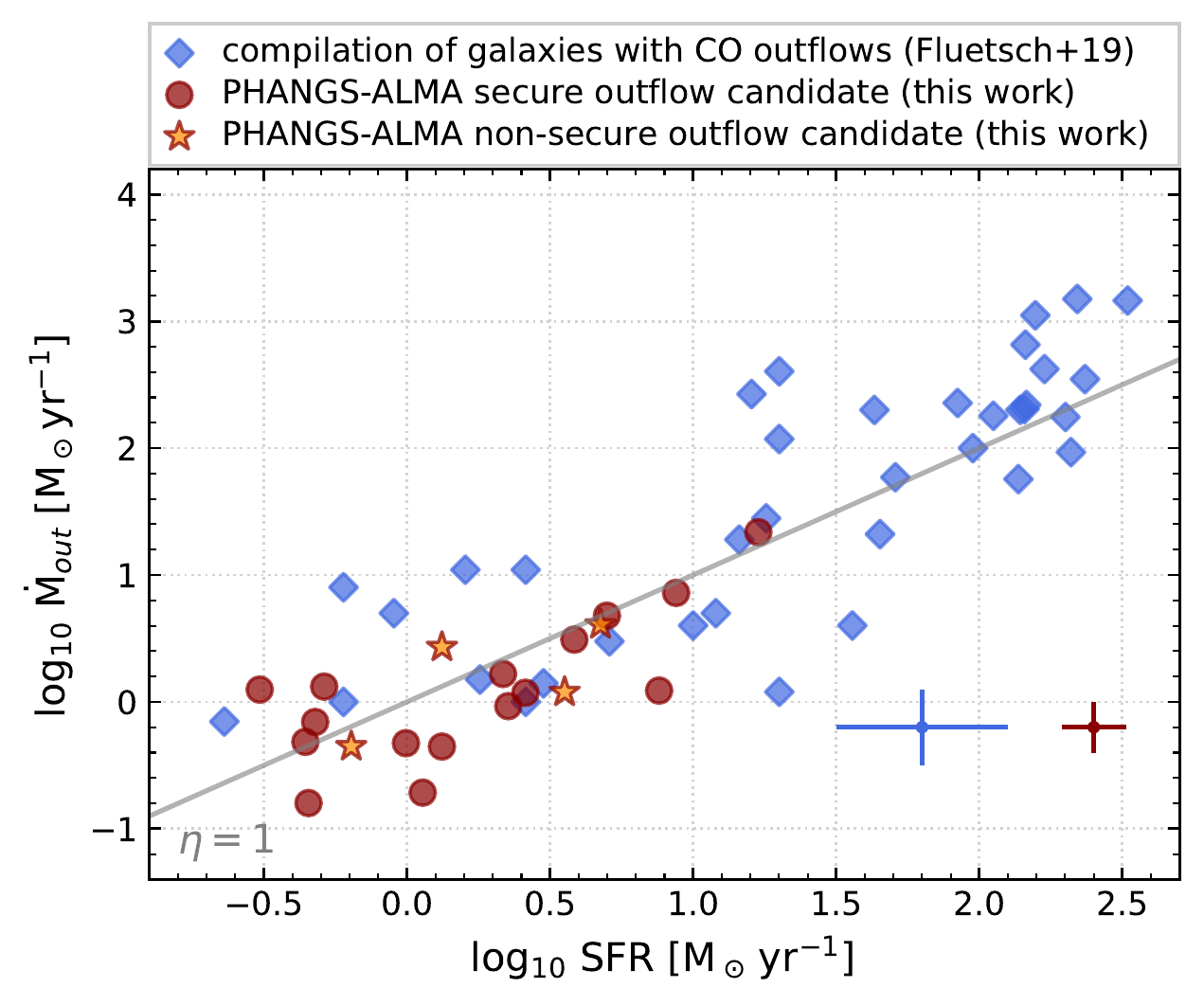}
    \caption{Mass outflow rates $\dot{M}_\mathrm{out}$ as function of SFR for the secure outflow candidates (red circle) and possible outflow candidates (orange star) from PHANGS as well as the outflows compiled by \citet{Fluetsch2019} (blue diamond, blue errorbar). The gray line corresponds to a mass loading factor $\eta$ of unity. PHANGS SFRs have an uncertainty of $0.11\,$dex and the mass outflow rates have statistical uncertainties of $0.2\,$dex when perturbing the radius the outflow mass is determined over, with a larger total uncertainty estimated to be $\sim1\,$dex given the unknown actual outflow geometries and properties.}
    \label{fig:Discussion:OutflrateSFR}
\end{figure}

Figure~\ref{fig:Discussion:OutflrateSFR} shows the mass outflow rates calculated according to Equation~\eqref{eq:Outflowrate} as function of SFR for PHANGS outflow candidates as well as 
a compilation of known molecular outflows by \citet{Fluetsch2019} that use the same $\alpha_\mathrm{CO}$ conversion factor as well as the same mass outflow rate calculation (Equation~\eqref{eq:Outflowrate}). 
We indicate where the mass loading factor $\eta = \dot{M}_\mathrm{outflow} / \mathrm{SFR}$ is unity (gray line). 
Overall, we find good agreement between both outflow samples, with the PHANGS outflow candidates further expanding the trend to lower mass outflow rates.

\begin{figure}
    \centering
    \includegraphics[width = 0.49\textwidth]{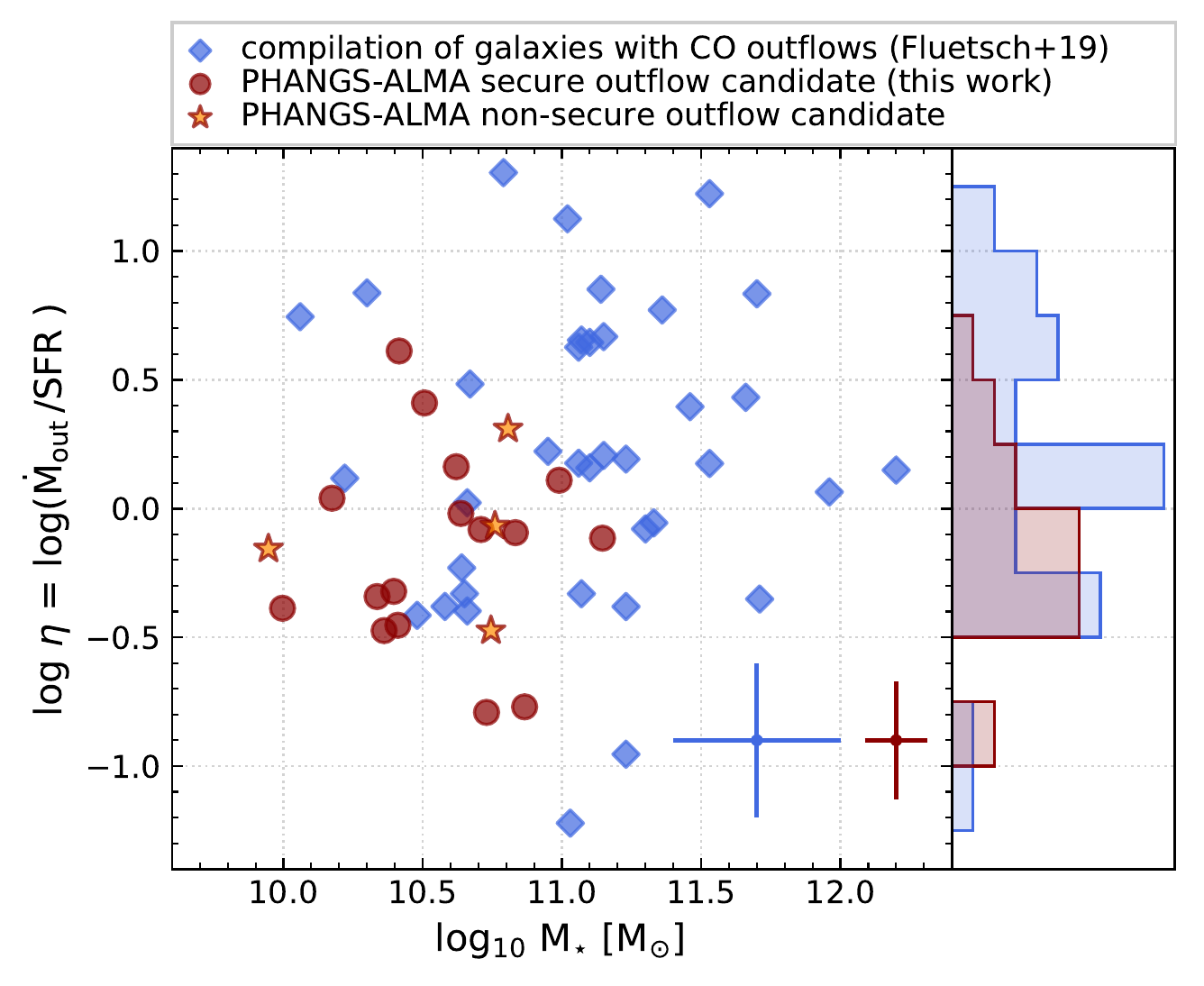}
    \caption{Mass loading factor $\eta$ as function of stellar mass for PHANGS outflow candidates (secure candidate: red circle, otherwise: orange star) and outflows compiled by \citet{Fluetsch2019} (blue diamond, blue errorbar). The PHANGS outflow candidates populate the lower stellar mass and lower mass loading part of the probed parameter space, 
    but overall no significant difference between both samples is found. PHANGS stellar masses have an uncertainty of $0.11\,$dex and the statistical uncertainty of the mass loading factor due to error propagation is estimated to be $0.23\,$dex (red errorbar) with a larger total uncertainty of up to $\sim1\,$dex.}
    \label{fig:Discussion:EtaMstarHist}
\end{figure}

In Figure~\ref{fig:Discussion:EtaMstarHist}, we show the mass loading factor $\eta$ as function of stellar mass for the PHANGS outflow candidates and the literature sample from \citet{Fluetsch2019}. The PHANGS outflow candidates have lower mass loading factors and lower stellar masses, but overall no significant difference between both samples is found.
The average mass loading factor of the PHANGS sample is $\log_{10} \langle \eta \rangle = 0.0$, with a statistical uncertainty of $0.23$\,dex and an estimated total uncertainty of $\sim1\,$dex, indicating that 
SFR and outflow rate are comparable. 
Put in another way, the outflows remove as much gas from the galaxy center as is consumed by central SFR when present, and so they can represent an important contribution to the overall future evolution of the gas.
We do not find a significant difference between PHANGS outflow candidate galaxies that have an AGN or a bar compared to those that do not. 
Using the central SFRs derived over the central $2\,$kpc instead of global SFRs will result in a significant increase of the mass loading factor.  
However, more meaningful estimates of this factor might be derived from resolution matched SFR maps that allow for tailored measurements in a future work.

In conclusion, while the outflow rates are subject to large uncertainties, our measurements suggest that these are modest molecular outflows (if the kinematic signatures that we observe indeed trace only outflowing molecular gas).
Such outflows might not be able to quench a whole galaxy, but can determine the fate of starbursts or the evolution of galaxy centers.
A detailed follow-up study with multi-wavelength data would be necessary to confirm which of these outflow candidates host true outflows and to refine the calculation of individual mass outflow rates.
Also, searching for differences in outflow properties such as extent, mass, energetics, and geometry between SF- and AGN-driven outflow will require a significant and well-characterized sample.
The molecular outflows of the PHANGS outflow candidates have an average extent of $\langle R^\mathrm{weighted}_\mathrm{outflow} \rangle \sim 260$\,pc, velocity of $\langle v^\mathrm{weighted}_\mathrm{outflow} \rangle \sim 130$\,\kms\ (see also Table~\ref{tab:outflowcandidatelist}), and integrated CO luminosity of $\langle L_\mathrm{outflow} \rangle \sim 8.6$\,K\,\kms. 
These numbers emphasize the importance of high resolution, high sensitivity studies such as PHANGS, that enable us to visually detect these outflows on sub-kpc scales. This must be considered when comparing this study to previous results from the literature (see Section~\ref{sec:Discussion:Literature})

\subsection{Literature comparison}
\label{sec:Discussion:Literature}

\subsubsection{Individual galaxies}
\label{sec:Discussion:Literature:single_galaxies}

To gauge the robustness of our outflow classification methods, we searched the literature for detected outflow signatures in molecular gas for our PHANGS galaxy sample. We separately discuss our outflow candidates and the rest of the sample.

For 7 out of 16 of our secure outflow candidates either molecular and/or ionized central gas outflows have been reported in the literature. 
Four studies on our candidates explicitly exclude outflows in cold molecular gas.
For the remaining secure candidates, no previous studies on molecular outflows were found. 

Out of the four candidate outflow galaxies that are marked with a star~($^\star$) in Table~\ref{tab:outflowcandidatelist}, thus not fulfilling the stricter criterion, none had a cold molecular outflow study in the literature and only some evidence for other gas phases (e.g., warm molecular gas or ionized gas).

For four of the remaining non-outflow candidate galaxies studies report evidence for molecular/\linebreak[0]{}ionized outflows. However, for three of them and for three additional non-outflow candidate galaxies also studies are available that present evidence against the presence of outflows. 
For the remaining galaxies in the non-outflow candidate sample, no studies on molecular outflows were found in the literature.

In summary, we find reasonable agreement for most of our outflow candidates, given that for nearly half of the galaxies no molecular outflow study has been made, and the existing differences in resolution, gas phase, and methodology applied to identify the outflows. 
Our study may have missed one potential outflow candidate among the PHANGS galaxies, namely NGC~1068, and for two of our candidates (NGC~1365, NGC~1672) high resolution \mbox{CO\,(3--2)} observations of the galaxy centers provided no clear evidence for
molecular outflows. 

\paragraph{PHANGS outflow candidates and literature matches}

\begin{itemize}[noitemsep,topsep=0pt,leftmargin=2\parindent]

    \item \textbf{NGC~0253}: Its central molecular outflow is long known and has been studied in detail \citep[e.g.,][]{Sakamoto2006, Bolatto2013Nature, Walter2017, Zschaechner2018, Krieger2019}.

    \item \textbf{NGC~1365}: \citet[][\mbox{CO\,(3--2)} emission]{Combes2019ALMATori} do not find clear evidence for a molecular outflow but rather for an inflow. However, NGC~1365 has long been known to exhibit a biconical outflow in, for example, ionized gas or \mbox{X-ray} \citep[as reported by e.g.,][]{Lena2016, Venturi2017, Davies2020}.
    Recent work by \citet{Gao2021} reports evidence for an outflow in the central 5.4\,kpc in molecular (\mbox{CO\,(2--1)}) and ionized gas.
    
    \item \textbf{NGC~1433}: Evidence for a central molecular outflow is reported by \citet[][in \mbox{CO\,(3--2)} emission]{Combes2013}, and re-analyzed with additional observations by \citet[][in molecular hydrogen and \mbox{CO\,(3--2)} emission]{Smajic2014_NGC1433}
    
    \item \textbf{NGC~1672}: \citet{Combes2019ALMATori} study the cold molecular gas torus in the central region and do not find evidence for an outflow. 
    \citet{Fazeli2020} rules out the presence of an outflow in hot molecular and ionized gas.
    
    \item \textbf{NGC~3351}: \cite{LeamanQuerejeta2019NGC3351} find that a molecular bubble exists in the center, and \citet{Swartz2006} interpret their results from \mbox{X-ray} as evidence for an outflow confined by cold ambient gas. 

    \item \textbf{NGC~4579}: This galaxy possesses a jet in the center  and radio emission indicates the presence of shocked material at $R \gtrsim 100$\,pc \citep{Contini2004}.
    \citet{GarciaBurillo2009} conclude the presence of an outflow to be unlikely based on their multiphase observations (\mbox{CO\,(1--0)}, \mbox{CO\,(2--1)}, \mbox{HI} and several optical, UV and IR images)

    \item \textbf{NGC~4941}: 
    Tentative evidence for outflowing  photoionized  gas is presented by \citet{Coccato2005_ngc4941}.
    Furthermore, high-resolution molecular \mbox{CO\,(3--2)} gas observations by \citet{Garcia-Burillo2021} find tentative evidence for an outflow.

    \item \textbf{NGC~5643}: 
    The presence of an outflow is recently confirmed by high-resolution molecular \mbox{CO\,(3--2)} gas observations by \citet{Garcia-Burillo2021}.
    Also, evidence for a cold molecular outflow is reported by \citet[][in \mbox{S(1)H$_{2}$\,(1--0)}]{Davies2014} and \citet[][in \mbox{CO\,(2--1)}]{Alonso-Herrero2018} and the outflow region is analysed in more detail by 
    \citet{GarciaBernadette2020} using CO and ionized gas emission. Several other studies of different gas phases find evidence for an outflow \citep[e.g.,][for ionized gas]{Cresci2015}.

\end{itemize}

    \noindent For the four galaxies that did \emph{not} fulfill the stricter \lwm\ criterion, and are not classified as secure candidates, we find the following: 
\begin{itemize}[noitemsep,topsep=0pt,leftmargin=2\parindent]
    \item \textbf{NGC~1097} is known to possess an inflow in molecular gas as well as other gas phase studies \citep[e.g.,][]{vandeVen2010, PinolFerrer2011}. 
    There is evidence for inflows as well as outflows between the central spiral arms seen in warm molecular hydrogen gas H$_2$ \citep{Davies2009}. 
    
    \item For \textbf{NGC~1637}, \textbf{NGC~4321}, and \textbf{NGC~4569} no cold molecular outflow studies are available, and only indications for hot gas outflow \citep{Immler_2003} for the former and detection of an ionized outflow \citep{Boselli2016} in the latter.

    \item For the remaining galaxies 
    \textbf{NGC~1317}, \textbf{NGC~2566}, \textbf{NGC~3507}, \textbf{NGC~3627}, \textbf{NGC~4293}, \textbf{NGC~4457}, \textbf{NGC~5134}, \textbf{NGC~7496}, no studies on molecular outflows were found in the literature.
\end{itemize}

\paragraph{PHANGS non-outflow candidate literature matches}

PHANGS galaxies that are known to possess molecular outflows in the central 1\,kpc region that we did not recognize as outflow candidates: 

\begin{itemize}[noitemsep,topsep=0pt,leftmargin=2\parindent]
    
    \item For \textbf{NGC~1566} several studies have analyzed the kinematics of molecular and other gas phases. \citet{Slater2019NGC1566} interpret their findings in \mbox{CO\,(2--1)} and ionized gas at a resolution of ${\sim}24$\,pc as outflowing gas with little evidence for an additional inflow.
    On the other hand, \citet{Combes2014NGC1566, Combes2019ALMATori} argue for the presence of an inflow in their molecular gas observations rather than an outflow in this galaxy.
    In this work we obtained an $\overline{\ocl}$ of $1.89$ for this galaxy.

    \item \citet{Davies2014} find molecular outflow signatures (\mbox{S(1)H$_{2}$\,(1--0)}) in \textbf{NGC~6300} and \textbf{NGC~7743}. Our $\overline{\ocl}$s are rather low with $1.56$ for NGC~6300 and $0.33$ for NGC~7743. 
    For NGC~6300, \citet{NGC6300Gaspar2019} interpret a broad component in ionized gas as a central outflow, but they find no outflow signatures in molecular hydrogen observations. 
    
    \item \citet{Stone2016} analysed molecular OH emission of several PHANGS galaxies (\textbf{NGC~1068}, \textbf{NGC~1365}, \textbf{NGC~1566}, \textbf{NGC~4579}, \textbf{NGC~4941}, \textbf{NGC~6300}, \textbf{NGC~7465}). 
    They find no evidence for inflows in these galaxies. 
    For the galaxies where OH absorption was available (NGC~6300), a search for galactic-scale outflows provided no evidence for an outflow.
    \item \citet{GarciaBurillo2014} find strong evidence for an molecular outflow in \textbf{NGC~1068} in \mbox{CO\,(2--1)}. The outflow is orientated along the inner stellar bar of the galaxy, as revealed by a careful modeling of the velocity distribution in their work. 
    \item \citet[][]{Yukita2012} find signs that suggest a potential outflow in X-ray emission for \textbf{NGC~2903}. 
\end{itemize}

\subsubsection{Sample comparison}
\label{sec:Discussion:Literature:Samplecomparison}

\begin{figure}[t!]
    \centering
    \includegraphics[width = 0.5\textwidth]{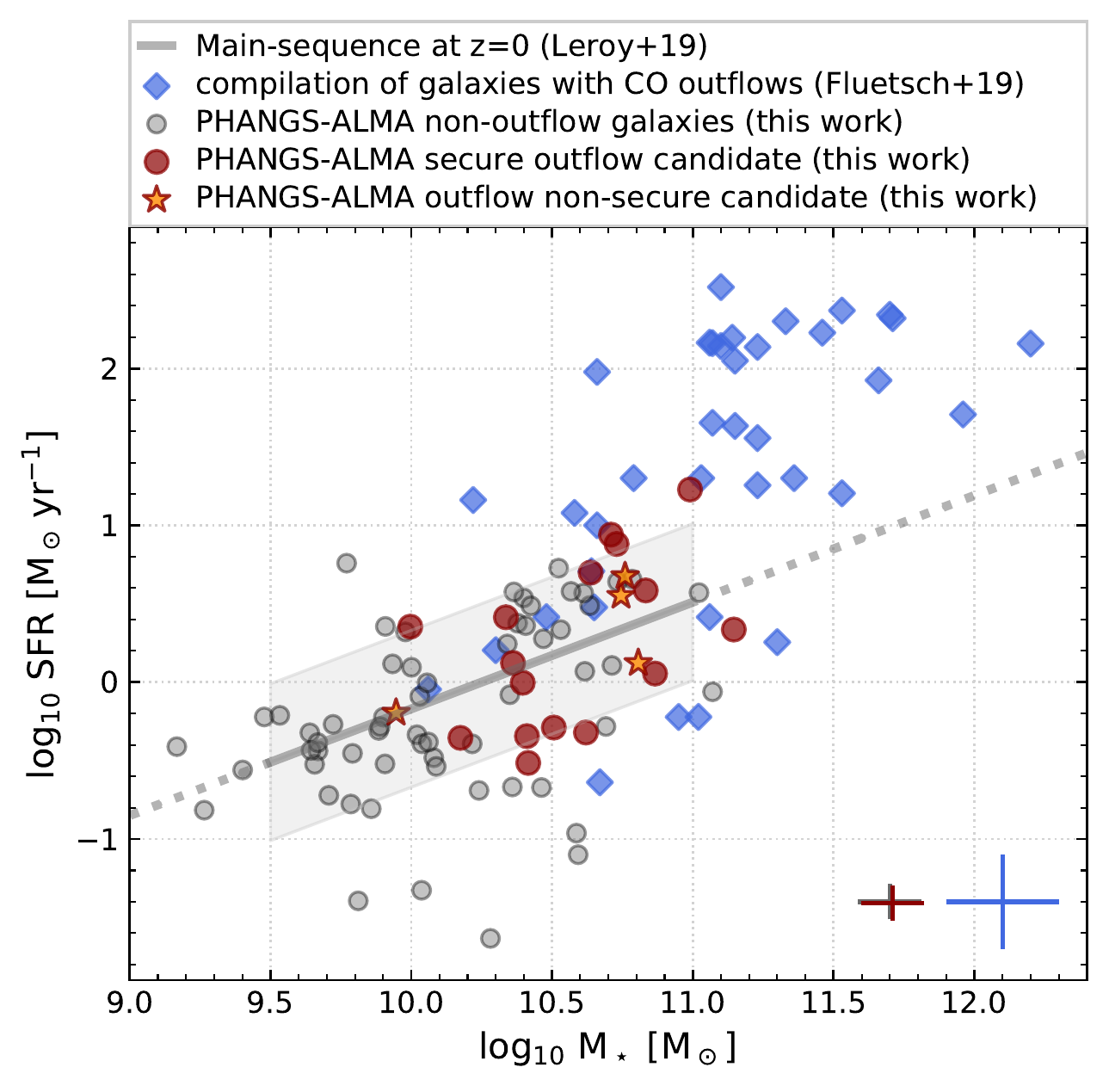}
    \caption{Outflow candidates (red circles, orange stars) compared to galaxies without outflows (gray circles) in the PHANGS sample and the compilation of outflows by \citet{Fluetsch2019} (blue diamonds). 
    The main sequence of star-forming galaxies from \citet[][grey line]{Leroy2019} with a width of $0.5$\,dex (gray shaded area) and extrapolated to higher and lower stellar masses (dotted line) is shown for reference.}
    \label{fig:Results:LitcompCandidates}
\end{figure}

The most comprehensive compilation of central molecular outflows has been assembled by \cite{Fluetsch2019}. In Figure~\ref{fig:Results:LitcompCandidates}, we compare our outflow candidates to their compiled sample in the SFR versus stellar mass plane. 
The galaxy NGC~1433 is in both the sample of \citeauthor{Fluetsch2019} as well as our outflow candidate sample, although SFR and stellar mass differ slightly.
On average our candidates overlap with the low-mass end of their sample. 
In this mass range, their galaxies are also mostly consistent with being on the main sequence of star-forming galaxies while most of their galaxies with stellar mass of $\log_{10} \left( M_{\star} / \mathrm{M}_{\odot} \right) \geq 11.0$ are in the starburst regime.
This is in line with our study where ${\sim}36$\% of the galaxies in the stellar mass range $10.0 \leq \log_{10}(M_{\star} / \mathrm{M}_{\odot}) \leq 11.0$ are outflow candidates, in fact 19 of our outflow candidates are in this mass regime.

\subsubsection{Literature outflow frequencies}
\label{sec:Discussion:Literature:Fractioncomparison}

Assuming that all of our (secure) outflow candidates correspond to actual outflows, the number of (secure) outflow candidates translates into an outflow frequency of $25\pm2\%$ ($20\pm2$\%) for the full sample of 80 galaxies used out of the 90 PHANGS galaxies. 
If we restrict it to the stellar mass range of $9.95 \leq \log_{10}(M_{\star}/\mathrm{M}_{\odot}) \leq 11.15$, where we have identified outflows, this increases to 36\%. 
This is the first time the frequency of central molecular outflows in high resolution and high sensitivity data has been robustly estimated as frequency determination in molecular gas are rare, and sample sizes are often small. For example, \citet{D-Fernandez2020} analyzed the \mbox{CO\,(2--1)} emission in five nearby Seyfert galaxies and find two of their targets being likely to host an outflow.

\citet{Wylezalek2020} used the MaNGA sample of galaxies at mean redshift of $z \sim 0.03$ to determine the outflow frequency in galaxies. Over matched stellar mass range of $9 \geq \log_{10} \left(M_{\star}/\mathrm{M}_{\odot}\right) \geq 11$, they find that 25\% of low- to intermediate luminosity AGN and 7\% of non-AGN MaNGA galaxies exhibit outflow-typical emission line widths.  
This is a lower fraction than what we expect based on our sample: as mentioned in Section~\ref{sec:Discussion:CandidatesVSPHANGS} we find 53\% (42\% secure) outflow candidates in the 19 AGN-active PHANGS galaxies and 17\% (13\% secure) outflow candidates in PHANGS galaxies without AGN. 
However, gas phase, resolution as well as detection methods are different. 

\citet{Veilleux2013} studied molecular gas outflows traced by OH in 37 local ULIRGs and QSOs, and find evidence for outflows for 70\% of their targets, which unsurprisingly is higher than for our less active sample. 

\citet{Lopez2019} analyzed ionized gas emission from 203 highly inclined disk galaxies in the CALIFA sample to identify outflow candidates over their full stellar mass range $9.5 < \log_{10}\left( M_{\star} / \mathrm{M}_\odot \right) < 11$ and in the SFR range of about $-0.7 < \log_{10} \left( \mathrm{SFR} / \mathrm{M}_\odot\,\mathrm{yr}^{-1}\right) < 0.7$, resulting in an ionized outflow frequency of 8\% for the highly inclined galaxies or 2\% for the extended CALIFA sample. 
This is in contrast to our galaxy sample, that selects galaxies of low to moderate inclination \citep[$i < 75$\,deg, see][]{Leroy20b}. This or the difference in gas phase and tracers may explain the discrepancy in frequency estimates.

\citet{Roberts-Borsani2020} studied neutral gas (traced by \NaI~D) in 405 local MaNGA galaxies with stellar masses ($M_{\star} \geq 10^{10}\,\mathrm{M}_{\odot}$) similar to our sample and find outflow evidence in the central regions of 78 galaxies (19\%). 

\citet{Stone2016} analyzed emission from OH as tracer for warm molecular gas for galactic-scale outflows in local AGN galaxies (BAT-AGN). 
They find that 24\% of their targets, in which an outflow analysis was possible, show evidence for outflows. One of the PHANGS galaxies (NGC~6300) was included in this sub-sample and no outflow was found, in agreement with our analysis.

A study of ionized gas in main sequence galaxies and galaxies near the main sequence with 
$9 < \log_{10} \left( M_{\star} / \mathrm{M}_\odot \right) < 11.7$ at redshift $0.6 < z < 2.7$ by \citet{ForsterSchreiber2019} reports a global ionized gas outflow frequency of ${\sim} 25\%$ (with ${\sim} 10\%$ being SF-driven and ${\sim} 15\%$ AGN-driven) with a strongly increasing fraction of AGN-driven outflows toward higher-mass galaxies (log$_{10}\left(M_{\star} / \mathrm{M}_{\odot} \right) \gtrsim 10.8$) to up to $3/5$. 
Although this frequency value agrees with our results, a direct comparison is not straight-forward, as the
outflow tracers, the galaxies, and their environment are very different. 
For example, our candidate molecular outflows have much smaller spatial extents and also lower outflow velocities.

In summary, our frequency of central molecular outflows of $(20{-}36)\% \pm2\%$ depending on the mass range considered agrees reasonably well with previous studies, though a direct comparison is significantly hindered by the use of different gas tracers, methods, galaxy populations, as well as the data quality.
This emphasizes the need for studies of more nearby, well resolved galaxies for a direct comparison to our results. 
Thus, in the context of galaxy evolution our numbers need to be interpreted carefully, as our frequency allows for two interpretations:
\begin{itemize}[noitemsep,topsep=0pt,leftmargin=2\parindent]
    \item[a)] One possible scenario would be to assume that \emph{all} galaxies possess outflows that are detectable with our methods. 
    This immediately implies a short outflow lifetime, so that we end up with our $20{-}36$\% frequency.
    \item[b)] Another scenario is that outflows are long-lived, and thus our frequency tells us that outflows are more rare events. 
\end{itemize}

\noindent Likely, the answer will be in-between. 
Total SFR and stellar mass play an important, but yet not fully understood role in outflow launching and future research might uncover unknown processes to answer these questions.

\subsection{Caveats and improvements}
\label{sec:Discussion:Caveats}

Although the methods used in this work are commonly applied in the literature, we note some caveats regarding the construction and evaluation of spectra, \pv--diagrams, and \lwm{}s, which add to the uncertainty of the outflow frequency estimate.

\paragraph{Construction and analysis of the method figures}
    
In general, projection effects, that make it impossible to detect outflow signatures with our methods, can lead to an underestimation of the outflow frequency.  
    
\begin{itemize}[noitemsep,topsep=0pt,leftmargin=2\parindent]
    \item Spectra and \pv--diagrams: 
    Instead of using fixed values for aperture and slit size, observationally motivated values for each galaxy (e.g., depending on galaxy size) may avoid evaluating misleading signals as outflow signs or enhance fainter outflow features. 
        
    \item \lwm{}s:
    The automated creation of \lwm{}s depends on the spectrum and its signal to noise (S/N). 
    If the spectrum has low S/N, the intersections with the 5\% and 20\% level of the peak will not contain the intended information, but only random velocity ranges between noise peaks. 
    In case of a noise-dominated spectrum, no significant amounts of cold molecular gas are present in the central galaxy region, thus a detection of a molecular outflows is also unlikely. 
    For sparsely distributed gas in the center, summed up noise can hide faint broad wings in the spectrum.
    Whereas with an optimized velocity range, these features may become visible in the \lwm{}s since the spatial information is retained.
\end{itemize}

\paragraph{Visual evaluation}
\ocl{}s were allocated by visual inspection of the figures  (Section~\ref{sec:methods:Spectrum}, \ref{sec:methods:Pv} and~\ref{sec:methods:lwm}). 

\begin{itemize}[noitemsep,topsep=0pt,leftmargin=2\parindent]
    \item Visual identification of potential outflow components is generally a less rigorously defined approach than fitting the spectra and investigating \pv--diagrams.
    However, fitting proved to be an unreliable method in tests due to the large variety of spectral shapes and S/N in our sample.

    \item Some of the \pv--diagrams and \lwm{}s showed potential low-level continuum emission or other artefacts, that may either overlap and hide actual outflow emission, or could be mistaken for outflow emission.
    As all inspectors were aware of this effect, the impact should be negligible. 

    \item In addition to the \lwm{}s, velocity fields and residual velocity fields could potentially improve the outflow identification process, However, this would result in a much more detailed and time-consuming analysis of each galaxy, which is beyond the scope of this paper. 
\end{itemize}

Overall we find a good agreement between the three authors' assigned \ocl{}s, which indicates that -- despite the presence of some caveats --  reproducibility is achieved.


\section{Summary and conclusion}
\label{sec:Summary}
We use ${\sim}1\arcsec \approx 100$\,pc resolution \mbox{CO\,(2--1)} imaging for 80 nearby, massive, star-forming galaxies from the PHANGS-ALMA survey \citep[][see also \citealt{Sun20}]{Leroy20a} to search for molecular outflows in the central regions of $300$\,pc to $2$\,kpc in diameter. 
Based on our applied methods, we identify 20 outflow candidates out of which 16 are classified as secure outflow candidates which translates into an estimated frequency of central molecular outflows in massive ($\rm 9.17 \leq \log_{10}(M_{\star}/M_{\odot}) \leq 11.15$) star-forming main sequence galaxies of $25\pm2$\% (or $20\pm2$\% for secure candidates; see Section~\ref{sec:Results:Abundance}).  

Further results are as follows: 
\begin{enumerate}
    \item We examine central spectra of two different apertures, \pv--diagrams along major and minor axes, and automatically created line-wing maps (\lwm) using velocity ranges determined from the spectra to identify kinematic signatures possibly tracing outflowing gas (Section~\ref{sec:methods}). Combining these methods ensures high robustness against other kinematic features, such as gas flows along bars. 
    The resulting outflow confidence labels (\ocl{}s) allocated via visual inspection agree well among different inspectors for each galaxy and can be used to determine the outflow candidates as well as secure outflow candidates (Section~\ref{sec:Results}, Figure~\ref{fig:Results:average_rankings}).
    Global host galaxy properties (stellar mass, SFR) show a significant correlation with these confidence labels, which could be pointing to a correlation of these labels with actual outflow incidence or a correlation of outflow properties (e.g., outflow mass and extent) with stellar mass and/or SFR, which enhance the detectability of outflow signatures (Section~\ref{sec:Discussion:HostgalCandidates}). 
    
    \item No outflow candidates are identified in galaxies with stellar masses below $\log_{10}(M_{\star}/M_{\odot}) = 9.95$ (Section~\ref{sec:Discussion:HostgalCandidates} and Figure~\ref{fig:Monsterplot}).
    The estimated frequency of central molecular outflows in the mass range $9.95 \leq \log_{10} \left( M_{\star}/M_{\odot} \right) \leq 11.15 $ is 36\%.
    This may indicate a possible stellar mass and/or SFR threshold needed to launch outflows or a threshold for our detection methods. 
    
    \item The fraction of galaxies hosting an AGN is ${\sim} 2\times$  higher among outflow candidates (50\%, secure candidates: 50\%) than the fraction in the entire sample (24\%). 
    The same applies to barred galaxies which are ${\sim} {1.5}\times$  more prominent among outflow candidates (89\%, secure candidates: 87\%) than the entire sample (61\%; Section~\ref{sec:Discussion:CandidatesVSPHANGS}, Figure~\ref{fig:Hist_CandidatesvsPHANGS}).
    This may imply that bars transport the gas necessary to build up an outflow towards the central region. 
    Similarly, AGN might provide the necessary energy to launch outflows.
    However, this may also be related to the fact that both AGN and stellar bars correlate with global host galaxy properties (Section~\ref{sec:Discussion:OCLReliability}). 
    
    \item The molecular outflows of our outflow candidates have a typical extent of ${\sim} 260\,$pc with velocities of ${\sim} 130\,$\kms as well as integrated CO luminosities of ${\sim} 8.6\,$K\,\kms, indicating that high resolution, sub-kpc, high sensitivity studies such as PHANGS are required to detect molecular outflows in main sequence galaxies (Section~\ref{sec:Discussion:Outflowrates}). 
    
    \item Our outflow candidates extend the trend between SFR and mass outflow rates down to lower outflow rates ($\log_{10} ({\dot{M}_\mathrm{out}} / \mathrm{M}_{\odot}\,\mathrm{yr}^{-1}) < 0$) (Figure~\ref{fig:Discussion:OutflrateSFR}). 
    Mass loading factors are on average near unity, thus indicating that these small and less-massive outflows remove as much gas from the centers as is consumed by SFR, and can represent an important correction to the overall future evolution of the gas.
    (Figure~\ref{fig:Discussion:EtaMstarHist}).

\end{enumerate}

Our outflow candidates significantly expand the previous compilation of molecular outflow galaxies by \citet{Fluetsch2020}, in particular for main sequence galaxies with intermediate stellar mass.
Our study suggests that central molecular outflows might be a common feature (${\sim} 1$ out of $3$~galaxies may possess an outflow) in star-forming main sequence galaxies with stellar mass $\log_{10}(M_{\star}/\mathrm{M}_{\odot}) \gtrsim 10.0$. 
Given that the outflow signatures are not very prominent in many of our galaxies, spatial resolution and sensitivity are important to identify faint outflows and to distinguish kinematic signatures from outflows and other non-circular motions in galaxies.

\begin{acknowledgements}
    We greatly appreciate the very helpful and constructive comments by the anonymous referee.

      This work was carried out as part of the PHANGS collaboration.

    ATB and FB would like to acknowledge funding from the European Research Council (ERC) under the European Union’s Horizon 2020 research and innovation programme (grant agreement No.726384/Empire).
    
    ES, HAP, TS, and TGW acknowledge funding from the European Research Council (ERC) under the European Union’s Horizon 2020 research and innovation programme (grant agreement No. 694343).
    
    RSK acknowledges financial support from the German Research Foundation (DFG) via the collaborative research center (SFB 881, Project-ID 138713538) “The Milky Way System” (subprojects A1, B1, B2, and B8). He also thanks for funding from the Heidelberg Cluster of Excellence ``STRUCTURES'' in the framework of Germany’s Excellence Strategy (grant EXC-2181/1, Project-ID 390900948) and for funding from the European Research Council via the ERC Synergy Grant ``ECOGAL'' (grant 855130). 
    
    JMDK gratefully acknowledges funding from the Deutsche Forschungsgemeinschaft (DFG, German Research Foundation) through an Emmy Noether Research Group (grant number KR4801/1-1) and the DFG Sachbeihilfe (grant number KR4801/2-1), as well as from the European Research Council (ERC) under the European Union's Horizon 2020 research and innovation programme via the ERC Starting Grant MUSTANG (grant agreement number 714907).
    
    HAP acknowledges funding from the European Research Council (ERC) under the European Union’s Horizon 2020 research and innovation programme (grant agreement No. 694343).
    
    MQ acknowledges support from the research project  PID2019-106027GA-C44 from the Spanish Ministerio de Ciencia e Innovaci\'on.
    
    ER acknowledges the support of the Natural Sciences and Engineering Research Council of Canada (NSERC), funding reference number RGPIN-2017-03987.
    
    The work of JS is partially supported by the National Science Foundation under Grants No. 1615105, 1615109, and 1653300.
    
    AU acknowledges support from the Spanish funding grants AYA2016-79006-P (MINECO/FEDER), PGC2018-094671-B-I00 (MCIU/AEI/FEDER), and PID2019-108765GB-I00 (MICINN). 
    
    This paper makes use of the following ALMA data: \linebreak
    ADS/JAO.ALMA\#2012.1.00650.S, \linebreak 
    ADS/JAO.ALMA\#2013.1.00803.S, \linebreak 
    ADS/JAO.ALMA\#2013.1.01161.S, \linebreak 
    ADS/JAO.ALMA\#2015.1.00121.S, \linebreak 
    ADS/JAO.ALMA\#2015.1.00782.S, \linebreak 
    ADS/JAO.ALMA\#2015.1.00925.S, \linebreak 
    ADS/JAO.ALMA\#2015.1.00956.S, \linebreak 
    ADS/JAO.ALMA\#2016.1.00386.S, \linebreak 
    ADS/JAO.ALMA\#2017.1.00392.S, \linebreak 
    ADS/JAO.ALMA\#2017.1.00766.S, \linebreak 
    ADS/JAO.ALMA\#2017.1.00886.L, \linebreak 
    ADS/JAO.ALMA\#2018.1.01321.S, \linebreak 
    ADS/JAO.ALMA\#2018.1.01651.S. \linebreak 
    ADS/JAO.ALMA\#2018.A.00062.S. \linebreak 
    ALMA is a partnership of ESO (representing its member states), NSF (USA) and NINS (Japan), together with NRC (Canada), MOST and ASIAA (Taiwan), and KASI (Republic of Korea), in cooperation with the Republic of Chile. The Joint ALMA Observatory is operated by ESO, AUI/NRAO and NAOJ.

\end{acknowledgements}

  \bibliographystyle{aa} 
  \bibliography{Bibliography.bib} 

\begin{appendix} 

\begin{appendix}
\label{sec:appendix}

\section{Outflow frequency threshold dependence}
\label{sec:Appendix:Frequency_threshold}

\begin{figure}
    \centering
    \includegraphics[width = 0.45\textwidth]{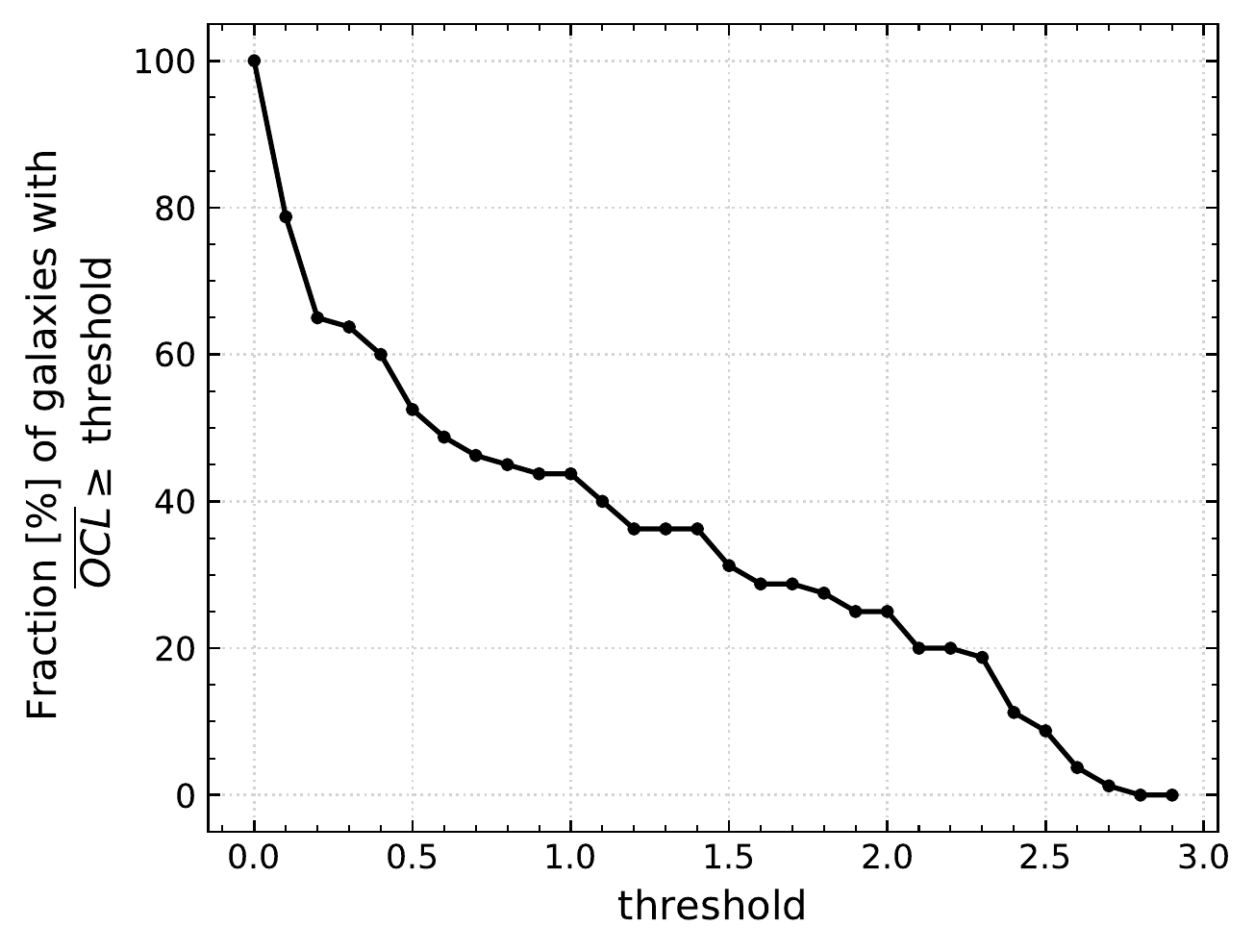}
    \caption{Frequency of galaxies with $\overline{\ocl} \geq$ threshold value, as a function of threshold in steps of 0.1 between 0.0 and 2.9 for the full sample of 80 galaxies.}
    \label{fig:frequencythreshold}
\end{figure}

We provide the outflow frequency as a function of its threshold in Figure~\ref{fig:frequencythreshold}. 
A steep decline at small thresholds ($\sim 0.0{-}0.5$), is followed by a linear trend for medium thresholds ($\sim 0.5{-}2.0$) and another steeper decline for large thresholds ($\gtrsim 2.0$). 
In this work, we set the threshold for galaxies to be considered an outflow candidate to $\overline{\ocl} \geq 2$ as well as for the secure outflow candidacy a threshold of $\overline{\ocl} \geq 2 \, \land \, \mbox{\lwm-\ocl} \geq 2$. 
We can see from this plot, that most of our galaxies ($>50\%$) have $\overline{\ocl} < 1$, thus exhibit no significant outflow signatures at all.  

\section{\ocl\ method analysis - the presence of AGN and bars}
\label{sec:Appendix:OCLreliabilityBarAGN}

In Section~\ref{sec:Discussion:OCLReliability} we found a significant ($>3\sigma$) correlation between $\overline{\ocl}$ and stellar mass as well as SFR. 
In this Section we search for secondary dependencies due to the presence of stellar bars or AGN. 

\paragraph{Bars}

\begin{figure*}[t!]
    \centering
    \textbf{Barred and unbarred sub-samples:}
    \includegraphics[width = 0.99\textwidth]{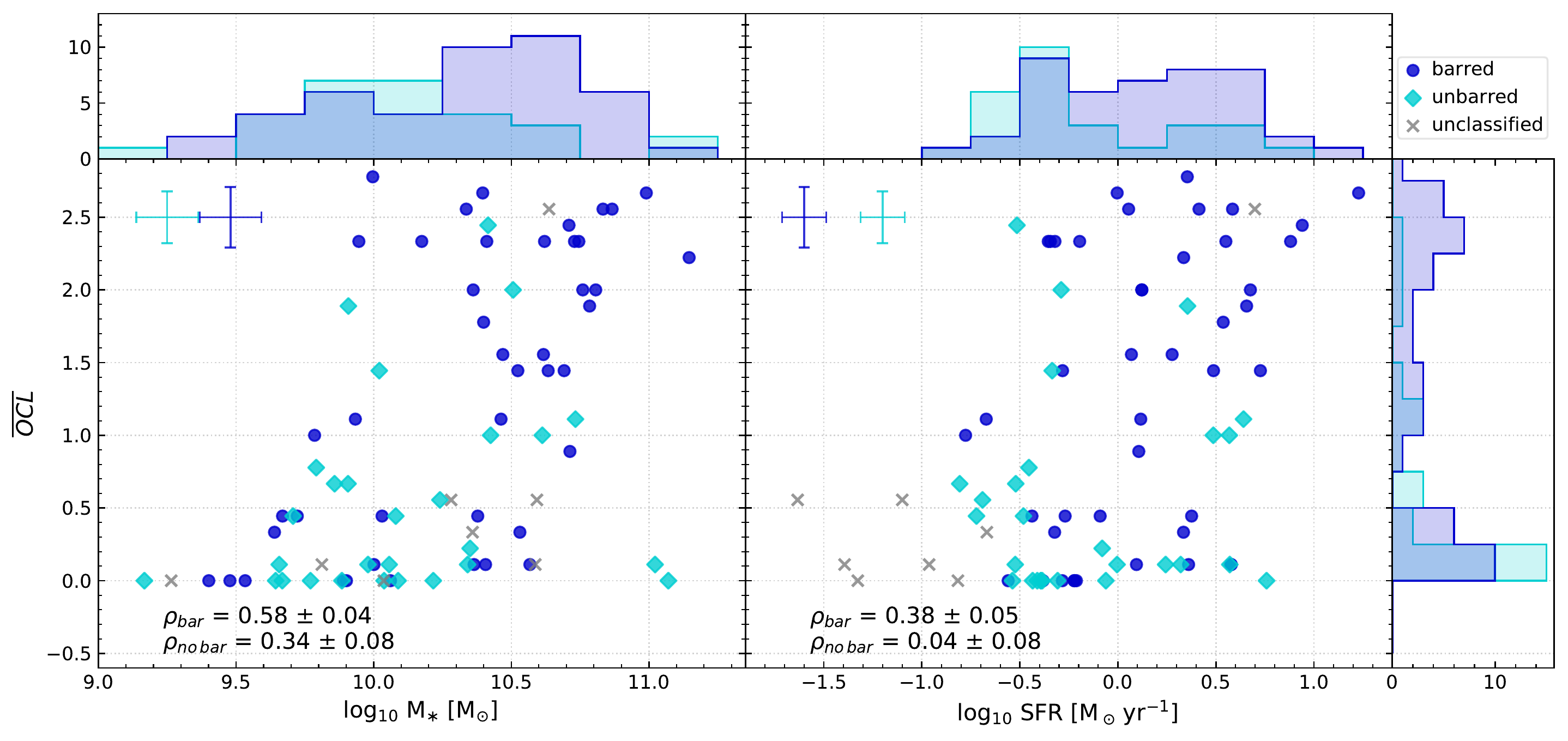}
    \textbf{AGN-active and inactive sub-samples:}
    \includegraphics[width=0.99\textwidth]{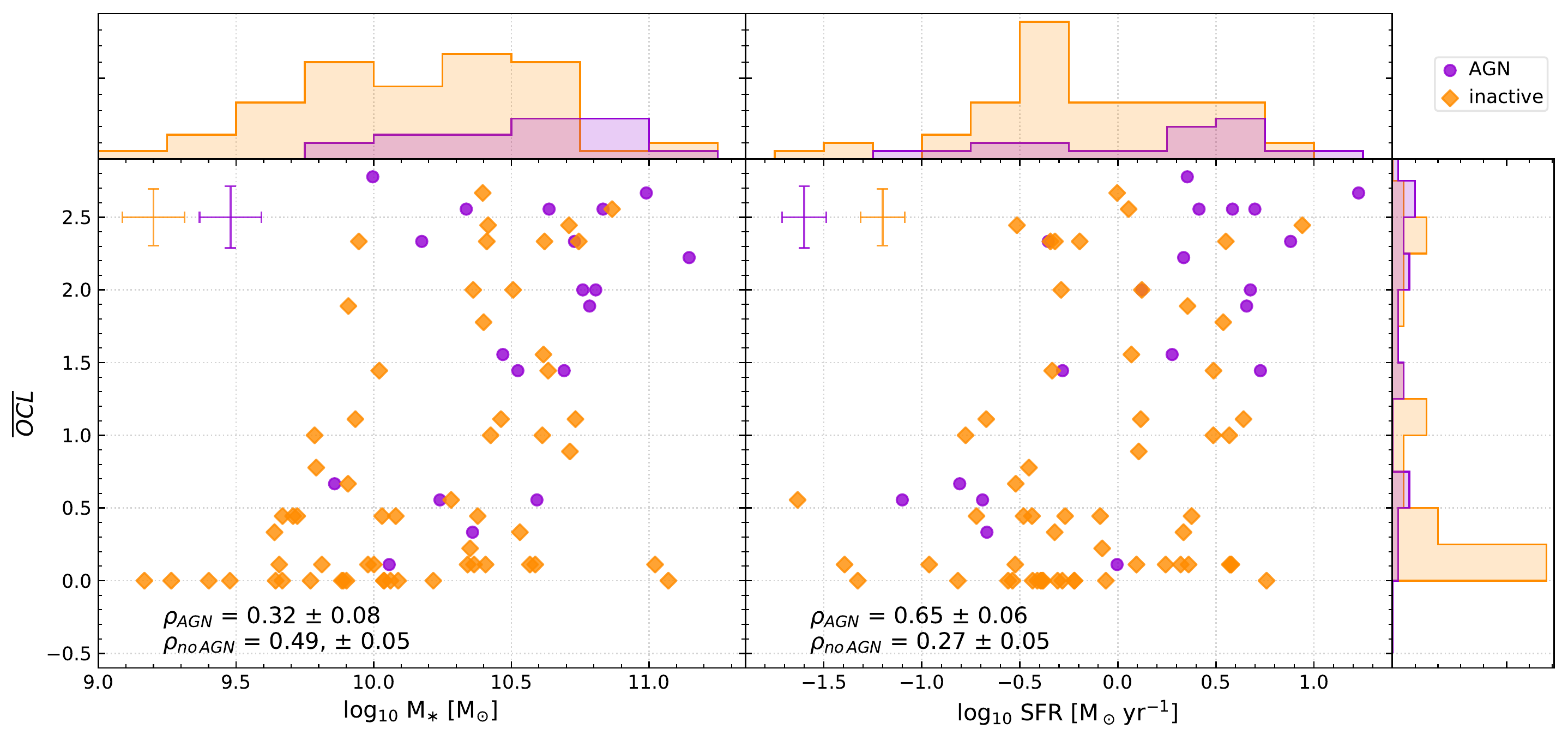}
    \caption{$\overline{\ocl}$ dependence on the presence of a stellar bar or AGN. Top row: $\overline{\ocl}$s of PHANGS galaxies as a function of the host galaxy's stellar mass (left) or SFR (right) for barred (blue circles) and unbarred galaxies (cyan diamonds) (galaxies without a bar classification: grey cross). Histograms of the barred (blue) and unbarred (cyan) sub-sample are added for stellar mass (top left), SFR (top right panel) and $\overline{\ocl}$ (far right panel). Bottom row: Analogous plot for galaxies hosting an AGN (purple circles) or not (orange diamonds). Histograms of the AGN (purple) and non-AGN (orange) sub-samples are added for stellar mass (top left), SFR (top right) and $\overline{\ocl}$s (far right panel). Spearman rank correlation coefficient between $\overline{\ocl}$ and galaxy property are listed at the bottom of each panel, for the respective sub-samples ($\rho_\mathrm{bar}$, $\rho_\mathrm{no\,bar}$, $\rho_\mathrm{AGN}$, $\rho_\mathrm{no\,AGN}$). For comparison, the Spearman rank correlation coefficient of the full sample between $\overline{\ocl}$ and stellar mass is $\rho_{M_\star} = 0.56 \pm 0.03$ and between $\overline{\ocl}$ and SFR is $\rho_\mathrm{SFR} = 0.40\pm 0.03$.
    Typical uncertainties for $\overline{\ocl}$ and the galaxy property are added in the upper left corner of each panel. }
    \label{fig:Results:MethodCompMSBar_and_AGN}
\end{figure*}

The upper panels in Figure~\ref{fig:Results:MethodCompMSBar_and_AGN} show the $\overline{\ocl}$s as a function of stellar mass (top left) and SFR (top right), divided into sub-samples of barred (blue circles) and unbarred galaxies (cyan diamonds). 
The Spearman rank correlation coefficients and their uncertainties are provided at the bottom of each panel (see also Table~\ref{tab:results:spearmanRMstarMethodsplit}), correlating $\overline{\ocl}$s with stellar mass and SFR of either barred or unbarred galaxies.
Uncertainties are derived by perturbing the $\overline{\ocl}$ and stellar mass (SFR) of all sub-samples by their errors, calculating the Spearman correlation coefficient for the perturbed values, and then repeating this process 1000 times. We infer the scatter as uncertainty.

We find equally high Spearman rank correlation coefficients between both stellar mass and $\overline{\ocl}$ as well as SFR and $\overline{\ocl}$ for barred galaxies compared to those of the full sample ($\rho_{M_{\star}} = 0.56 \pm 0.03$ and $\rho_\mathrm{SFR} = 0.40 \pm 0.03$). 
The unbarred systems have an  insignificantly ($< 3\sigma$) smaller correlation coefficient between $\overline{\ocl}$ and stellar mass. 
Also, unbarred systems exhibit a significantly different rank correlation coefficient between $\overline{\ocl}$ and SFR compared to both those of the full sample and the barred sub-sample. 
This indicates that SFR increases the likelihood of outflow signatures in barred galaxies, but not in unbarred galaxies, where the correlation coefficient is close to zero. 
However, most of the unbarred galaxies are clustered at lower SFR and low $\overline{\ocl}$s, thus, if the likelihood of outflows to occur increases with SFR, this correlation is not due to the presence of a bar, but rather due to enhanced bar fraction at higher SFRs. 
We interpret this in the following way: 
\begin{itemize}[noitemsep,topsep=0pt,leftmargin=2\parindent]
    \item[a)] Kinematic patterns of bars can be mistaken for outflows, since both outflows and streaming motions in bars exhibit substantial radial velocity components.
    However, by combining different methods we minimize the impact on the $\overline{\ocl}$s. 
    In particular, the \ocl{}s of \lwm{}s tend to show the least trends as described above. 
    \item[b)]The higher $\overline{\ocl}$ values are only the result of higher SFR and the presence of a bar is a coincidence, as bar fraction increases with SFR in our sample as seen in the histograms of Figure~\ref{fig:Results:MethodCompMSBar_and_AGN}.
    This could imply that SFR itself is the driver for the enhanced outflow probability. 
    With a higher fraction of barred galaxies to unbarred galaxies at higher SFR, the chances are higher to find more outflow candidates in barred galaxies than in unbarred galaxies, resulting in a putative correlation. 
    \item[c)] Bar instabilities can be an efficient means to transport gas into the center of galaxies \citep[e.g., bar-fed nuclear starburst in NGC~253, e.g.][]{Chown2019}. The larger gas reservoirs in the centers of barred galaxies can then lead to a higher fraction of detectable outflows relative to unbarred galaxies as well as a higher SFR. 
    This is similar to the previously mentioned enhanced correlation between central molecular gas density and $\overline{\ocl}$ (compare Section~\ref{sec:Discussion:OCLReliability}).
\end{itemize}

We find that this correlation, that barred galaxies show stronger trends between $\overline{\ocl}$ and SFR than galaxies without a bar can be explained by the properties of host galaxy parameters, but we cannot exclude that other mechanisms are also at play.
The trends for $\overline{\ocl}$ with stellar mass are not significantly different for barred and unbarred galaxies, indicating that stellar mass might directly correlate with the likelihood of outflows to occur, rather than with the presence of a bar.

We also tested for correlations between stellar mass (SFR) and outflows for the individual identification methods for the sub-samples and found no significant differences (see Table~\ref{tab:results:spearmanRMstarMethodsplit} in Appendix~\ref{sec:Appendix:OCLMethodsplit}).

\paragraph{AGN}

Here we test if the $\overline{\ocl}$ depends on the presence or absence of an AGN.
The bottom panels of Figure~\ref{fig:Results:MethodCompMSBar_and_AGN} show $\overline{\ocl}$ of host galaxies with (purple circle) and without (orange diamond) AGN as a function of stellar mass (bottom left panel) and SFR (bottom right panel). Again the Spearman rank correlation coefficients  between $\overline{\ocl}$ and stellar mass (SFR) are provided in each panel for the sub-samples. 
Uncertainties are derived in the same way as done for the barred and unbarred galaxies.

Both active and inactive sub-sample reveal significant ($>3\sigma$) correlations between stellar mass and $\overline{\ocl}$, which are in agreement within $3\sigma$ with each other and with the correlation of stellar mass and $\overline{\ocl}$ found in the full sample. 
For galaxies with an AGN the correlation coefficient between SFR and $\overline{\ocl}$ is significantly ($>3\sigma$) higher than the one for the inactive galaxies and also significantly ($>3\sigma$) higher than the values of the full sample. 
This implies that a higher SFR leads to a higher outflow confidence when an AGN is present.
Such a trend is not seen for the inactive sub-sample, which shows a slightly weaker trend between $\overline{\ocl}$ and SFR, in agreement with the one of the full sample.
The inactive and active galaxy rank correlation coefficients differ by $5\sigma$. 

However, the AGN fraction of 24\% (19 galaxies) in the
PHANGS sample of 80 galaxies is too small to perform proper statistical tests.  
Figure~\ref{fig:Results:MethodCompMSBar_and_AGN} shows the spare distribution of galaxies with AGN. 
The uncertainty of the Spearman correlation coefficient utilizes the statistical uncertainties of stellar mass and $\overline{\ocl}$, which is likely much smaller than the actual systemic uncertainties.
The observed trend is thus tentative and could also be the result of two clusters of galaxies with AGN in the SFR--$\overline{\ocl}$ plane.

On the other hand, AGN activity might actually help boost the correlation between $\overline{\ocl}$ and SFR. 
This is not unexpected, as there is a relation between SFR and AGN activity due to stellar feedback-driven feeding \citep[e.g.,][]{Davies2007}. Also, while stellar mass is distributed over large radii and varies slowly in time, SFR and AGN activity can vary significantly in time and can be more centrally concentrated, which is key for our study of central outflows.

We can interpret this trend in the following ways: 
\begin{itemize}[noitemsep,topsep=0pt,leftmargin=2\parindent]
    \item[a)] Bias in WISE SFR estimates: global SFRs are determined using WISE data. In cases where a luminous AGN is present, this can lead to a significant overestimation of the SFR, which might falsify the observed trend. 
    Among the PHANGS galaxies, such a luminous AGN is only present in NGC~1365, which should not strongly contribute to the observed trend.
    
    \item[b)] SFR correlates with AGN strength \citep[compare e.g.][]{Davies2007}: The trend between SFR and \ocl\ is actually the trend that higher SFR implies higher AGN strength, and our \ocl{}s are biased towards higher AGN strength, which could again be a trend with stellar mass,  because SFR and stellar mass are clearly correlated in this MS-sample.

    \item[c)] A recent ionized gas study by \citet{Woo2020} find that AGNs with stronger outflow strengths are hosted by galaxies with higher SFR.
    If some of the AGN galaxies possess outflows, they might show stronger outflow signatures at higher SFR, and thus we expect higher \ocl{}s at higher SFR. 
    As our final outflow candidate sample is small and only half are hosted by an AGN galaxy, this can not explain the correlation seen. 
\end{itemize}
In short, the presence of an AGN or bar seems to affect the $\overline{\ocl}$, either because of outflow kinematics, revealing an actual correlation, or due to underlying correlations with other galaxy properties instead of outflows, which we can not exclude.

Similarly as done for the correlation between bars and $\overline{\ocl}$, we perform tests whether the correlation between AGN and \ocl\ depends on the particular method used to identify and classify the outflows.
The resulting Spearman rank correlation coefficients can be found in Table~\ref{tab:results:spearmanRMstarMethodsplit} in Appendix~\ref{sec:Appendix:OCLMethodsplit}. 
Except for one value, we do not find a significant difference.

\subsection{\ocl\ method analysis -- Impact on individual methods}
\label{sec:Appendix:OCLMethodsplit}

To test if all methods are equally impacted by the presence of bars,
we calculate Spearman correlation coefficients for \ocl{}s of the individual methods (Spectra with 300\,pc diameter, spectra with annulus with outer 2\,kpc diameter, \pv--diagrams and \lwm, each averaged over the different inspectors) and stellar mass (SFR) for all galaxies (see Table~\ref{tab:results:spearmanRMstarMethodsplit} Columns (2),~(3)) as well as barred and unbarred galaxies (Columns (4)--(7)). 
We do the same for AGN and inactive galaxies (Columns (8)--(11)). 

Except for one value we do not find any significant ($>3\sigma$) difference between the Spearman rank correlation coefficients of individual methods compared to the previously presented coefficients calculated with the $\overline{\ocl}$. 
Still, we observe that \lwm-\ocl\ tend to have the least correlation coefficient among all values. 
The 2\,kpc spectra-\ocl\ on the other hand tend to have the highest correlation coefficient. 

The only significant difference can be found for the Spearman rank correlation coefficient between the 2\,kpc spectra-\ocl{} and stellar mass at $3\sigma$. 
One possible explanation might be, that some CO velocity curves of our galaxies peak at ${\sim} 1\,$kpc radius from the center \citep{Lang20}, thus roughly at the outer edge of the 2\,kpc (diameter) aperture. 
This peak can correspond to a small amount of emission which broadens the spectrum and makes it harder to identify potential outflows. This effect then results in a larger \ocl.
As more massive galaxies tend to have central disks or bulges, they also have higher rotation velocities in their central 2\,kpc. With stellar mass correlating with SFR, this could explain the found deviation. 
This is, however, a tentative assumption, as many galaxies do not show a clear peak in their rotation curve and the peaks can stretch over a range of galactic radii.

\begin{table*}
\caption{Spearman correlation coefficient $\rho_\mathrm{S}$ and $p$-values of the \ocl{}s of all 4 methods divided into sub-samples of barred and unbarred galaxies and AGN-active and inactive galaxies. \label{tab:results:spearmanRMstarMethodsplit}}
\centering
\begin{small}
\begin{tabular}{l|ll|ll|ll||ll|ll}
\hline\hline
\noalign{\smallskip}
Method  & $\rho_\mathrm{all}$ & $p_\mathrm{all}$ & $\rho_\mathrm{bar}$ & $p_\mathrm{bar}$ & $\rho_\mathrm{no\,bar}$ & $p_\mathrm{no\,bar}$ & $\rho_\mathrm{AGN}$ &  $p_\mathrm{AGN}$ & $\rho_\mathrm{no\,AGN}$ & $p_\mathrm{no\,AGN}$\\
\noalign{\smallskip}

(1) & (2) & (3) & (4) & (5) & (6) & (7) & (8) & (9) & (10) & (11)\\ 
\noalign{\smallskip}
\hline
\noalign{\smallskip}
\multicolumn{11}{c}{Stellar mass correlated with}\\
\noalign{\smallskip}
Spectrum 300$\,$pc & 0.53 $\pm$ 0.05 & $<$0.001  & 0.55 $\pm$ 0.06 & $<$0.001  & 0.32 $\pm$ 0.15 & 0.10 & 0.33 $\pm$ 0.11 & 0.17 & 0.44 $\pm$ 0.08 & $<$0.001 \\ 
Spectrum 2$\,$kpc & 0.55 $\pm$ 0.06 & $<$0.001  & 0.62 $\pm$ 0.07 & $<$0.001  & 0.44 $\pm$ 0.12 & 0.02 & 0.68 $\pm$ 0.08 & 0.001 & 0.45 $\pm$ 0.09 & $<$0.001 \\ 
\pv--diagram & 0.53 $\pm$ 0.04 & $<$0.001  & 0.57 $\pm$ 0.05 & $<$0.001  & 0.30 $\pm$ 0.10 & 0.12 & 0.40 $\pm$ 0.10 & 0.09 & 0.45 $\pm$ 0.05 & $<$0.001 \\ 
\lwm\ & 0.44 $\pm$ 0.05 & $<$0.001  & 0.46 $\pm$ 0.06 & 0.002 & 0.19 $\pm$ 0.11 & 0.34 & 0.18 $\pm$ 0.1 & 0.47 & 0.39 $\pm$ 0.06 & 0.002\\ 
\noalign{\smallskip}
\multicolumn{11}{c}{SFR correlated with}\\
\noalign{\smallskip}
Spectrum 300$\,$pc & 0.44 $\pm$ 0.05 & $<$0.001  & 0.33 $\pm$ 0.06 & 0.03 & 0.19 $\pm$ 0.14 & 0.34 & 0.55 $\pm$ 0.09 & 0.015 & 0.3 $\pm$ 0.08 & 0.02\\ 
Spectrum 2$\,$kpc & 0.47 $\pm$ 0.06 & $<$0.001  & 0.44 $\pm$ 0.08 & 0.003 & 0.28 $\pm$ 0.14 & 0.15 & 0.64 $\pm$ 0.09 & 0.003 & 0.35 $\pm$ 0.09 & 0.007\\ 
\pv--diagram & 0.38 $\pm$ 0.04 & 0.001 & 0.37 $\pm$ 0.06 & 0.01 & -0.04 $\pm$ 0.11 & 0.83 & 0.56 $\pm$ 0.08 & 0.01 & 0.22 $\pm$ 0.06 & 0.09\\ 
\lwm\ & 0.33 $\pm$ 0.05 & 0.003 & 0.26 $\pm$ 0.06 & 0.09 & 0.04 $\pm$ 0.12 & 0.82 & 0.46 $\pm$ 0.09 & 0.05 & 0.2 $\pm$ 0.06 & 0.12\\ 
\noalign{\smallskip}
\hline
\end{tabular}
\end{small}
\tablefoot{Spearman rank correlation coefficient ($\rho$) and $p$-value ($p$) for all combinations between stellar mass and the individual method \ocl{}s (top half) and SFR and the individual method \ocl{}s used (bottom half).  
Columns (2) and (3) refer to the coefficients calculated for the full sample. 
Columns (4)--(7) present the Spearman coefficients for the sub-sample of barred galaxies ($R, \, p_\mathrm{bar}$) and unbarred galaxies ($R, \, p_\mathrm{no\,bar}$), as described in Section~\ref{sec:Appendix:OCLreliabilityBarAGN}.
Columns (8)--(11) present the Spearman coefficients when dividing the full sample into sub-samples of active galaxies ($R, \, p_\mathrm{AGN}$) and inactive galaxies ($R, \, p_\mathrm{no\,AGN}$), see Section~\ref{sec:Appendix:OCLreliabilityBarAGN}. Uncertainties of $\rho$ are achieved via the standard deviation when repeatedly (${\sim}1000$ times) perturbing the \ocl{}s and re-calculating the Spearman coefficients. }.

\end{table*}

\section{PHANGS galaxy properties}
\label{sec:Appendix:PHANGSproperties}
We list the global properties of the 80 PHANGS galaxies used in this analysis in Table~\ref{tab:PHANGSPropertyTable}.

\longtab[1]{
\begin{landscape}
\begin{small}
\begin{longtable}{lrrrrrrrrrrr}

\caption{PHANGS galaxy properties \label{tab:PHANGSPropertyTable}}\\
\noalign{\smallskip}
\hline\hline 
\noalign{\smallskip}
Name & Distance & RA & DEC & PA &  i & v$_{\mathrm{sys}}$ & Bar& AGN & log$_{10}$\,M$_{\ast}$ & log$_{10}$\,SFR & log$_{10}$\,SFR$_{\mathrm{center}}$\\ 
\noalign{\smallskip}
 & (Mpc) & (deg) & (deg) & (deg) & (deg) &(\kms) & y/n &y/n & ($\log{\mathrm{M}_\odot}$) & ($\log{\mathrm{M}_\odot\, \mathrm{yr}^{-1}}$) & ($\log{\mathrm{M}_\odot\, \mathrm{yr}^{-1}}$)\\
\noalign{\smallskip}
(1) & (2) & (3) & (4) & (5) & (6) & (7) & (8) & (9) & (10) & (11) & (12)\\ 
\noalign{\smallskip}
\hline
\noalign{\smallskip}
IC~1954 & 12.80 $\pm$ 0.07 & 52.879707 & -51.90486 & 63.4 $\pm$ 0.2 & 57.1 & 1039.12 & 1 & 0& 9.67 $\pm$ 0.11 & -0.44 $\pm$ 0.11 & -1.668 $\pm$ 0.002\\
IC~5273 & 14.18 $\pm$ 0.06 & 344.86118 & -37.70284 & 234 $\pm$ 2 & 52.0 & 1285.98 & 1 & 0&9.72 $\pm$ 0.11 & -0.27 $\pm$ 0.11 & -1.555 $\pm$ 0.003\\
IC~5332 & 9.01 $\pm$ 0.02 & 353.61453 & -36.10108 & 74 $\pm$ 10 & 26.9 & 699.30 & 0 & 0&9.67 $\pm$ 0.11 & -0.39 $\pm$ 0.11 & -2.77 $\pm$ 0.02\\
NGC~0253 & 3.70 $\pm$ 0.01 & 11.887966 & -25.288443 & 52 $\pm$ 10 & 75.0 & 235.36 & ... & 1 &10.64 $\pm$ 0.11 & 0.70 $\pm$ 0.11 & -0.145 $\pm<$0.001 \\
NGC~0300 & 2.09 $\pm$ 0.02 & 13.723024 & -37.684475 & 114 $\pm$ 10 & 39.8 & 155.46 & ... & 0& 9.26 $\pm$ 0.11 & -0.82 $\pm$ 0.11 & -2.744 $\pm$ 0.011\\
NGC~0628 & 9.84 $\pm$ 0.03 & 24.173855 & 15.783643 & 20.7 $\pm$ 1.0 & 8.9 & 650.75 & 0 & 0& 10.34 $\pm$ 0.11 & 0.24 $\pm$ 0.11 & -1.971 $\pm$ 0.004\\
NGC~0685 & 19.94 $\pm$ 0.06 & 26.928452 & -52.76198 & 101 $\pm$ 3 & 23.0 & 1346.65 & 1 & 0& 10.06 $\pm$ 0.11 & -0.38 $\pm$ 0.11 & -2.164 $\pm$ 0.018\\
NGC~1087 & 15.85 $\pm$ 0.06 & 41.60492 & -0.498717 & 359.1 $\pm$ 1.2 & 42.9 & 1501.53 & 1 & 0&9.93 $\pm$ 0.11 & 0.12 $\pm$ 0.11 & -1.063 $\pm$ 0.001\\
NGC~1097 & 13.58 $\pm$ 0.06 & 41.578957 & -30.274675 & 122 $\pm$ 4 & 48.6 & 1257.52 & 1 & 1& 10.76 $\pm$ 0.11 & 0.68 $\pm$ 0.11 & -0.230 $\pm<$ 0.001\\
NGC~1300 & 18.99 $\pm$ 0.06 & 49.920815 & -19.411114 & 278.0 $\pm$ 1.0 & 31.8 & 1545.35 & 1 & 0 & 10.62 $\pm$ 0.11 & 0.07 $\pm$ 0.11 & -1.576 $\pm$ 0.003\\
NGC~1317 & 19.11 $\pm$ 0.02 & 50.68454 & -37.10379 & 222 $\pm$ 3 & 23.2 & 1930.53 & 1 & 0& 10.62$\pm$ 0.11 & -0.32 $\pm$ 0.11 & -1.457 $\pm$ 0.0012\\
NGC~1365 & 19.57 $\pm$ 0.02 & 53.40152 & -36.140404 & 201 $\pm$ 8 & 55.4 & 1613.31 & 1 & 1& 10.99 $\pm$ 0.11 & 1.23 $\pm$ 0.11 & 0.296 $\pm<$ 0.001\\
NGC~1385 & 17.22 $\pm$ 0.06 & 54.369015 & -24.501162 & 181 $\pm$ 5 & 44.0 & 1476.80 & 0 & 0& 9.98 $\pm$ 0.11 & 0.32 $\pm$ 0.11 & -0.932 $\pm<$ 0.001\\
NGC~1433 & 18.63 $\pm$ 0.04 & 55.506195 & -47.221943 & 199.7 $\pm$ 0.3 & 28.6 & 1057.38 & 1 & 0& 10.87 $\pm$ 0.11 & 0.05 $\pm$ 0.11 & -1.823 $\pm$ 0.002\\
NGC~1511 & 15.28 $\pm$ 0.06 & 59.90246 & -67.63393 & 297 $\pm$ 2 & 72.7 & 1331.01 & 0 & 0&9.91 $\pm$ 0.11 & 0.36 $\pm$ 0.11 & -0.729 $\pm<$ 0.001 \\
NGC~1512 & 18.83 $\pm$ 0.04 & 60.975574 & -43.348724 & 262 $\pm$ 4 & 42.5 & 871.43 & 1 & 0& 10.71 $\pm$ 0.11 & 0.11$\pm$ 0.11 & -1.426 $\pm$ 0.001\\
NGC~1546 & 17.69 $\pm$ 0.05 & 63.65122 & -56.060898 & 147.8 $\pm$ 0.4 & 70.3 & 1243.81 & 0 & 0& 10.35 $\pm$ 0.11 & -0.08 $\pm$ 0.11 & -1.034 $\pm<$ 0.001\\
NGC~1559 & 19.44 $\pm$ 0.01 & 64.40238 & -62.78341 & 245 $\pm$ 3 & 65.4 & 1275.22 & 1 & 0& 10.36 $\pm$ 0.11 & 0.58 $\pm$ 0.11 & -1.165 $\pm$ 0.001\\
NGC~1566 & 17.69 $\pm$ 0.05 & 65.00159 & -54.93801 & 215 $\pm$ 4 & 29.5 & 1483.27 & 1 & 1& 10.78  $\pm$ 0.11 & 0.66 $\pm$ 0.11 & -1.020 $\pm<$ 0.001\\
NGC~1637 & 11.70 $\pm$ 0.04 & 70.36741 & -2.857962 & 21 $\pm$ 10 & 31.1 & 698.91 & 1 & 0&9.95 $\pm$ 0.11 & -0.19 $\pm$ 0.11 & -1.061 $\pm<$ 0.001\\
NGC~1672 & 19.40 $\pm$ 0.06 & 71.42704 & -59.247257 & 134.3 $\pm$ 0.4 & 42.6 & 1318.26 & 1 & 1& 10.72 $\pm$ 0.11 & 0.88 $\pm$ 0.11 & -0.086 $\pm<$ 0.001\\
NGC~1792 & 16.20 $\pm$ 0.06 & 76.30969 & -37.98056 & 318.9 $\pm$ 0.9 & 65.1 & 1175.94 & 0 & 0& 10.61 $\pm$ 0.11 & 0.57 $\pm$ 0.11 & -0.976 $\pm<$ 0.001\\
NGC~1809 & 19.95 $\pm$ 0.11 & 75.52066 & -69.56794 & 138 $\pm$ 9 & 57.6 & 1290.40 & 0 & 0& 9.77 $\pm$ 0.11 & 0.76 $\pm$ 0.11 & -2.023 $\pm$ 0.015\\
NGC~2090 & 11.75 $\pm$ 0.03 & 86.75787 & -34.2506 & 192.5 $\pm$ 0.6 & 64.5 & 898.19 & 0 & 0& 10.04 $\pm$ 0.11 & -0.39 $\pm$ 0.11 & -1.930 $\pm$ 0.005\\
NGC~2283 & 13.68 $\pm$ 0.06 & 101.46997 & -18.2108 & -4.1 $\pm$ 1.0 & 43.7 & 821.90 & 1 & 0&9.89 $\pm$ 0.11 & -0.28 $\pm$ 0.11 & -1.841 $\pm$ 0.006\\
NGC~2566 & 23.44 $\pm$ 0.06 & 124.69003 & -25.49952 & 312 $\pm$ 2 & 48.5 & 1609.59 & 1 & 0& 10.71 $\pm$ 0.11 & 0.94 $\pm$ 0.11 & 0.099 $\pm<$ 0.001\\
NGC~2775 & 23.15 $\pm$ 0.06 & 137.58395 & 7.038066 & 156.50 $\pm$ 0.10 & 41.2 & 1339.22 & 0 & 0& 11.07  $\pm$ 0.11 & -0.06 $\pm$ 0.11 & -2.281 $\pm$ 0.038\\
NGC~2835 & 12.22 $\pm$ 0.03 & 139.47044 & -22.35468 & 1.0 $\pm$ 1.0 & 41.3 & 867.28 & 1 & 0& 10.00 $\pm$ 0.11 & 0.09 $\pm$ 0.11 & -2.289 $\pm$ 0.013\\
NGC~2903 & 10.00 $\pm$ 0.08 & 143.04211 & 21.500841 & 204 $\pm$ 2 & 66.8 & 546.96 & 1 & 0& 10.63  $\pm$ 0.11 & 0.49 $\pm$ 0.11 & -0.661 $\pm<$ 0.001\\
NGC~2997 & 14.06 $\pm$ 0.08 & 146.41164 & -31.19109 & 108.1 $\pm$ 0.7 & 33.0 & 1076.92 & 0 & 0& 10.73$\pm$ 0.11 & 0.64 $\pm$ 0.11 & -1.039 $\pm<$ 0.001\\
NGC~3059 & 20.23 $\pm$ 0.08 & 147.534 & -73.922195 & -15 $\pm$ 3 & 29.4 & 1236.55 & 1 & 0& 10.38 $\pm$ 0.11 & 0.38 $\pm$ 0.11 & -0.867 $\pm<$ 0.001\\
NGC~3137 & 16.37 $\pm$ 0.06 & 152.28116 & -29.0643 & -0.3 $\pm$ 0.5 & 70.3 & 1086.58 & 0 & 0& 9.88 $\pm$ 0.11 & -0.31 $\pm$ 0.11 & -2.172 $\pm$ 0.014\\
NGC~3239 & 10.86 $\pm$ 0.04 & 156.27031 & 17.163702 & 73 $\pm$ 10 & 60.3 & 748.34 & 0 & 0& 9.17 $\pm$ 0.11 & -0.41 $\pm$ 0.11 & -2.547 $\pm$ 0.019\\
NGC~3351 & 9.96 $\pm$ 0.01 & 160.99065 & 11.70367 & 193 $\pm$ 2 & 45.1 & 774.74 & 1 & 0& 10.36 $\pm$ 0.11 & 0.12 $\pm$ 0.11 & -0.869 $\pm<$ 0.001\\
NGC~3489 & 11.86 $\pm$ 0.06 & 165.07736 & 13.90123 & 70 $\pm$ 10 & 63.68 & 692.10 & ... & 0&10.28 $\pm$ 0.11 & -1.63 $\pm$ 0.11 & -2.053 $\pm$ 0.007\\
NGC~3507 & 23.55 $\pm$ 0.07 & 165.85573 & 18.13552 & 55.8 $\pm$ 1.3 & 21.7 & 969.42 & 1 & 0& 10.40$\pm$ 0.11 & -0.004 $\pm$ 0.11 & -1.506 $\pm$ 0.007\\
NGC~3511 & 13.94 $\pm$ 0.06 & 165.84921 & -23.086714 & 256.8 $\pm$ 0.8 & 75.1 & 1096.72 & 1 & 0& 10.03  $\pm$ 0.11 & -0.09 $\pm$ 0.11 & -1.448 $\pm$ 0.003\\
NGC~3521 & 13.24 $\pm$ 0.06 & 166.4524 & -0.035949 & 343.0 $\pm$ 0.6 & 68.8 & 797.96 & 0 & 0& 11.02 $\pm$ 0.11 & 0.57 $\pm$ 0.11 & -1.146 $\pm$ 0.001\\
NGC~3596 & 11.30 $\pm$ 0.04 & 168.7758 & 14.787066 & 78.4 $\pm$ 1.0 & 25.1 & 1187.92 & 0 & 0&9.66 $\pm$ 0.11 & -0.52 $\pm$ 0.11 & -1.824 $\pm$ 0.004\\
NGC~3599 & 19.86 $\pm$ 0.06 & 168.86229 & 18.110376 & 42 $\pm$ 10 & 23.0 & 836.76 & ... & 0&10.04 $\pm$ 0.11 & -1.33 $\pm$ 0.11 & -1.970 $\pm$ 0.014\\
NGC~3621 & 7.06 $\pm$ 0.02 & 169.56792 & -32.8126 & 343.8 $\pm$ 0.3 & 65.8 & 724.31 & 0 & 1& 10.06 $\pm$ 0.11 & -0.004 $\pm$ 0.11 & -1.544 $\pm$ 0.001\\
NGC~3626 & 20.05 $\pm$ 0.05 & 170.01588 & 18.356846 & 165 $\pm$ 2 & 46.6 & 1470.73 & 1 & 0& 10.46 $\pm$ 0.11 & -0.67 $\pm$ 0.11 & -1.362 $\pm$ 0.004\\
NGC~3627 & 11.32 $\pm$ 0.02 & 170.06252 & 12.9915 & 173 $\pm$ 4 & 57.3 & 715.36 & 1 & 1& 10.83$\pm$ 0.11 & 0.59 $\pm$ 0.11 & -1.086 $\pm<$ 0.001\\
NGC~4207 & 15.78 $\pm$ 0.06 & 183.87682 & 9.584928 & 122 $\pm$ 2 & 64.5 & 606.63 & 0 & 0&9.71 $\pm$ 0.11 & -0.72 $\pm$ 0.11 & -1.51 $\pm$ 0.003\\
NGC~4254 & 13.10 $\pm$ 0.06 & 184.7068 & 14.416412 & 68.1 $\pm$ 0.5 & 34.4 & 2388.19 & 0 & 0& 10.42 $\pm$ 0.11 & 0.49 $\pm$ 0.11 & -1.061 $\pm<$ 0.001\\
NGC~4293 & 15.76 $\pm$ 0.06 & 185.30347 & 18.382574 & 48 $\pm$ 2 & 65.0 & 926.20 & 0 & 0& 10.51$\pm$ 0.11 & -0.29 $\pm$ 0.11 & -0.942 $\pm<$ 0.001\\
NGC~4298 & 14.92 $\pm$ 0.04 & 185.3865 & 14.60611 & 313.9 $\pm$ 0.7 & 59.2 & 1138.12 & 0 & 0&10.02$\pm$ 0.11 & -0.34 $\pm$ 0.11 & -1.638 $\pm$ 0.003\\
NGC~4303 & 16.99 $\pm$ 0.07 & 185.47888 & 4.473744 & 312 $\pm$ 3 & 23.5 & 1559.84 & 1 & 1& 10.52 $\pm$ 0.11 & 0.73 $\pm$ 0.11 & -0.913 $\pm$ 0.001\\
NGC~4321 & 15.21 $\pm$ 0.01 & 185.72887 & 15.822304 & 156.2 $\pm$ 1.7 & 38.5 & 1572.31 & 1 & 0& 10.75 $\pm$ 0.11 & 0.55 $\pm$ 0.11 & -0.827 $\pm<$ 0.001\\
NGC~4424 & 16.20 $\pm$ 0.02 & 186.7982 & 9.420637 & 88 $\pm$ 2 & 58.2 & 447.38 & 0 & 0&9.91 $\pm$ 0.11 & -0.52 $\pm$ 0.11 & -1.265 $\pm$ 0.002\\
NGC~4457 & 15.10 $\pm$ 0.05 & 187.24593 & 3.57062 & 79 $\pm$ 2 & 17.4 & 886.02 & 0 & 0& 10.42 $\pm$ 0.11 & -0.52 $\pm$ 0.11 & -1.282 $\pm$ 0.002\\
NGC~4476 & 17.54 $\pm$ 0.06 & 187.49622 & 12.348649 & 27 $\pm$ 10 & 60.14 & 1962.67 & ... & 0& 9.81 $\pm$ 0.11 & -1.39 $\pm$ 0.11 & -2.086 $\pm$ 0.013\\
NGC~4477 & 15.76 $\pm$ 0.06 & 187.50917 & 13.636418 & 26 $\pm$ 10 & 33.51 & 1362.18 & ... & 1& 10.59 $\pm$ 0.11 & -1.10 $\pm$ 0.11 & -2.325 $\pm$ 0.018\\
NGC~4496 & 14.86 $\pm$ 0.03 & 187.91354 & 3.939608 & 51 $\pm$ 4 & 53.8 & 1721.78 & 1 & &9.53 $\pm$ 0.11 & -0.21 $\pm$ 0.11 & -2.132 $\pm$ 0.012\\
NGC~4535 & 15.77 $\pm$ 0.01 & 188.5846 & 8.197973 & 179.7 $\pm$ 1.6 & 44.7 & 1953.60 & 1 & 0& 10.53 $\pm$ 0.11 & 0.34 $\pm$ 0.11 & -1.018 $\pm$ 0.0014\\
NGC~4536 & 16.25 $\pm$ 0.03 & 188.61278 & 2.188243 & 306 $\pm$ 2 & 66.0 & 1794.59 & 1 & 0&10.40$\pm$ 0.11 & 0.54$\pm$ 0.11 & -0.253 $\pm<$ 0.001\\
NGC~4540 & 15.76 $\pm$ 0.06 & 188.71193 & 15.551724 & 13 $\pm$ 4 & 28.7 & 1286.53 & 1 & 0&9.79 $\pm$ 0.11 & -0.78 $\pm$ 0.11 & -1.956 $\pm$ 0.008\\
NGC~4548 & 16.22 $\pm$ 0.01 & 188.86024 & 14.496331 & 138 $\pm$ 2 & 38.3 & 482.71 & 1 & 1&10.69 $\pm$ 0.11 & -0.28 $\pm$ 0.11 & -1.992 $\pm$ 0.009\\
NGC~4569 & 15.76 $\pm$ 0.06 & 189.2076 & 13.162875 & 18 $\pm$ 2 & 70.0 & -225.60 & 1 & 1& 10.81 $\pm$ 0.11 & 0.12 $\pm$ 0.11 & -0.821 $\pm<$ 0.001\\
NGC~4571 & 14.90 $\pm$ 0.03 & 189.23492 & 14.217327 & 217.5 $\pm$ 0.6 & 32.7 & 342.97 & 0 & 0&10.09 $\pm$ 0.11 & -0.54 $\pm$ 0.11 & -2.368 $\pm$ 0.016\\
NGC~4579 & 21.00 $\pm$ 0.04 & 189.43138 & 11.818217 & 91.3 $\pm$ 1.6 & 40.22 & 1516.75 & 1 & 1& 11.15 $\pm$ 0.11 & 0.34 $\pm$ 0.11 & -1.126 $\pm$ 0.003\\
NGC~4596 & 15.76 $\pm$ 0.06 & 189.98308 & 10.176163 & 120 $\pm$ 10 & 36.56 & 1883.34 & ... & 0& 10.59 $\pm$ 0.11 & -0.96 $\pm$ 0.11 & -2.206 $\pm$ 0.014\\
NGC~4654 & 21.98 $\pm$ 0.02 & 190.98575 & 13.126715 & 123.2 $\pm$ 1.0 & 55.6 & 1051.51 & 1 & 0& 10.57 $\pm$ 0.11 & 0.58 $\pm$ 0.11 & -1.007 $\pm$ 0.002\\
NGC~4689 & 15.00 $\pm$ 0.06 & 191.9399 & 13.762724 & 164.1 $\pm$ 0.3 & 38.7 & 1614.18 & 0 & 0& 10.22 $\pm$ 0.11 & -0.39 $\pm$ 0.11 & -1.780 $\pm$ 0.005\\
NGC~4694 & 15.76 $\pm$ 0.06 & 192.0627 & 10.983726 & 143 $\pm$ 2 & 60.7 & 1168.38 & 0 & 1&9.86 $\pm$ 0.11 & -0.81 $\pm$ 0.11 & -1.659 $\pm$ 0.004\\
NGC~4731 & 13.28 $\pm$ 0.06 & 192.75504 & -6.392839 & 255 $\pm$ 2 & 64.0 & 1483.60 & 1 & 0&9.48 $\pm$ 0.11 & -0.22 $\pm$ 0.11 & -1.967 $\pm$ 0.007\\
NGC~4781 & 11.31 $\pm$ 0.04 & 193.59917 & -10.537116 & 290.0 $\pm$ 1.3 & 59.0 & 1248.30 & 1 & 0&9.64 $\pm$ 0.11 & -0.32 $\pm$ 0.11 & -1.627 $\pm$ 0.003\\
NGC~4826 & 4.41 $\pm$ 0.02 & 194.18184 & 21.683083 & 293.6 $\pm$ 1.2 & 59.1 & 409.68 & 0 & 1& 10.24 $\pm$ 0.11 & -0.69 $\pm$ 0.11 & -1.450 $\pm$ 0.002\\
NGC~4941 & 15.00 $\pm$ 0.13 & 196.05461 & -5.551536 & 202.2 $\pm$ 0.6 & 53.4 & 1116.02 & 1 & 1& 10.17 $\pm$ 0.11 & -0.36 $\pm$ 0.11 & -1.229 $\pm$ 0.002\\
NGC~4951 & 15.00 $\pm$ 0.11 & 196.28214 & -6.493824 & 91.2 $\pm$ 0.5 & 70.2 & 1176.12 & 0 & 0&9.79 $\pm$ 0.11 & -0.45 $\pm$ 0.11 & -1.449 $\pm$ 0.002\\
NGC~5042 & 16.78 $\pm$ 0.06 & 198.8792 & -23.983883 & 190.6 $\pm$ 0.8 & 49.4 & 1385.60 & 1 & 0&9.90 $\pm$ 0.11 & -0.22 $\pm$ 0.11 & -2.176 $\pm$ 0.016\\
NGC~5068 & 5.20 $\pm$ 0.02 & 199.72807 & -21.038744 & 342 $\pm$ 3 & 35.7 & 667.21 & 1 & 0&9.40 $\pm$ 0.11 & -0.56 $\pm$ 0.11 & -2.315 $\pm$ 0.012\\
NGC~5134 & 19.92 $\pm$ 0.06 & 201.32726 & -21.134195 & 312 $\pm$ 2 & 22.7 & 1749.12 & 1 & 0& 10.41 $\pm$ 0.11 & -0.34 $\pm$ 0.11 & -1.962 $\pm$ 0.010\\
NGC~5248 & 14.87 $\pm$ 0.04 & 204.38336 & 8.885195 & 109 $\pm$ 4 & 47.4 & 1163.05 & 1 & 0& 10.41 $\pm$ 0.11 & 0.36 $\pm$ 0.11 & -0.908 $\pm<$ 0.001\\
NGC~5530 & 12.27 $\pm$ 0.06 & 214.6138 & -43.38826 & 305.4 $\pm$ 1.0 & 61.9 & 1183.20 & 0 & 0& 10.08 $\pm$ 0.11 & -0.48 $\pm$ 0.11 & -1.822 $\pm$ 0.005\\
NGC~5643 & 12.68 $\pm$ 0.02 & 218.1699 & -44.17461 & 319 $\pm$ 2 & 29.9 & 1191.34 & 1 & 1& 10.34 $\pm$ 0.11 & 0.41 $\pm$ 0.11 & -0.569 $\pm<$ 0.001\\
NGC~6300 & 11.58 $\pm$ 0.06 & 259.2478 & -62.82055 & 105 $\pm$ 2 & 49.6 & 1102.15 & 1 & 1& 10.47 $\pm$ 0.11 & 0.28 $\pm$ 0.11 & -0.669 $\pm<$ 0.001\\
NGC~7456 & 15.70 $\pm$ 0.06 & 345.54306 & -39.569412 & 16 $\pm$ 3 & 67.3 & 1192.28 & 0 & 0&9.64 $\pm$ 0.11 & -0.43 $\pm$ 0.11 & -2.537 $\pm$ 0.029\\
NGC~7496 & 18.72 $\pm$ 0.06 & 347.44702 & -43.42785 & 194 $\pm$ 4 & 35.9 & 1639.16 & 1 & 1& 10.00 $\pm$ 0.11 & 0.35 $\pm$ 0.11 & -0.437$\pm<$ 0.001\\
NGC~7743 & 20.32 $\pm$ 0.06 & 356.08804 & 9.934028 & 86 $\pm$ 10 & 37.1 & 1687.27 & ... & 1& 10.36 $\pm$ 0.11 & -0.67 $\pm$ 0.11 & -1.570 $\pm$ 0.006\\
\end{longtable}
\end{small}
\tablefoot{List of PHANGS galaxy properties extracted from intern data release sample table version 1.6 with more details in \citet{Leroy20b}. 
For 80 PHANGS-ALMA galaxies (1), we list the best distances with their errors \citep[][and references therein]{Anand20} (2), galaxy central positions on the sky (3),(4) with constant uncertainties of 1\,deg, as well as galaxy disk position angles (5), inclinations (6) and galaxy systemic velocities in the LSRK frame and radio convention (7). These values are estimated by the kinematics modeling method described in \citet{Lang20}, and the up-dated version available in the sample table.
Columns~(8) and~(9) list morphological properties, such as the presence of a stellar bar \citep[][]{Querejeta2021} or AGN\footnote{We used the 13$^{th}$ edition of the Catalog of Quasars and Active Galactic Nuclei (\url{https://heasarc.gsfc.nasa.gov/W3Browse/all/veroncat.html}). Based on the \texttt{Object$\_$Type} column of the catalog, we classified galaxies with the following flags as AGN-host galaxies in this paper: S, S1h, S1i, S1n, S1.0--S1.9, S2, S3, S3b, and S3h.} \citep{VeronCatalogue13thedition}, coded with 1 $\widehat{=}$ yes and 0 $\widehat{=}$ no.
Stellar mass (10) is derived using the z=0 Multiwavelength Galaxy Synthesis catalog \citep[z0MGS;][]{Leroy2019}. They employed the \textit{WISE} 3.4$\mu$m flux and a mass-to-light ratio derived from the GALEX-SDSS-WISE Legacy Catalog \citep[GSWLC;][]{Salim20}.
The stellar masses have a characteristic $0.1\,$dex uncertainty added in quadrature with the statistical uncertainty from the maps, resulting in a $0.112$\,dex uncertainty.
Column~(11) gives extinction-corrected SFRs derived from \textit{GALEX} UV and \textit{WISE} fluxes using the prescription described in \citet{Leroy2019}.
The SFRs have a systematic uncertainty of $\sim 0.2\,$dex added in quadrature to the observational errors (e.g., the highly covariant distance uncertainty).
Column~(12) gives SFRs within the central 2\,kpc resolution element of the \textit{GALEX} and \textit{WISE} maps (i.e., central SFRs).
The central SFR or central SFR divided by total SFR are more directly linked to the properties of nuclear molecular outflows.}
\end{landscape}}

\section{PHANGS galaxy \ocl\ results}
\label{sec:Appendix:PHANGS_OCL_ALL}
We list all inspector-averaged \ocl{}s per method and per galaxy in Table~\ref{tab:PHANGSOCLtable} a well as all total averages ($\overline{\ocl}$. 
Outflow candidates are marked in column~(2).

\longtab[1]{
\begin{landscape}
\begin{longtable}{lcrrrrr}

\caption{PHANGS galaxy \ocl\ results \label{tab:PHANGSOCLtable}}\\
\noalign{\smallskip}
\hline\hline 
\noalign{\smallskip}
Name & Outfl. candidate & $\overline{\ocl}$ & Spec. 300\,pc  & Spec. 2\,kpc & pv & lwm  \\ 
 & (y/n) &  &  &  &  &  \\
\noalign{\smallskip}
(1) & (2) & (3) & (4) & (5) & (6) & (7)\\ 
\hline 
\noalign{\smallskip}
IC~1954 & 0 & 0.44 & 0.00 & 0.00 & 0.33 & 1.00\\
IC~5273 & 0 & 0.44 & 0.67 & 0.00 & 0.67 & 0.00\\
IC~5332 & 0 & 0.00 & 0.00 & 0.00 & 0.00 & 0.00\\
NGC~0253 & 1 & 2.56 & 3.00 & 2.00 & 2.33 & 2.33\\
NGC~0300 & 0 & 0.00 & 0.00 & 0.00 & 0.00 & 0.00\\
NGC~0628 & 0 & 0.11 & 0.00 & 0.00 & 0.00 & 0.33\\
NGC~0685 & 0 & 0.00 & 0.00 & 0.00 & 0.00 & 0.00\\
NGC~1087 & 0 & 1.11 & 0.67 & 0.67 & 1.33 & 1.33\\
NGC~1097 & * & 2.00 & 1.33 & 2.00 & 3.00 & 1.00\\
NGC~1300 & 0 & 1.56 & 1.33 & 0.33 & 2.00 & 1.33\\
NGC~1317 & 1 & 2.33 & 2.67 & 2.33 & 2.00 & 2.33\\
NGC~1365 & 1 & 2.67 & 0.67 & 2.00 & 3.00 & 3.00\\
NGC~1385 & 0 & 0.11 & 0.00 & 0.00 & 0.00 & 0.33\\
NGC~1433 & 1 & 2.56 & 1.67 & 0.33 & 3.00 & 3.00\\
NGC~1511 & 0 & 1.89 & 1.00 & 1.00 & 1.67 & 3.00\\
NGC~1512 & 0 & 0.89 & 0.33 & 1.00 & 1.67 & 0.00\\
NGC~1546 & 0 & 0.22 & 0.00 & 0.00 & 0.67 & 0.00\\
NGC~1559 & 0 & 0.11 & 0.00 & 0.00 & 0.33 & 0.00\\
NGC~1566 & 0 & 1.89 & 2.67 & 2.33 & 2.33 & 0.67\\
NGC~1637 & * & 2.33 & 2.33 & 0.67 & 3.00 & 1.67\\
NGC~1672 & 1 & 2.33 & 2.33 & 2.67 & 2.00 & 2.33\\
NGC~1792 & 0 & 1.00 & 0.67 & 0.67 & 1.67 & 0.67\\
NGC~1809 & 0 & 0.00 & 0.00 & 0.00 & 0.00 & 0.00\\
NGC~2090 & 0 & 0.00 & 0.00 & 0.00 & 0.00 & 0.00\\
NGC~2283 & 0 & 0.00 & 0.00 & 0.00 & 0.00 & 0.00\\
NGC~2566 & 1 & 2.44 & 2.00 & 2.00 & 3.00 & 2.33\\
NGC~2775 & 0 & 0.00 & 0.00 & 0.00 & 0.00 & 0.00\\
NGC~2835 & 0 & 0.11 & 0.00 & 0.00 & 0.33 & 0.00\\
NGC~2903 & 0 & 1.44 & 1.00 & 0.00 & 1.33 & 2.00\\
NGC~2997 & 0 & 1.11 & 0.33 & 1.00 & 1.67 & 0.67\\
NGC~3059 & 0 & 0.44 & 0.00 & 0.00 & 0.33 & 1.00\\
NGC~3137 & 0 & 0.00 & 0.00 & 0.00 & 0.00 & 0.00\\
NGC~3239 & 0 & 0.00 & 0.00 & 0.00 & 0.00 & 0.00\\
NGC~3351 & 1 & 2.00 & 0.00 & 0.67 & 2.33 & 3.00\\
NGC~3489 & 0 & 0.56 & 0.00 & 0.00 & 0.67 & 1.00\\
NGC~3507 & 1 & 2.67 & 2.67 & 0.00 & 2.33 & 3.00\\
NGC~3511 & 0 & 0.44 & 0.00 & 0.67 & 0.67 & 0.00\\
NGC~3521 & 0 & 0.11 & 0.00 & 0.33 & 0.00 & 0.00\\
NGC~3596 & 0 & 0.11 & 0.00 & 0.00 & 0.33 & 0.00\\
NGC~3599 & 0 & 0.00 & 0.00 & 0.00 & 0.00 & 0.00\\
NGC~3621 & 0 & 0.11 & 0.00 & 0.33 & 0.00 & 0.00\\
NGC~3626 & 0 & 1.11 & 0.00 & 0.00 & 2.00 & 1.33\\
NGC~3627 & 1 & 2.56 & 2.67 & 2.00 & 3.00 & 2.00\\
NGC~4207 & 0 & 0.44 & 0.67 & 0.00 & 0.67 & 0.00\\
NGC~4254 & 0 & 1.00 & 0.33 & 1.67 & 1.00 & 0.33\\
NGC~4293 & 1 & 2.00 & 1.67 & 1.00 & 2.00 & 2.33\\
NGC~4298 & 0 & 1.44 & 0.33 & 0.00 & 2.00 & 2.00\\
NGC~4303 & 0 & 1.44 & 1.67 & 0.00 & 2.00 & 0.67\\
NGC~4321 & * & 2.33 & 2.67 & 0.33 & 2.67 & 1.67\\
NGC~4424 & 0 & 0.67 & 0.00 & 1.00 & 0.00 & 1.00\\
NGC~4457 & 1 & 2.44 & 2.33 & 1.33 & 3.00 & 2.00\\
NGC~4476 & 0 & 0.11 & 0.00 & 0.00 & 0.33 & 0.00\\
NGC~4477 & 0 & 0.56 & 0.00 & 0.00 & 0.67 & 1.00\\
NGC~4496 & 0 & 0.00 & 0.00 & 0.00 & 0.00 & 0.00\\
NGC~4535 & 0 & 0.33 & 0.67 & 0.00 & 0.33 & 0.00\\
NGC~4536 & 0 & 1.78 & 1.33 & 0.00 & 2.00 & 2.00\\
NGC~4540 & 0 & 1.00 & 0.33 & 0.00 & 1.00 & 1.67\\
NGC~4548 & 0 & 1.44 & 2.00 & 2.00 & 0.67 & 1.67\\
NGC~4569 & * & 2.00 & 2.00 & 1.33 & 2.33 & 1.67\\
NGC~4571 & 0 & 0.00 & 0.00 & 0.00 & 0.00 & 0.00\\
NGC~4579 & 1 & 2.22 & 1.67 & 2.00 & 2.67 & 2.00\\
NGC~4596 & 0 & 0.11 & 0.00 & 0.00 & 0.33 & 0.00\\
NGC~4654 & 0 & 0.11 & 0.00 & 0.00 & 0.00 & 0.33\\
NGC~4689 & 0 & 0.00 & 0.00 & 0.00 & 0.00 & 0.00\\
NGC~4694 & 0 & 0.67 & 0.00 & 0.33 & 0.33 & 1.33\\
NGC~4731 & 0 & 0.00 & 0.00 & 0.00 & 0.00 & 0.00\\
NGC~4781 & 0 & 0.33 & 0.00 & 0.00 & 1.00 & 0.00\\
NGC~4826 & 0 & 0.56 & 0.00 & 0.00 & 1.67 & 0.00\\
NGC~4941 & 1 & 2.33 & 2.00 & 0.00 & 3.00 & 2.00\\
NGC~4951 & 0 & 0.78 & 0.00 & 0.00 & 1.00 & 1.33\\
NGC~5042 & 0 & 0.00 & 0.00 & 0.00 & 0.00 & 0.00\\
NGC~5068 & 0 & 0.00 & 0.00 & 0.00 & 0.00 & 0.00\\
NGC~5134 & 1 & 2.33 & 2.00 & 0.00 & 2.67 & 2.33\\
NGC~5248 & 0 & 0.11 & 0.00 & 0.00 & 0.33 & 0.00\\
NGC~5530 & 0 & 0.44 & 0.00 & 0.00 & 0.33 & 1.00\\
NGC~5643 & 1 & 2.56 & 2.33 & 0.67 & 2.67 & 2.67\\
NGC~6300 & 0 & 1.56 & 0.67 & 0.33 & 1.67 & 2.33\\
NGC~7456 & 0 & 0.00 & 0.00 & 0.00 & 0.00 & 0.00\\
NGC~7496 & 1 & 2.78 & 2.33 & 1.00 & 3.00 & 3.00\\
NGC~7743 & 0 & 0.33 & 1.00 & 0.33 & 0.00 & 0.00\\
\noalign{\smallskip}
\hline 
\noalign{\smallskip}
\end{longtable}
\tablefoot{
Inspector-averaged \ocl s per method ((4)-(7)), as well as total average \ocl{} ($\overline{\ocl}$, (3)) per galaxy (1).  
Outflow candidates according to the criteria $\overline{\ocl} \geq 2$ and $\ocl_{\lwm} \geq 2 $ are marked with yes (1) and no (0). 
Galaxies that full-filled the first criterion but not the line-wing map criterion are marked $*$}

\end{landscape}}

\section{Additional \pv-diagrams}
\label{sec:Appendix:AdditionalPlots}
We present \pv-diagrams as in Figure~\ref{fig:methods:pvexample} for all remaining outflow candidates. 

\begin{figure*}[h!]
    \centering
    \textbf{NGC~7496 and NGC~3507}\\
    \includegraphics[width = 0.49\textwidth]{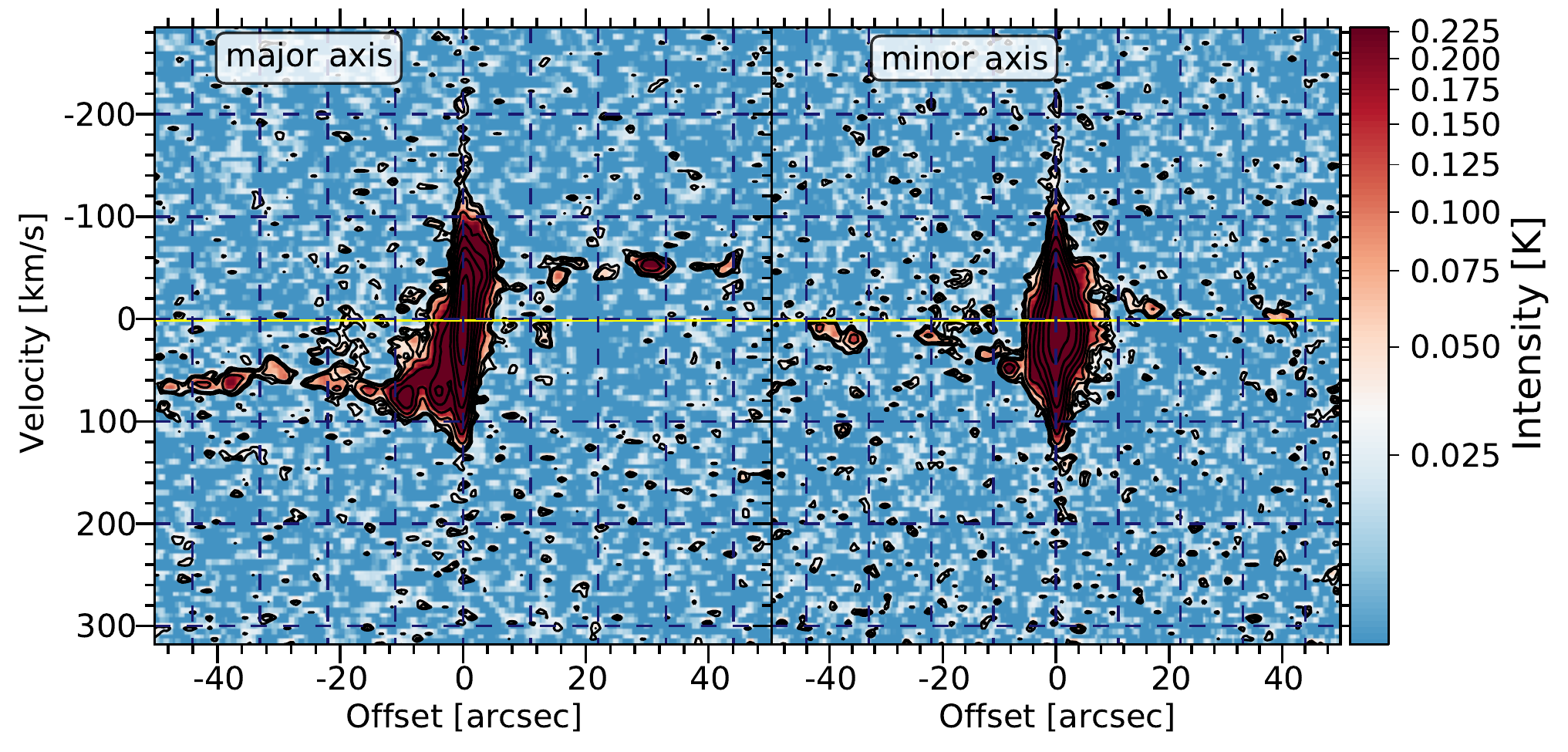}
    \includegraphics[width = 0.49\textwidth]{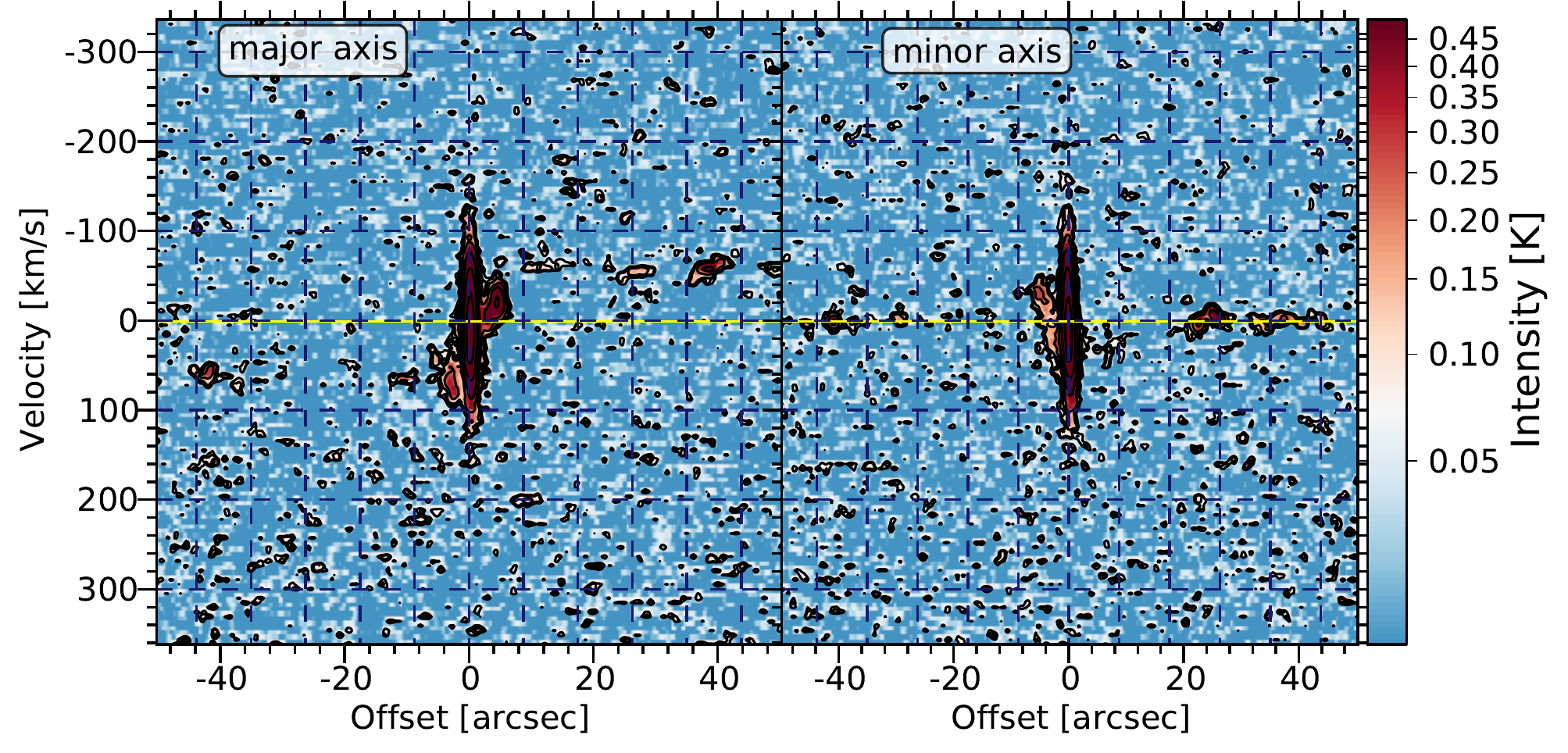}
    \textbf{NGC~1365 and NGC~1433}\\
    \includegraphics[width = 0.49\textwidth]{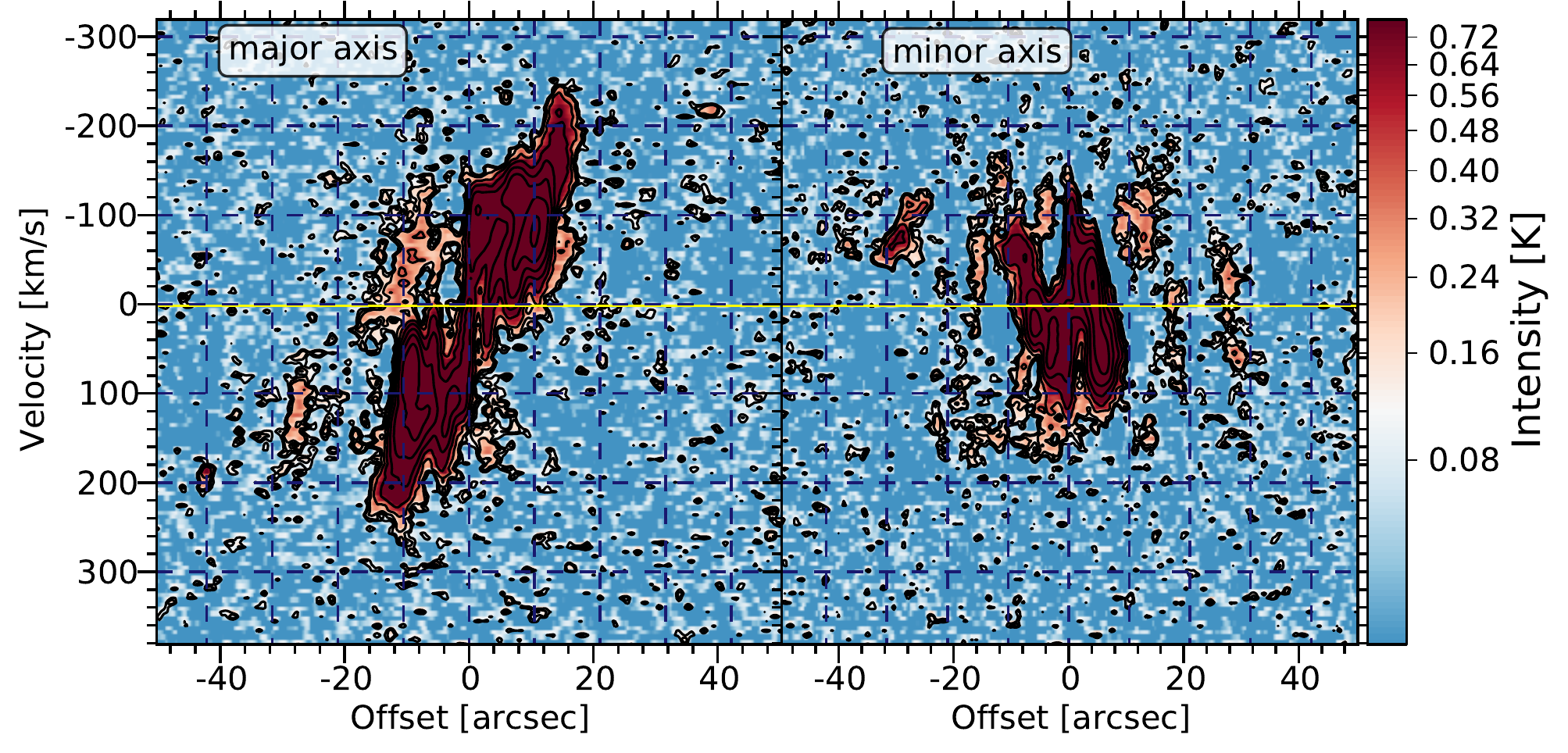}
    \includegraphics[width = 0.49\textwidth]{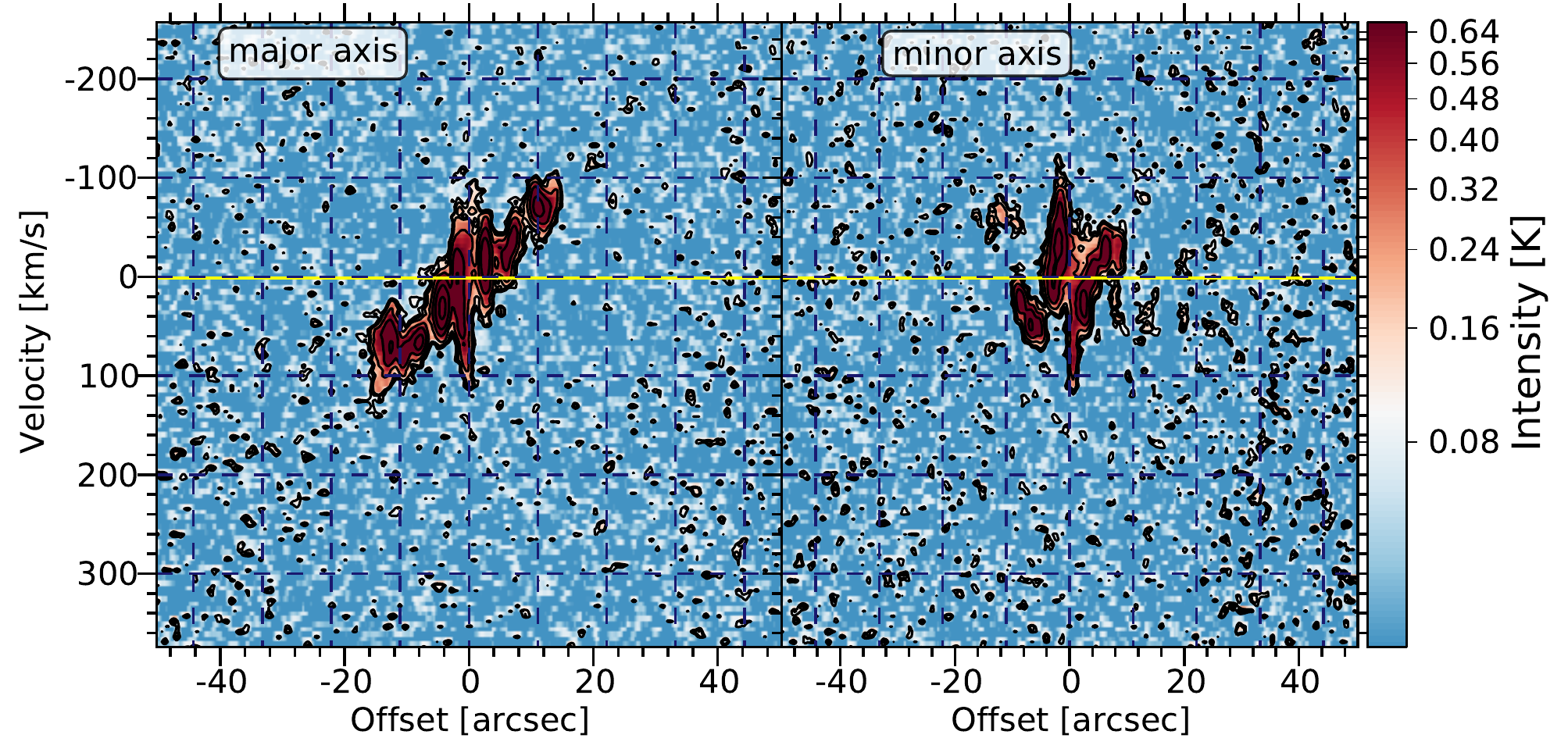}
    \textbf{NGC~0253$^\ast$ and NGC~5643}\\
    \includegraphics[width = 0.49\textwidth]{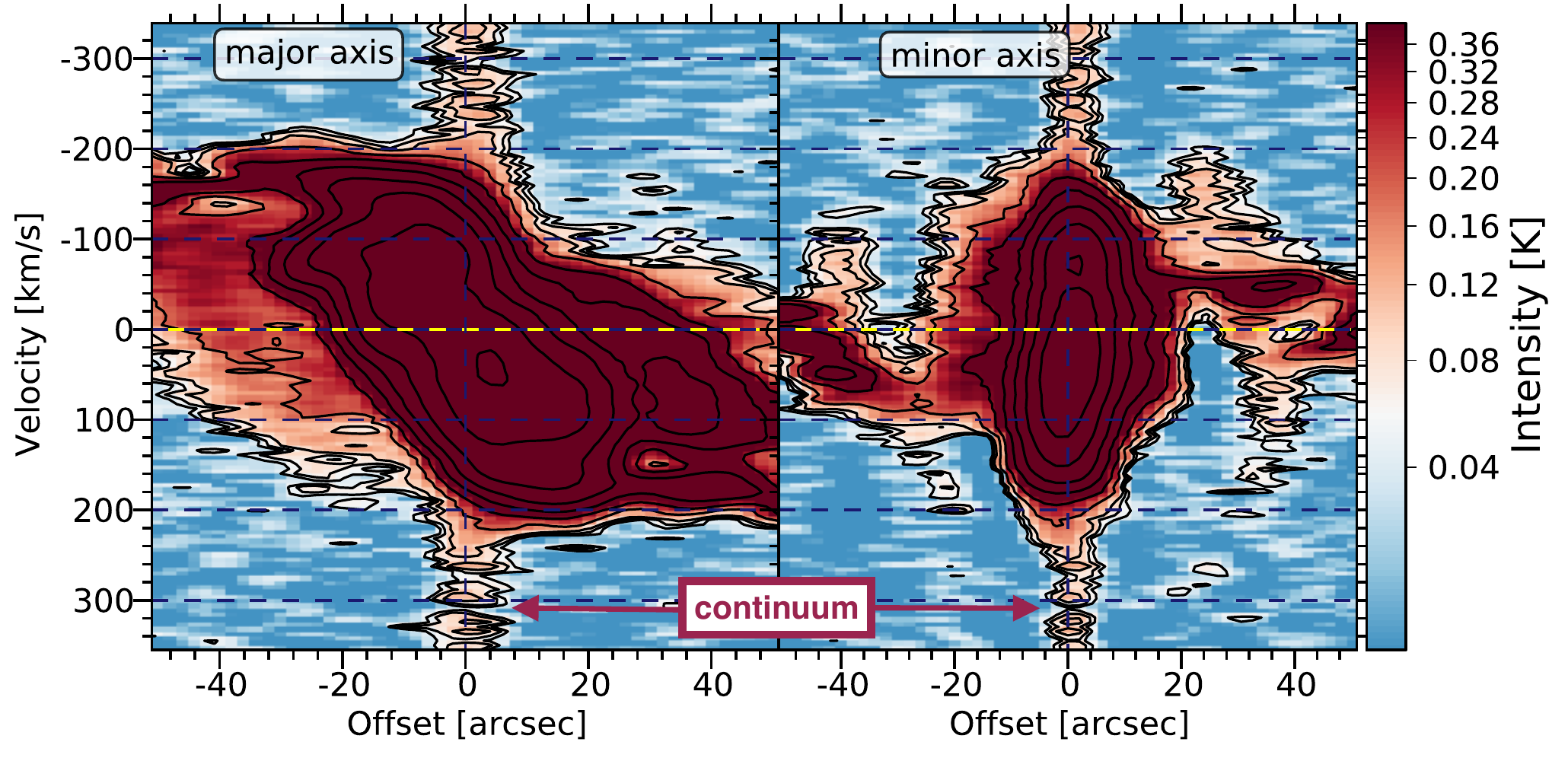}
    \includegraphics[width = 0.49\textwidth]{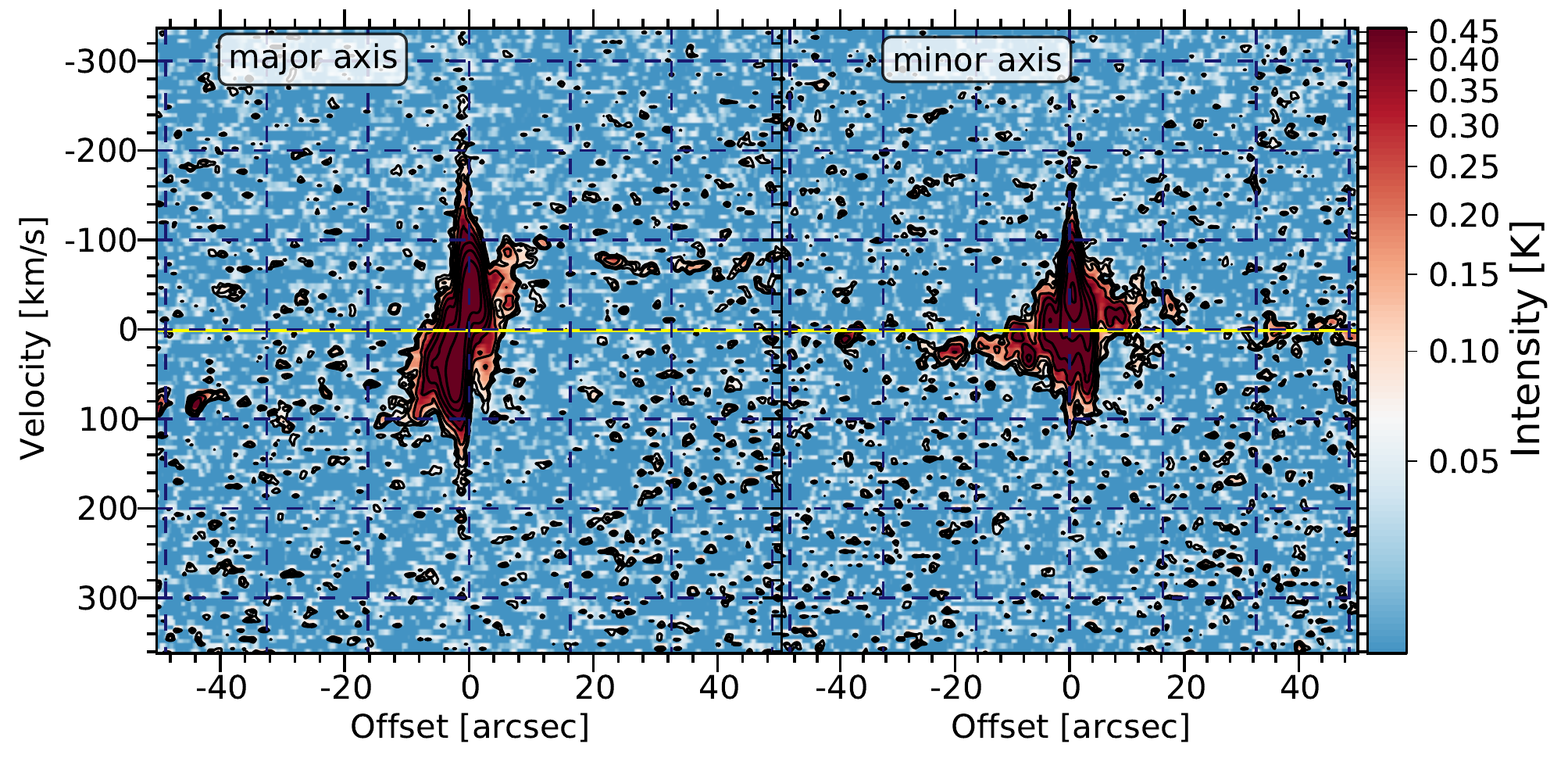}
    \textbf{NGC~4457 and NGC~2566}\\
    \includegraphics[width = 0.49\textwidth]{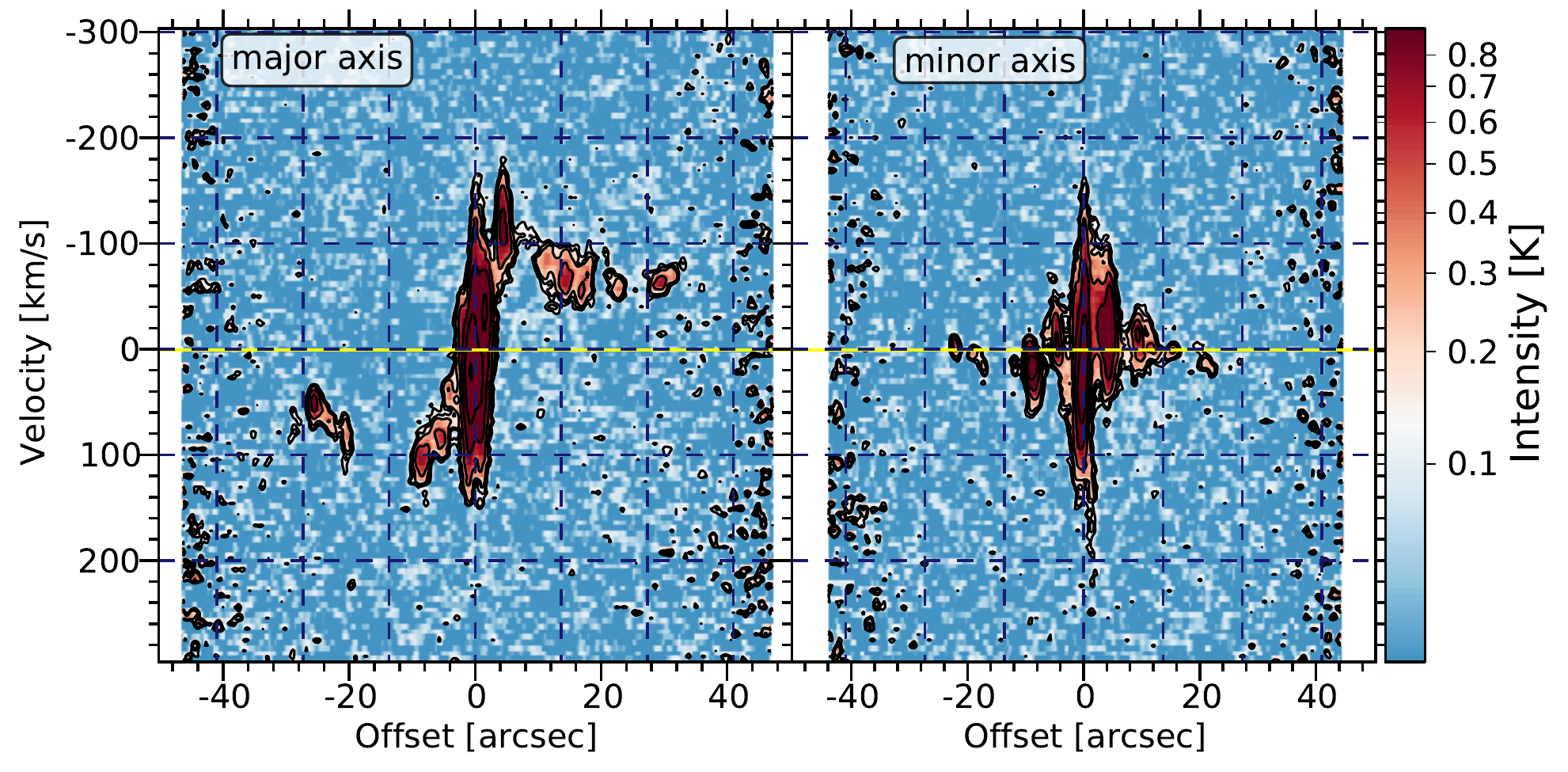}
    \includegraphics[width = 0.49\textwidth]{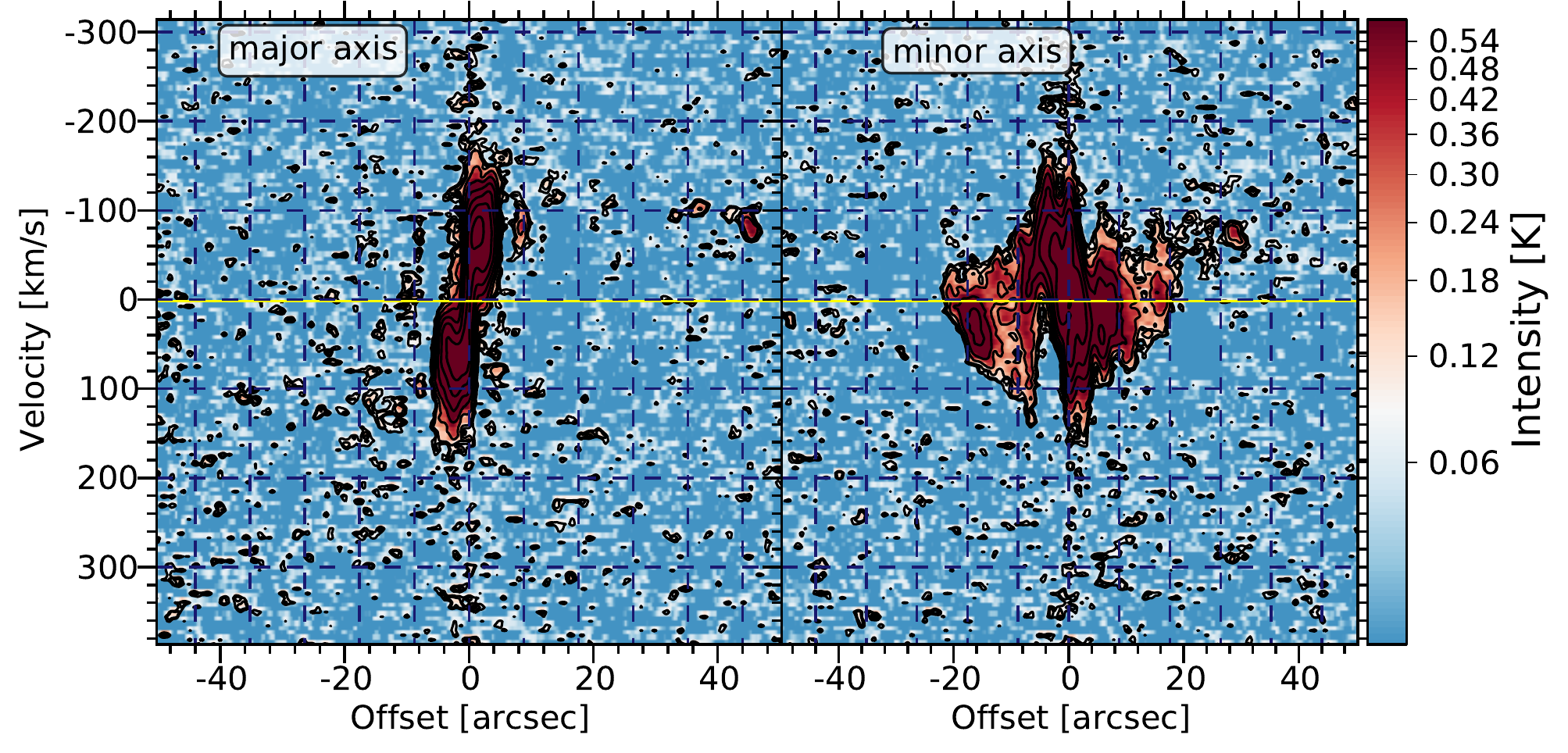}
    \textbf{NGC~5134 and NGC~4941}\\
    \includegraphics[width = 0.49\textwidth]{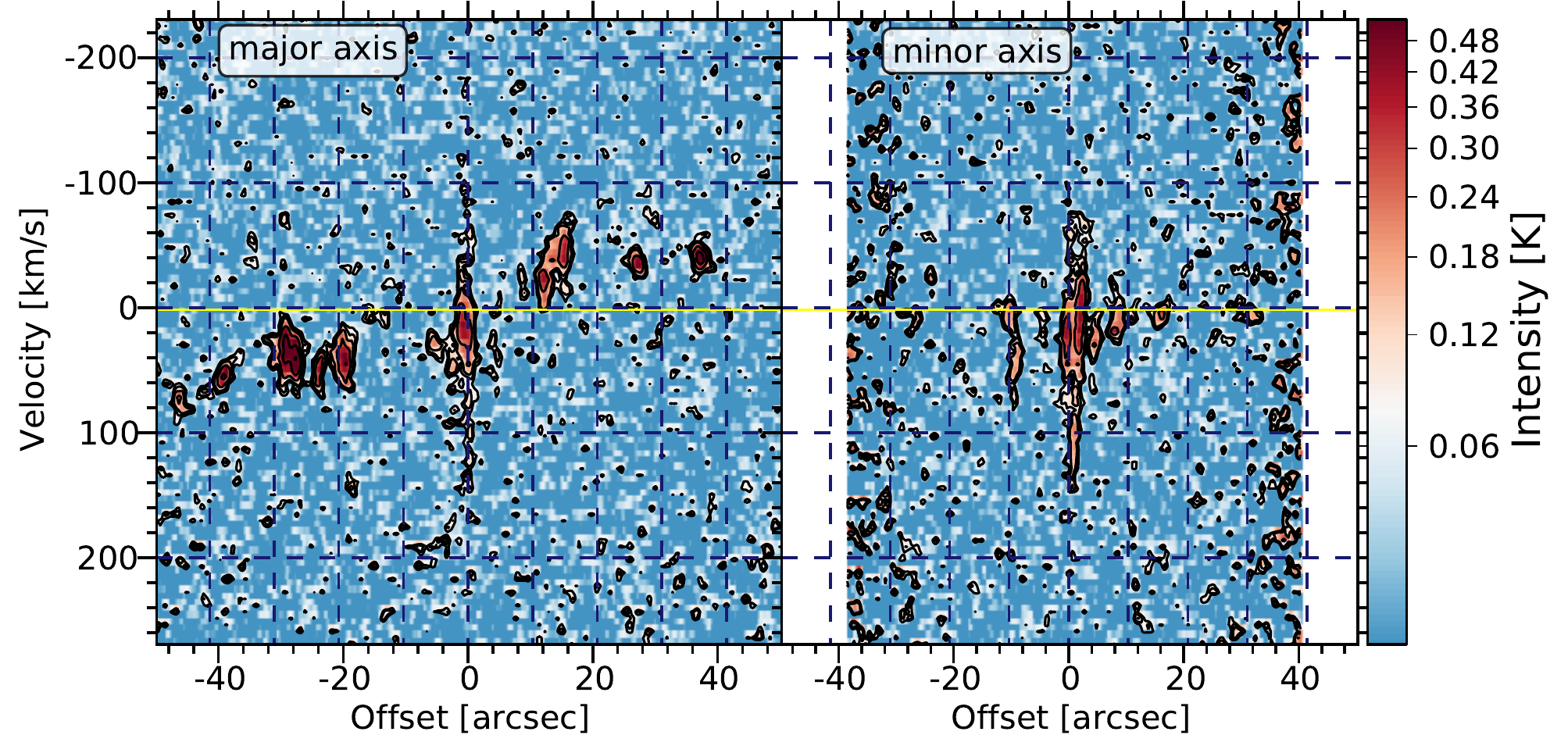}
    \includegraphics[width = 0.49\textwidth]{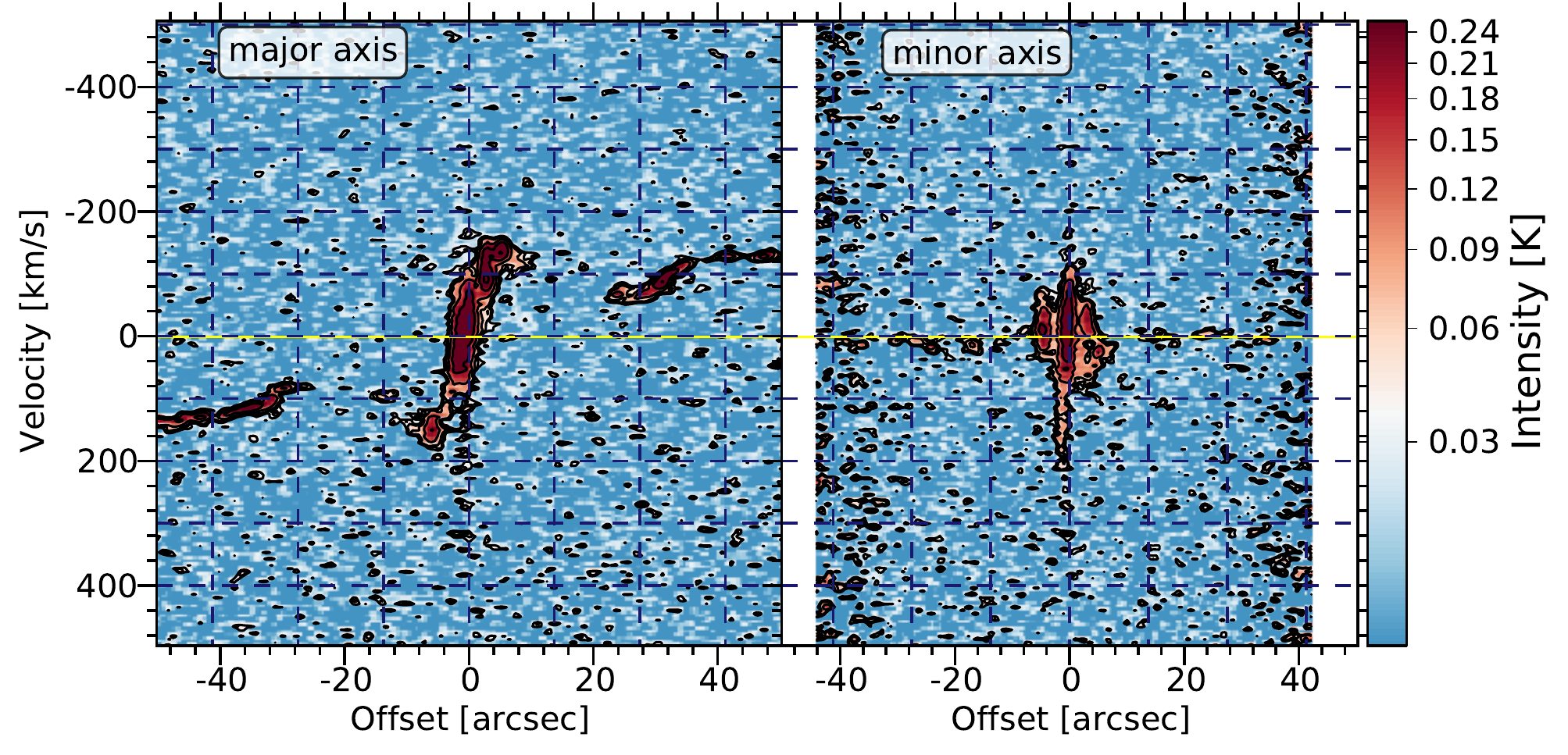}
    \caption{same as Figure~\ref{fig:methods:pvexample} for the remaining outflow candidates. Galaxies marked with $^\ast$ have continuum emission at offset = 0 due to imperfect subtraction in the data pipeline. This issue is fixed in the publicly available data release.}
    \label{fig:AdditionalPVDiagramsI}
\end{figure*}
\begin{figure*}\ContinuedFloat
    \centering
    \textbf{NGC~4321 and NGC~1672}\\
    \includegraphics[width = 0.49\textwidth]{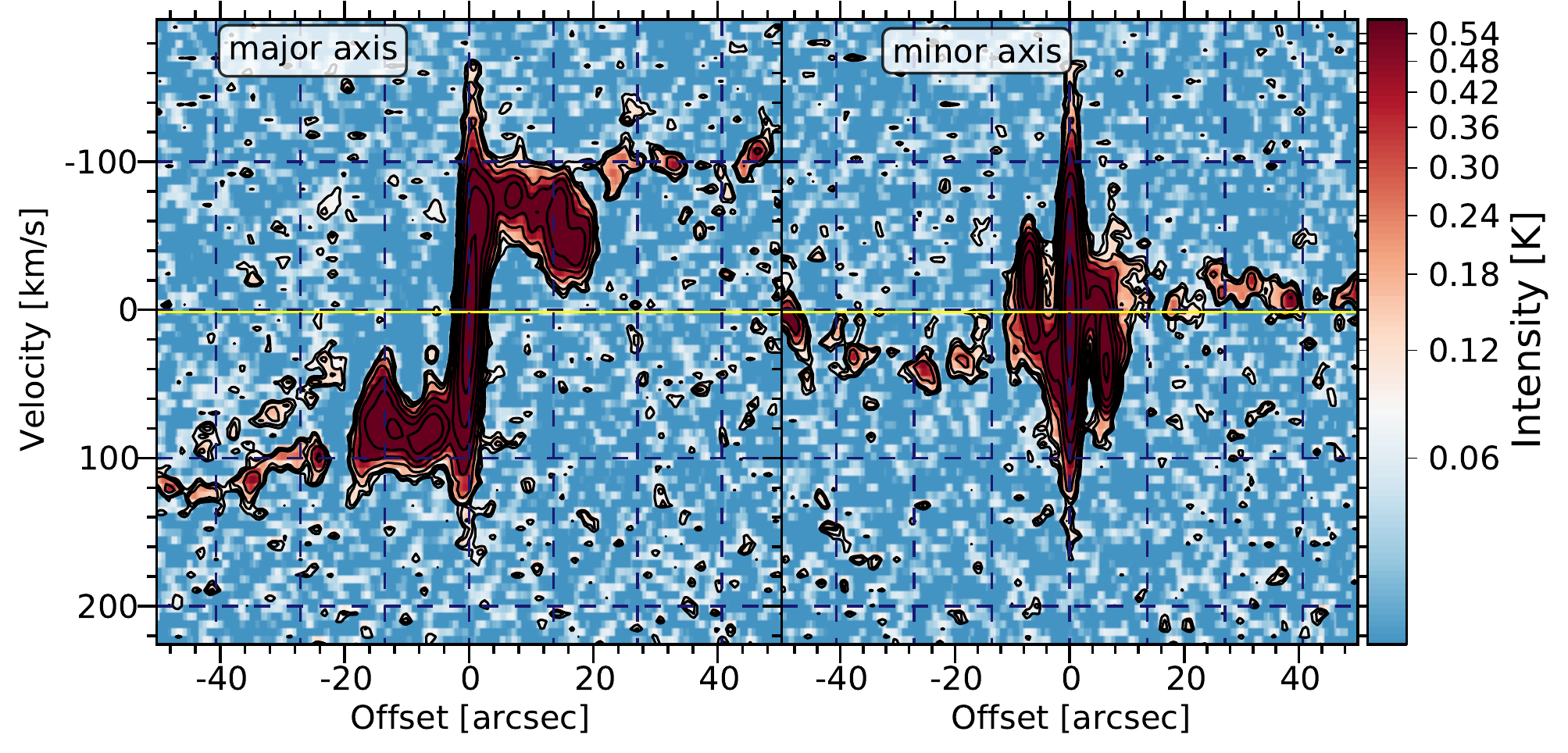}
    \includegraphics[width = 0.49\textwidth]{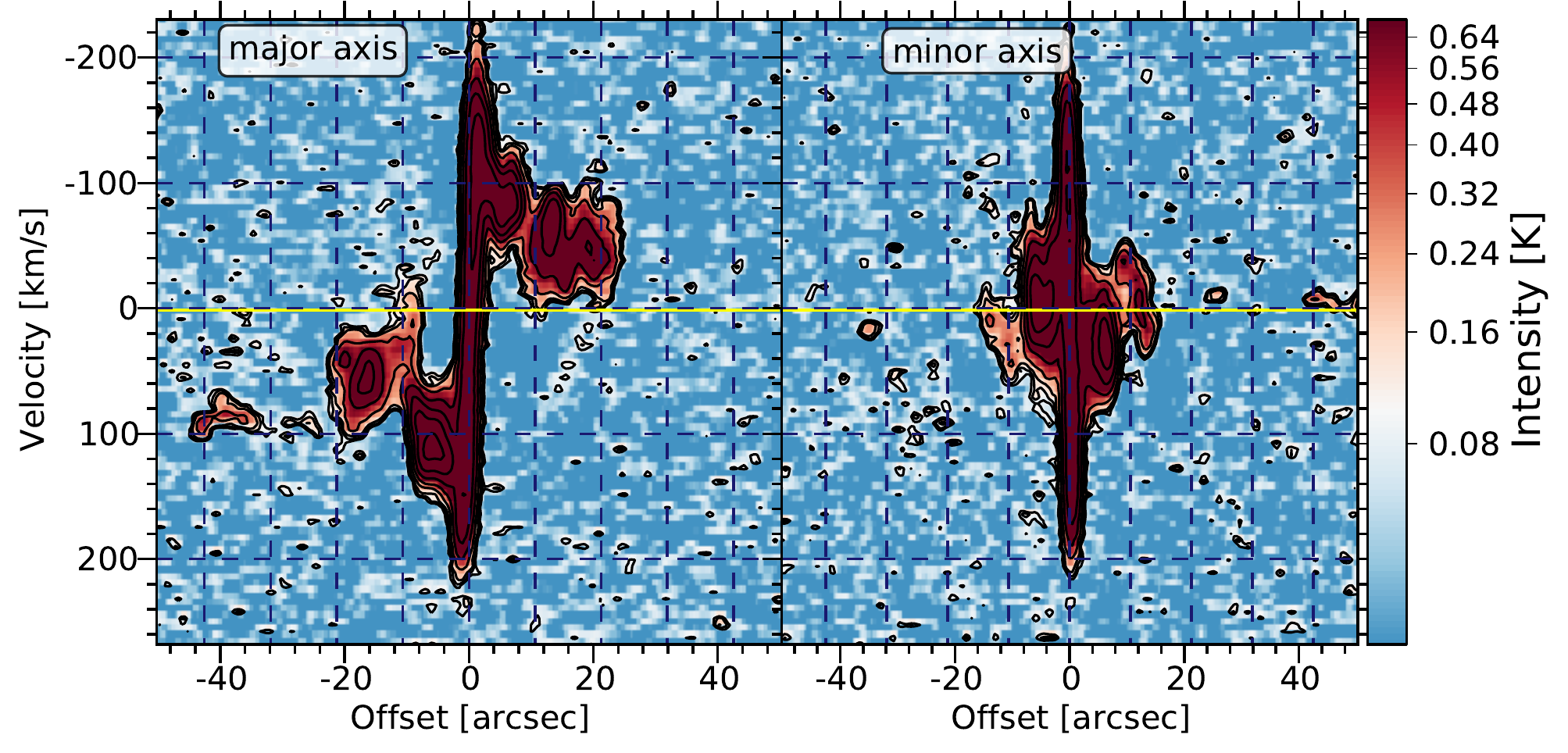}
    \textbf{NGC~1637 and NGC~1317}\\
    \includegraphics[width = 0.49\textwidth]{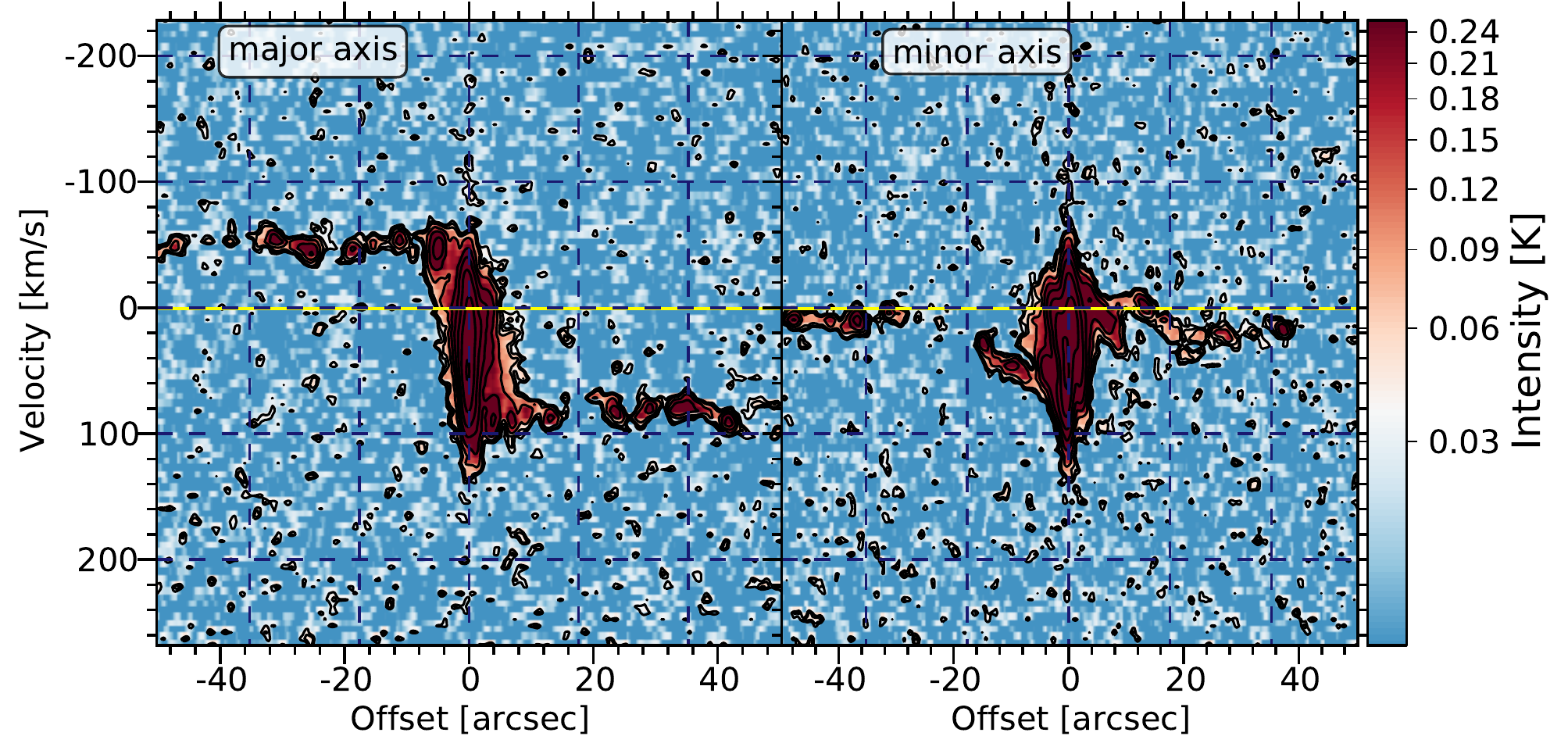}
    \includegraphics[width = 0.49\textwidth]{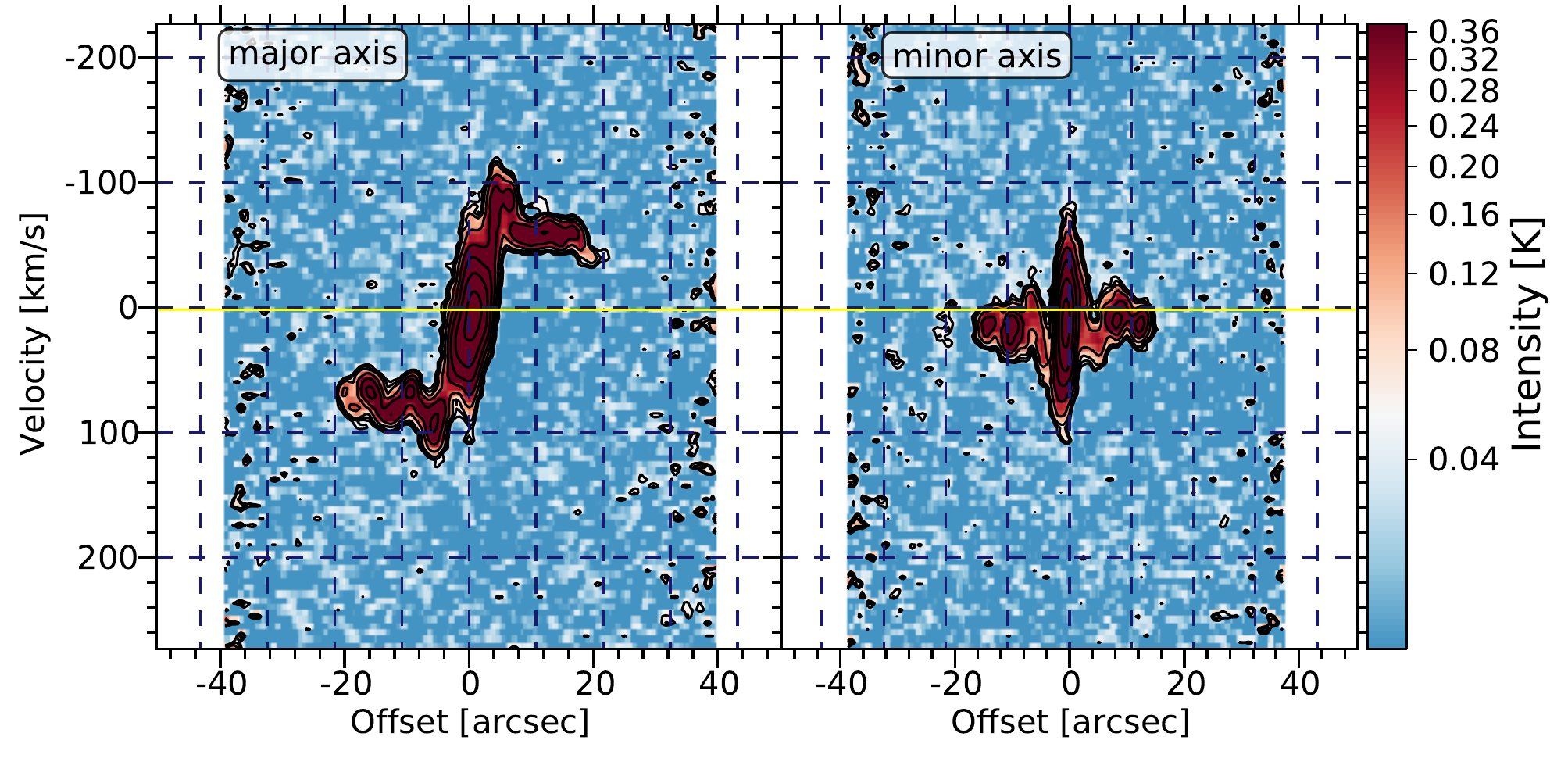}
    \textbf{NGC~4579$^\ast$ and NGC~4569}\\
    \includegraphics[width = 0.49\textwidth]{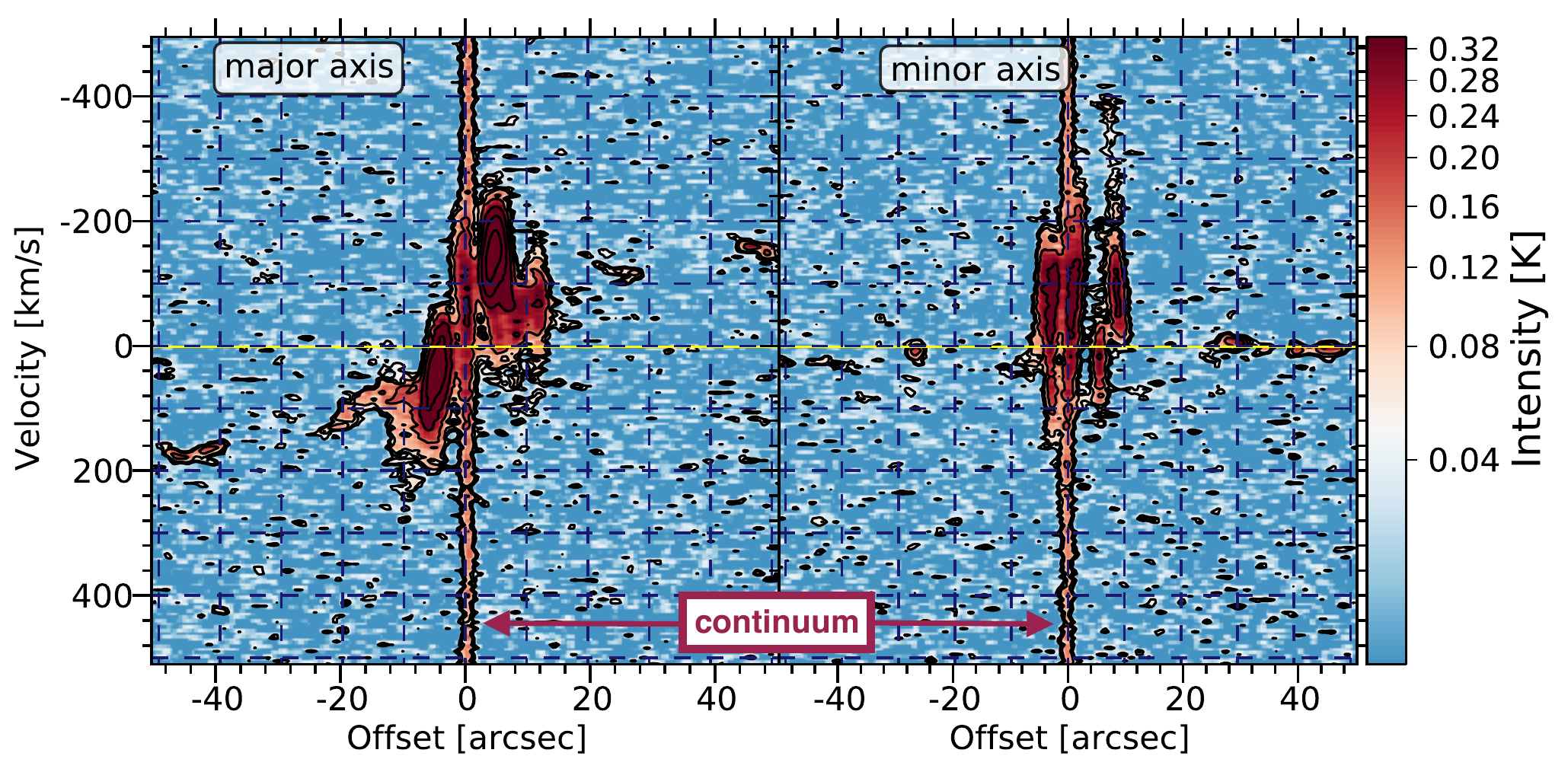}
    \includegraphics[width = 0.49\textwidth]{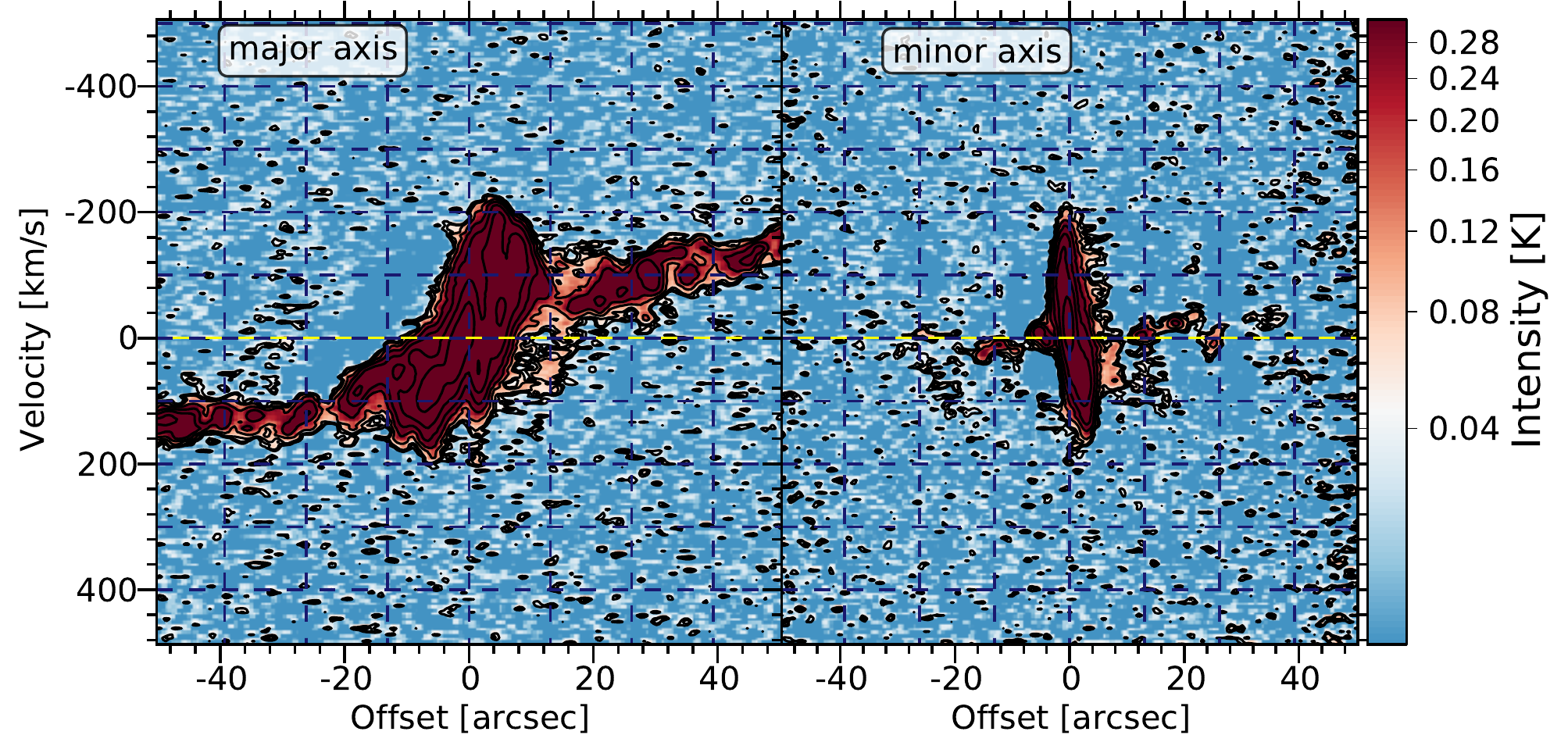}
    \textbf{NGC~2903 and NGC~3351}\\
    \includegraphics[width = 0.49\textwidth]{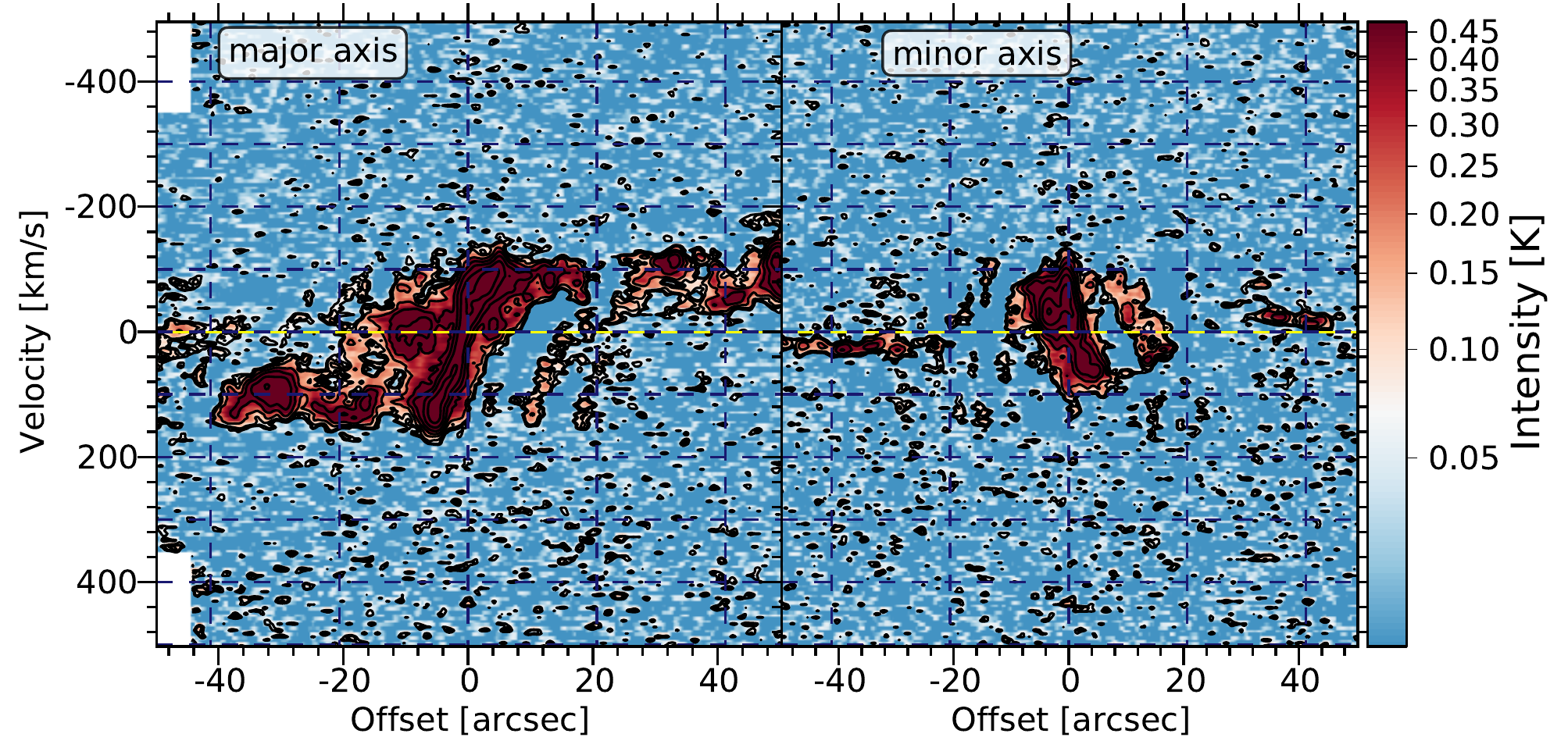}
    \includegraphics[width = 0.49\textwidth]{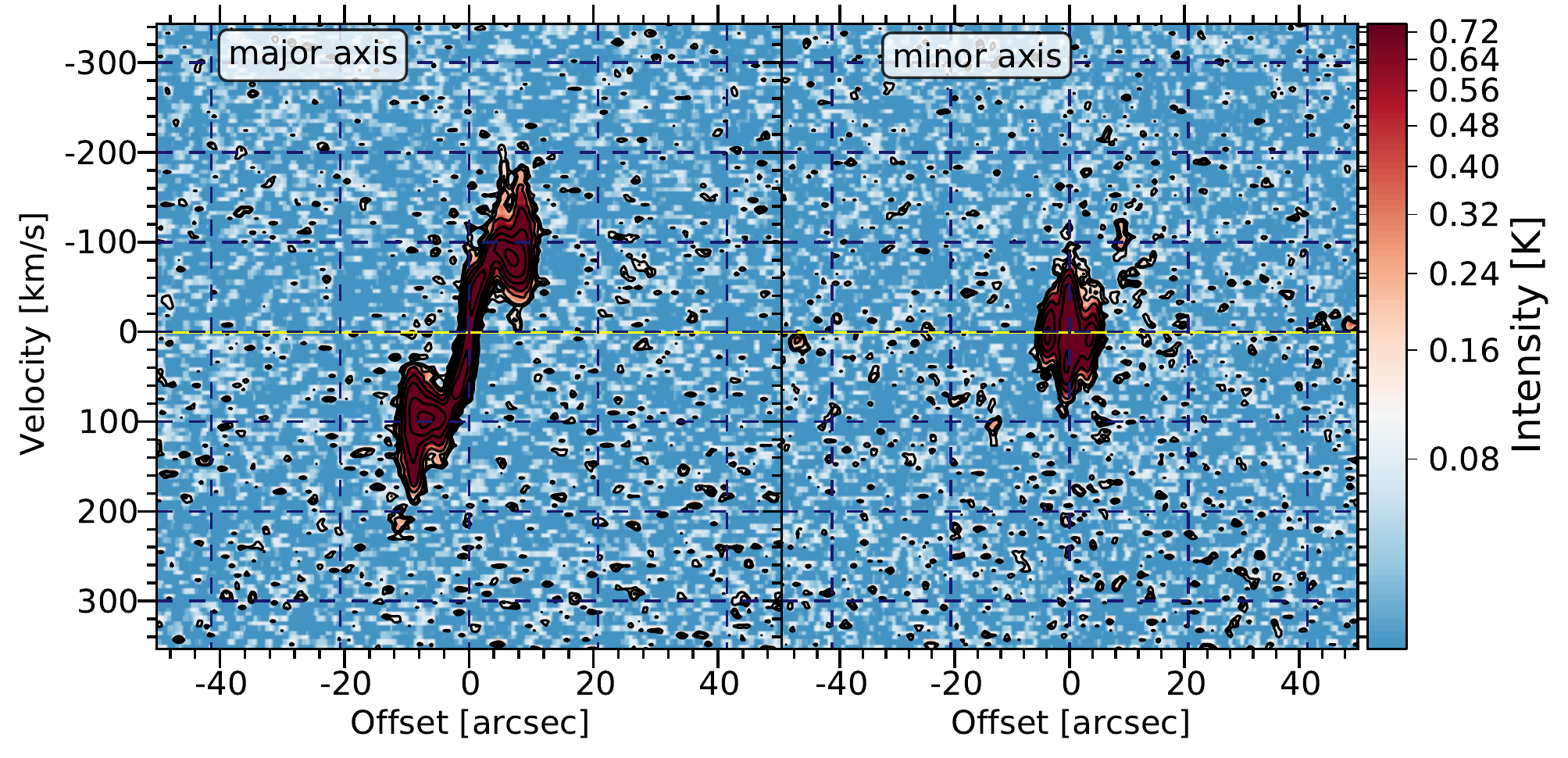}
    \textbf{NGC~1097$^\ast$}\\
    \includegraphics[width = 0.49\textwidth]{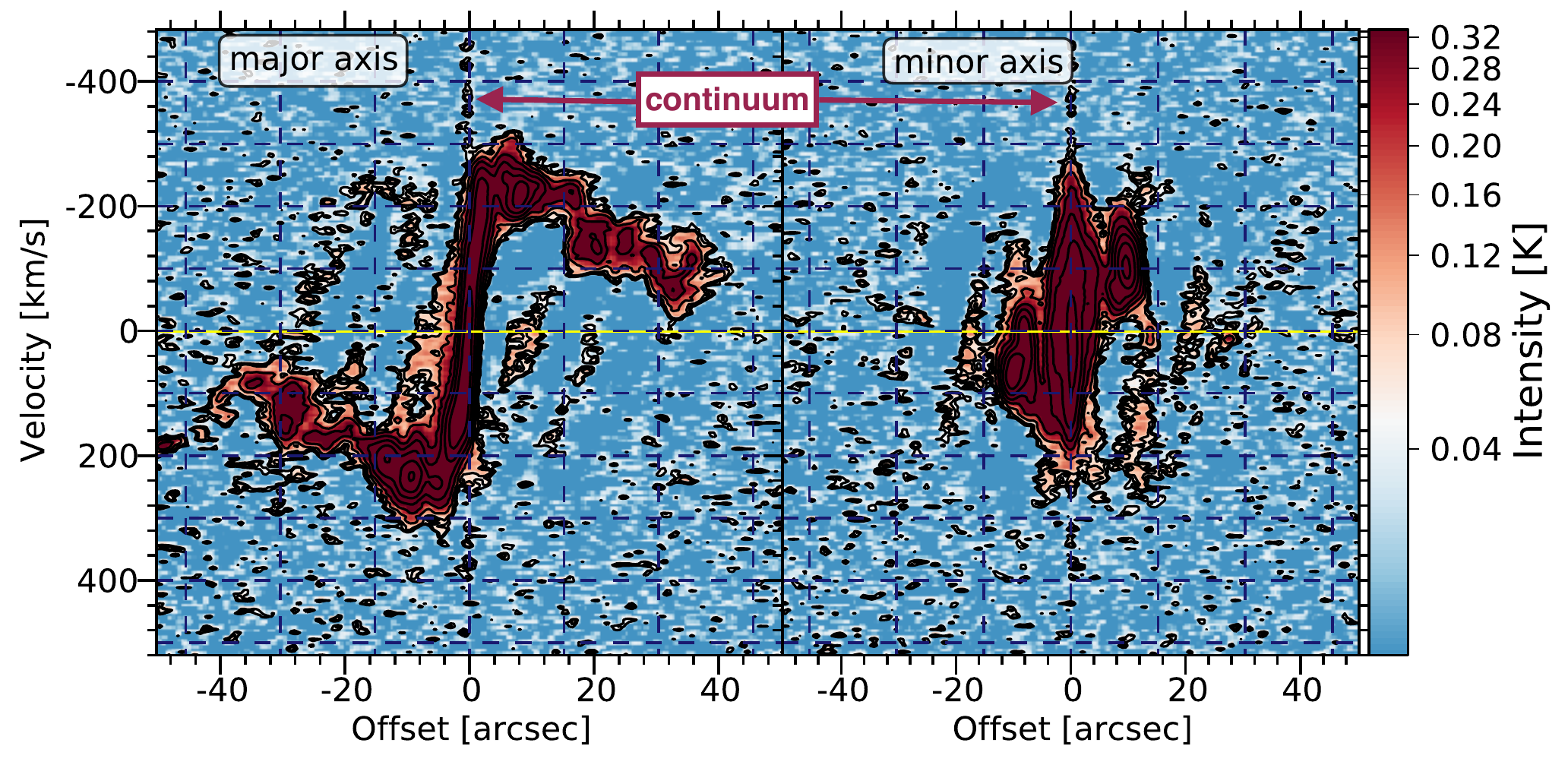}
    \caption{same as Figure~\ref{fig:methods:pvexample} for the remaining outflow candidates. Galaxies marked with $^\ast$ have continuum emission at offset = 0 due to imperfect subtraction in the data pipeline. This issue is fixed in the publicly available data release.}
    \label{fig:AdditionalPVDiagramsII}
\end{figure*}

\end{appendix}

\end{appendix}

\end{document}